# QUATERNIONS: A HISTORY OF COMPLEX NONCOMMUTATIVE ROTATION GROUPS IN THEORETICAL PHYSICS

by

Johannes C. Familton

A thesis submitted in partial
fulfillment of the requirements for the
degree of

Ph.D

Columbia University

2015

Approved by ______________________________
Chairperson of Supervisory Committee

______________________________

______________________________

______________________________

Program Authorized
to Offer Degree ______________________________

Date ______________________________

COLUMBIA UNIVERSITY

# QUATERNIONS: A HISTORY OF COMPLEX NONCOMMUTATIVE ROTATION GROUPS IN THEORETICAL PHYSICS

By Johannes C. Familton

Chairperson of the Supervisory Committee:

Dr. Bruce Vogeli and Dr Henry O. Pollak
Department of Mathematics Education

TABLE OF CONTENTS

LIST OF FIGURES

LIST OF TABLES

# ACKNOWLEDGMENTS

This thesis is dedicated to Dr. Richard Friedberg Professor emeritus of Barnard College, Columbia University who first person to introduce me to quaternions as an undergraduate in an independent study course on complex analysis.

The author wishes to express sincere appreciation to Dr. Bruce Vogeli who was there for me through thick and thin, and there was a lot of thin, throughout the Teacher's College program. I would also like to thank Dr. Henry Pollak for his advice and concern along the way.

I would also like to thank my readers for their time and suggestions on the source book part of this thesis: Dr. Margaret (Peggy) Dean, Dr. Avraham Goldstein, Dr. Chris McCarthy and Dr. Marcos Zyman, colleagues from Borough of Manhattan Community College (BMCC); Dr. Terence Blackman of the University of Denver who did his PhD. from a pure mathematics point of view in quaternions, and Dr. Richard Friedberg to whom this thesis is dedicated.

# CHAPTER I

## INTRODUCTION

In recent years the introduction of quaternions on various, perhaps in disguised forms, has become part of the education of undergraduate mathematics and physics students; yet most of these undergraduate students learn this subject without encountering any historical or scientific motivation from which the mathematics evolved. Despite the importance of 'Quaternion type Rotation Groups' in modern physics--usually in the guise of the rotation groups: SU(2), SO(4), and Grassmann Algebras--students of modern physics have little or no exposure about how these mathematical objects came about. The purpose of this dissertation is to clarify the emergence of quaternions in order to make the history of quaternions less opaque to teachers and students in mathematics and physics.

The journey of quaternions started as a geometric and algebraic curiosity in the mid-19$^{th}$ century. Soon they were found to have applications in mechanics; then later that century they were applied to electromagnetism via Maxwell's equations. The physics of the past century: quantum mechanics and relativity, string theory, super-symmetry and quantum gravity, also found uses for quaternion type-rotation groups. For example, students taking a course in undergraduate quantum mechanics or modern physics encounter complex non-commutative rotation groups in the guise of Pauli matrices (Griffiths,1994, p.156; Sakurai, 1994, p.168). In mathematics this subject is generally skipped over. If mathematics students do encounter quaternions or their associated rotation groups it would be in the form of exercises or part of an appendix in a modern algebra course (Birkhoff & Maclane, 1977, p. 258 ; Artin, 1991, p.123 and p.155).

**A. Need for the study**

Although the history of the quaternion rotation group is well documented, often it is not made clear how rotation groups were introduced into physics in the first place and why they took such deep a hold, especially in modern physics.

While there are many authoritative works for physics professionals, little is available that is appropriate for others. Moreover, the rich history of quaternions is rarely mentioned, even in advanced technical works, due to the limited availability of this information. The history of quaternions and their associated rotation groups as it relates to physics does not have a unified source from which college instructors of mathematics and physics can draw. The need for this sourcebook was brought to the author's attention by a couple of the readers of the sourcebook. Thus this sourcebook would allow instructors to incorporate historical materials into their teaching of rotation groups that would be useful addition to curricular resources.

B. **Purpose of the study**

The purpose of this study is to prepare a sourcebook for mathematics and physics instructors providing an historical perspective of quaternions and their cousins the rotation groups SU(2), Special Unitary Group 2, SO(4), Special Orthogonal group 4, both types of rotation groups occur frequently in advanced physics courses, especially in quantum mechanics or particle physics courses. Other subjects that come up in physics courses such as Clifford algebras and Grassmann algebras also incorporate quaternion structures. This study will seek to provide:

1. The historical development of rotation groups in their original guise of quaternions and how they developed into the more familiar rotation groups students encounter today.

2. The history of the incorporation of quaternions and rotation groups into classical physics: mechanics and electromagnetism.

3. The history of the incorporation of quaternions and rotation groups into modern physics: quantum mechanics and relativity theory.

A source book of this kind should be useful to historians of science; as well as mathematicians, physicists and educators who want to integrate the historical development of complex non-commutative rotation groups into their teaching in order to give a deeper understanding of the subject matter.

The historical goals of the sourcebook required an extensive review of original literature. Some of this literature was originally written in Latin, the language of educated people up through the 19th century, and English. The people who discovered and worked with quaternions, William Rowan Hamilton and his followers, were English speakers and wrote most of their work in English. No translations are needed in these cases, and direct quotations will be used. The translations of non-English sources presented here are the available academic translations and noted in the Bibliography.

The philosophical perspective, the technical language, and the notation changes as quaternions evolve into Grassmann algebras and into the theory of non-commutative complex rotational groups, so following the historical development is especially challenging in the absence of a source book.

**C. Procedures of the study**

In order to fulfill the purpose of this study, the following strategies were followed:

1. Present the history of the development of complex non-commutative rotation groups as a purely mathematical development.

2. Show how these developments initially became incorporated into theoretical physics in the first place.

3. Show how these developments became incorporated into current theoretical physics literature and textbooks.

4. The source book has been submitted to a selection of mathematicians and physicists who both teach and do research in quaternions and associated areas.

5. The evaluations obtained were summarized, as well as suggestions on how the source book can be amended.

6. This dissertation includes the initial questionnaire and answers. The author questioned the participants further and included a summary of their responses and anecdotal assessments in Chapter V.

7. The final amended version of the source book is included in this dissertation in Chapters II-IV

All sourcebook technical exposition will be at a level appropriate for a well-prepared instructor of undergraduate mathematics and physics who is familiar with these types of rotation groups and wants to have a better historical sense about what they are teaching. Far too often instructors are only aware of the polished version what they are teaching, thus giving students the illusion that this is the way these structures were from the beginning, when they were first discovered. Students are rarely exposed to in their textbooks or by their instructors the years of struggle that had gone into what they are learning.

The purpose of this sourcebook was to initially prepare a draft for further review by a 'jury' of readers who are familiar with the subject or related subjects in order to insure the quality and usefulness of this sourcebook. This 'field trial' consisted of six instructors who gave encountered this subject matter in their teaching or research. The reports of the jurors was the basis for refinement of the preliminary version of the source book.

# Chapter II
# HISTORICAL DEVELOPMENT OF QUATERNIONS

Quaternions have become a common part of mathematics and physics culture, but little is discussed about how quaternions came into being. In this chapter the author will discuss the history of quaternions and their relatives. These types of rotation groups have become a part of the mathematics and physics diet of the $21^{st}$ century.

The obvious question is "What are quaternions?" The next question asked is "How and why quaternions were invented in the first place?" In this chapter the author answers some of these questions.

## A. What Are Quaternions?

A complex number, is defined by $a + bi$, where $a$, $b$ are real numbers and $i$ is called an imaginary number. The number $i$ is not part of the real number system and has the property $i^2 = -1$. The idea of a square root of $-1$ as it comes up in an equation of the form $x^2 + a = 0$ when solving for x. This is the reason that complex numbers came into existence in the first place. Complex numbers are defined as being a two dimensional vector space over the real numbers. A basis vectors of this vector space is $\{1, i\}$. Complex numbers are added component-wise: $(a+bi) + (c+di) = (a+c) + (b+d)i$; multiplication is determined by the distributive law and $i^2 = -1$: $(a+bi)(c+di) = ac + adi + bci+bdi^2 = (ac-bd) + (ad+bc)i$.

A quaternion is an extension of the complex number system. Quaternions are called this because they have four basis vectors $(1, i, j, k)$. The word quaternion itself however was not coined by Hamilton. In fact according to the Merriam-Webster dictionary quaternion is a middle-English word, referring to a set of four parts/persons/things. Hamilton, according to his biographer Robert Graves, discussed the entomology of the word quaternion

> The word "Quaternion" requires no explanation, since, although not now very commonly used, it occurs in the Scriptures and in Milton. Peter was delivered to "four quaternions of soldiers" to keep him; Adam, in his morning hymn, invokes air and the elements "which in quaternion run." The word (like, the Latin "quaternio," from which it is derived) means simply a set of four, whether those "four" be persons or things.(Graves, 1889, p.635)

Quaternions have three imaginary components called $i, j, k$ that are three different square roots of $-1$, meaning $i^2=j^2=k^2=-1$. Quaternions can be also written as $(q_0, q_1, q_2, q_3)$ or $q_0+q_1i+q_2j+q_3k$ are complex coordinates and $q_0$, $q_1$, $q_2$, $q_3$ are real coordinates.

Now that quaternions have been defined the next question would be how to add quaternions. Addition with quaternions is done component-wise the same way it is done with complex numbers. With the multiplication of quaternions things become a bit more complicated. This is what held Hamilton up for more than 10 years. The key to Multiplying quaternions turned out to be: $ij=k=-ji$, $jk=i=-kj$, $ki=j=-ik$. This rule makes multiplication non-commutative since *ij* is not the same as *ji*, meaning that $ij \neq ji$ etc. As far as is known, William Rowan Hamilton was one of the first to look at a non-commutative system of numbers (Lambek, 1995, p. 8).

**B. William Rowan Hamilton and the discovery of Quaternions:**

The main difficulty, as mentioned, in the development of quaternions was how to define multiplication. Hamilton was looking for a way to formalize 3 points in 3-space in the same way that 2 points can be defined in the complex field, **C** (Lam, 2003, p. 230). Hamilton thought about this problem for over 10 years. As legend goes, on October 16, 1843, Hamilton solved the multiplication problem while taking a walk with his wife on what is today the Broom Bridge in Dublin, Ireland. His insight was that $i^2 = j^2 = k^2 = ijk = -1$. Hamilton was so excited about his insight that it is said that he stopped and carved this formula on Broom Bridge. The carving can no longer be seen, but where it is believed he carved the equation, a plaque has been placed. (Hamilton, 1854, p. 492-499, p. 125-137, p. 261-269, p. 446-451, p. 280-290,).

It was known in Hamilton's time that complex numbers correspond to points in a two dimensional space or plane. When in the complex number field multiplying them causes a rotation of the plane. The idea of points in a 2-dimensional space was already known in the 17th century by Descartes. The problem that Hamilton encountered for over a decade was that he was trying to find a more general kind of number that could extend points into a 3-dimensional space, where there was not a need for coordinates to describe them.

Recall that complex numbers are written $(a+bi)$ where $i^2 = -1$. Hamilton had tried to extend this by first writing $(a+bi+cj)$ where $i^2 = j^2 = -1$. The issue that came up was that in order to get a formula for multiplication he would have to decide what the product is, say, of $ij$ . He knew that the formula would be inconsistent unless it satisfied what he called 'the law of moduli'. This means that that when complex numbers are multiplied $(a + bi)(c + di) = (ac-bd) + (ad+bc)i$. Hamilton was looking for a 3-dimensional analogy to what occurs in the 2-dimensional complex plane. As it turned out what Hamilton was really looking for was something analogous to 'triplets', where if $(a+bi+cj)(d+ei+fj)=u+vi+wj$ then the 'law of moduli' would require that $(a^2+b^2+c^2)(d^2+e^2+f^2)=u^2+v^2+w^2$. This was the issue that Hamilton was confronted with. No doubt he tried many different possible configurations but couldn't find one that would work for his particular system (Leng, 2011, p. 63).

About his struggle trying to find a solution to 'triplets' Hamilton wrote in 1865 to his son "Every morning in the early part of the above-cited month (October 1843), on my coming down to breakfast, your (then) little brother William Edwin, and yourself, used to ask me: `Well, Papa, can you multiply triplets?' Where to I was always obliged to reply, with a sad shake of the head: `No, I can only add and subtract them.'" (Halberstam, 1967). Hamilton's difficulty was that he couldn't find a three-square identity. The reason being is that a three-square identity doesn't exist.

Hamilton was, no doubt, aware of the four-square identity discovered by Euler in 1749. This identity states $(a_1^2+a_2^2+a_3^2+a_4^2)(b_1^2+b_2^2+b_3^2+b_4^2)=(a_1b_1-a_2b_2-a_3b_3-a_4b_4)^2+(a_1a_2+a_2b_1+a_3b_4-a_4b_3)^2+(a_1b_3-a_2b_4+a_3b_1+a_4b_2)^2+(a_1b_4+a_2b_3-a_3b_2+a_4b_1)^2$ (Weisstein, 2002, 952). What could have occurred to Hamilton while walking with his wife on October 16 is that he might have better success if he used 'quadruples', $(a+bi+cj+dk)$, rather than 'triples'.

If he let $i^2 = j^2 = k^2 = -1$, he then needed to determine the products $ij$, $jk$, etc. He might have tried to do it in such a way that the multiplication formula would correspond to the expressions in Euler's identity. Perhaps that is what could have gone through Hamilton's mind at the time, is that he could accomplish what he was looking for by letting these products be non-commutative, in other words: *ij=-ji=k, jk=-kj=i, ki=-ik=j*, leading to 'the law of moduli'. Thus in order to preserve distance, absolute value, Hamilton had to give up commutativity. It was these equations that he is believed to have carved on Broom Bridge that October day (Hamilton , 1854, p. 492-499, p. 125-137, p. 261-269, p. 446-451, p. 280-290).

Hamilton's interest in quaternions is said to have developed from his interest in algebra and by reading Kant's *Essay on Algebra as the Science of Pure Time* written 1835.(Steffens, 1981, p.843-844) Hamilton wrote "Time is said to have only one dimension, and space to have three dimensions…The mathematical quaternion partakes of both these elements; in technical language it may be said to be 'time plus space', or 'space plus time': and in this sense it has, or at least involves a reference to, four dimensions. And how the One of Time, of Space the Three, Might in the Chain of Symbols girdled be." (Graves, 1889, p.635). It appears by this quote that Hamilton may have had some kind of insight into a connection between space and time, and its relationship to quaternions, but this may not be how we understand space-time today.

The influence of Kant's work on Hamilton has been debated. This debate is discussed by Michael Crowe in his book *A History of Vector Analysis*

> "It is generally believed that Hamilton's stress on time was derived from Kant. Such may not be the case, for Kant's name is never

> mentioned in the paper…In 1835 Hamilton wrote: ‘and my own convictions, mathematical and meta-physical, have been so long and so strongly converging to this point (confirmed no doubt of late by the study of Kant's Pure Reason), that I cannot easily yield to the authority of those other friends who stare at my strange theory.’ (Graves, 1835, p. 142) It thus seems that at most Kant served as a catalyst for the development of his ideas and as a confirmation of them.” (Crowe, 1967, pp.24-25)

This is an interesting observation, but it is the opinion of the author of this thesis that time as some kind of special parameter unto itself is not a necessary component for understanding the concepts related to quaternions and their association with rotation groups. Thus Kant’s influence on Hamilton’s work is possible, but not necessary to understand the development of Hamilton’s work.

Soon after discovering quaternions, Hamilton connected quaternion algebra to spatial rotations. It is an interesting fact that this relationship had been discovered earlier by Olinde Rodrigues. Hamilton was unaware of Olinde Rodrigues’ discovery since Rodrigues’ work was rather obscure. It appears as though Rodrigues had a better understanding of the algebra of rotations then Hamilton did (Chapter 3 Section D). It is said that Rodrigues also had the beginnings of what would later become Lie algebras (Altmann, 1986, p.201). (Chapter 3 Section G)

Hamilton was so pleased with the outcome of his quaternion system that he founded a school devoted to the study of quaternions that he called

‘quaternionists’ (Ebbinghaus, H.-D. et al,1991, 193). Hamilton wrote a long treatise, *Elements of Quaternions*, in an attempt to popularize them. The book is 800 pages. It was published shortly after his death in 1866. It is a rambling work that is not easy to read, as the author of this thesis discovered. It appears that Hamilton was never able to find a satisfactory interpretation how quaternions were related to ‘vectors’. This would later be clarified by Gibbs and Heaviside.

Peter Tait, became Hamilton’s successor in continuing the quaternion crusade. Quaternions became an examination topic in Dublin University (Ebbinghaus, H.-D. et al,1991, 192). American students in the late 19th century were introduced to quaternions via ‘Topics Courses' as they are called today. For example, during this time, the University of Michigan and Harvard offered quaternions as a part of their mathematical curriculum. (Tucker,2013,p.690) . Benjamin Peirce taught this type of course in Harvard. He included as one of his ‘topics’ quaternions. This was a part of a larger mathematics course that was offered it that time (Kennedy, 1979, p. 423).

Benjamin Peirce’s son James Mills Peirce was one of the impetuses behind the ‘cult of quaternions’. He was attracted to quaternions through his father who according to Crowe did more to promote an interest in quaternions than anyone else in the U.S. (Crowe 1967, p.125). James Mills Peirce along with Thomas Hill, also from Harvard University and inspired by Benjamin Peirce’s lectures and enthusiasm for quaternions, helped the founding and promotion of this ‘cult’ (Kennedy, 1979, p.424). It was Yale educated, Shunkichi Kimura of Japan, who coined the professional society “International Association for Promoting the Study of Quaternions and Allied Systems of Mathematics.” This ‘cult’ was dedicated to the study and promotion of quaternions (Kennedy, 1979, p. 425; Struik, 1967, p.172). For example this fellowship presented material and published in prestigious journals such as Nature (Kimura & Molenbroek ,1895, pp.545–6) and Science (Kimura & Molenbroek ,1895, pp.524–25).

By the mid 1880’s quaternions were being replaced by the vector analysis that Gibbs and Heaviside developed. Subjects that would have been described in terms of quaternions, now use vectors. Vector analysis is conceptually easier, and the notation is clearer then quaternions. This was fine for 19th century classical physics, but with the advent of subjects like quantum mechanics the limitations of vector analysis became more apparent and quaternions were rediscovered in the form of Pauli spin matrices.

Today, few students and professionals would feel comfortable with, or be able to, comprehend quaternions because they think in terms of the vector analysis that they learned in school. Hamilton's original definitions would be both unfamiliar and fundamentally different from what modern students or instructors would be familiar with. Although Hamilton believed his work on quaternions was his most important contribution to mathematics. Thus towards the end of the 19th century quaternions took a 'back seat' to other methods i.e. vector analysis for many classical applications.

One of the interesting reasons for the reappearance of quaternions in the later 20th century was due to computer animation and other applications involving computer programing. This appears to have happened because quaternions use algebra to describe spatial rotations. This makes quaternions more 'compact' than matrices, thus when programed into computers they can be computed 'faster' than matrices. This makes them more useful in computer applications (Shoemake, 1985, pp.245-254; Chi, 1998). The reason that quaternions are faster than matrices is that rotation matrices contain sines and cosines, but quaternions do not. Quaternions also have only 4 scalars, where the matrices that are usually used for these types of programs have 9. This also increases the speed of quaternion multiplication. (Gruber, 2000).

Hamilton himself was not sure of how to apply quaternions in the context of his own era. He originally conceived of them as a geometric curiosity. It appears that he thought about their applications but only after he had worked out the actual quaternion mathematical system. It does not appear the physics applications were an integral part of their development. Hamilton did put some thought into possible applications for quaternions since Hamilton was also interested in physics and applications. He wrote in a letter to Graves about his quaternion application musings,

> "There seems to me to be something analogous to polarized intensity in the pure imaginary part; and to unpolarized energy (indifferent to direction) in the real part of a quaternion: and thus we have some slight glimpse of a future Calculus of Polarities. This is certainly very vague,…" (Hamilton, 1844, pp.489-495)

Ultimately this task was left to his student Tait.

### C. Grassmann: A more Algebraic approach to Quaternion Geometry

While Hamilton was developing quaternions another mathematician, Hermann Günter Grassmann (1809 – 1877), was working on a more algebraic approach to conceptually similar issues. Grassmann was perhaps more motivated by applications then Hamilton was. Grassmann developed a system that put geometry into an algebraic form. This algebraic approach, unlike quaternions, is not bounded by 3-dimensional space. This was an unusual idea for Grassmann's time, since most mathematicians and physicists were bounded by 2 or 3 dimensional space. He incorporated non-commutative multiplication into his system, a cutting-edge idea for his time, making Grassmann a man truly ahead of his times.

In order to do this Grassmann developed a language that was rather esoteric by contemporary standards. As Carl Friedrich Gauss put it in a letter to Grassmann in 1844: "… in order to discern the essential core of your work it is first necessary to become familiar with your special terminology. …however that will require of me a time free of other duties …" (Grassmann, 1844, p.331).

Grassmann was one of the tragedies of mathematics due to the neglect his work received during his own lifetime. Grassmann used ideas that are part of modern vector analysis before Hamilton developed quaternions. Grassmann also developed a many-dimensional analog to dot product which he called inner products and to what are called cross products today, he called the outer products (Crowe, 1967, p. 65). In his work Grassmann included topics that are now considered part of vector analysis. For example he included vector addition, subtraction, differentiation, and function theory. (O'Connor and E F Robertson, 2005).

Once the esoteric nature of Grassmann's mathematical language is overcome, it turns out that his algebras were for applications, simpler to use, in general, then Hamilton's quaternions. For example, a vector is usually associated with a point P or a line from 0 to P. What Grassmann did was discuss situations where the lines didn't necessarily go through 0. This idea allowed for greater generality. A line can now be 'offset', where an 'offset' line segment from P to Q, say **b**, can be represented as **a**, what Grassmann referred to as '2-blade' or 'bivector'. This '2-blade' is called the exterior product of **a** and **b** (Figure 1) If another 'offset' vector is introduced to the '2-blade' this will add another dimension that is represented by a '3-blade' and so on. Thus dimensions can be added, so that with each 'offset' an extra degree of freedom is added. By adding and subtracting these 'offsets' new geometric objects can be developed.

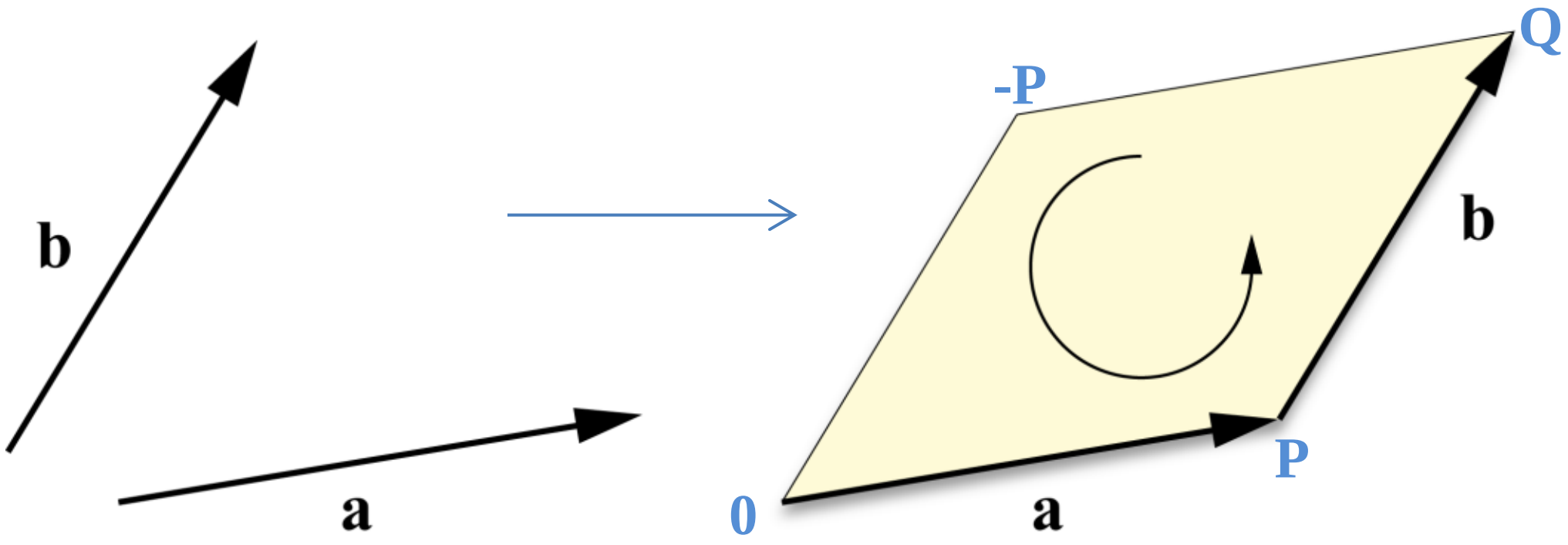

**Figure 1 :** A "2-blade" (adopted from : Lengyel, 2012, Slide 14)

Grassmann first began to apply his ideas in his *Theorie der Ebbe und Flut: Prüfungsarbeit 1840* (Theory of tides) (Grassmann, 1840). He then saw that he could go beyond his own initial applications, in 1844 he published, *Die Lineale Ausdehnungslehre* which contains vector analysis still in use today. (Crowe, 1967, p. 65). In 1877 Grassmann published *Die Mechanik nach den Principien der Ausdehnungslehre*, with the help of his brother. He compared this work to his earlier works writing that:

> "Without exception (except for changes of notation here and there), the methods which I use in this paper and the equations to which I get by means of them, I have already submitted, in a work about the theory of the tides, at Pfingsten 1840 as an examination paper at the scientific examination commission in Berlin…Very little of Extension Theory has changed since 1844." (Grassmann, 1877, p. 222).

He also says in his introduction that he was encouraged by what he read in Kirchhoff, to go back and continue his research in extension theory and its applications "The newer textbooks and papers in mechanics, namely, G. Kirchhoff's Lectures (1875, 1876) show me that the presentation of these methods will also be useful today, as it was thirty-seven years ago…" (Grassmann, 1877, p. 222). In this paper he elaborates on how his 'calculus' can be applied to mechanics in general. Unfortunately Grassmann's work is not easy to read.

In his preface to one of the early text books using Grassmanns's methods E.W. Hyde, a great admirer of Grassmann's work stated "… the great generality of Grassmann's processes – all results being obtained for n-dimensional space - has been one of the main hindrances to the general cultivation of his system…" (Hyde, 1890,p.v). This may have been one of the main reasons that mathematicians of the time didn't have the patience to delve deeper into Grassmann's work and truly appreciate what he was doing.

Grassmann came from a well-educated family, and went to good private schools in Stettin, Germany the town where he was born. He remained in Stettin most of his life except when he went to the University of Berlin. Despite the fact that he had more academic and personal advantages then most, he was not well appreciated as a scholar, not even by his own father. His father, master of the Gymnasium that he attended, thought that Hermann should be a gardener, craftsman or some other type of laborer. Grassmann went the University of Berlin in 1827 where he studied mainly linguistics and theology. According to O'Connor & Robertson, it does not appear that Grassmann studied or took any formal courses in mathematics or physics (O'Connor & Robertson, 2005).

Grassmann was a self-taught, self-made mathematical scientist in the truest sense. He was one of the few mathematicians who successfully applied his own mathematical discoveries to problems in physics.

In the fall of 1830 when Grassmann returned from college to his home town, Stettin. There he decided that he would earn a living as a school teacher and do research on his own time.

In the spring of 1832 Grassmann got a job as an 'assistant teacher' in his former Gymnasium at Stettin. It was about this time that Grassmann started applying algebra to geometry. He wrote in the preface of *Die Lineale Ausdehnungslehre, ein neuer Zweig der Mathematik,* (Linear Extension Theory, a New Branch of Mathematics) that he encountered, while working on his thesis on the theory of the tides, La Grange work *Méchanique analytique* (Analytical Mechanics). This inspired him to go "…back to the ideas of the analysis." (Grassmann, 1844, p.vii) He also mentions on the same page, in a footnote that he got some of his basic ideas from Laplace's *Méchanique celeste* (Celestial Mechanics).

In 1861 he published the second edition *Die Lineale Ausdehnungslehre, ein neuer Zweig der Mathematik* (Linear Extension Theory, a New Branch of Mathematics). This is one of the true mathematical masterpieces of the era. It essentially was a rewritten version of his earlier 1844 work. He expanded his earlier work, and took out the philosophical comments. He also changed it to the standard definition, theorem and proof way of writing. He probably hoped that by making these changes his work would attract more established mathematicians and scientists. Unfortunately this was not the case.

### D. Gibbs and Heaviside: Combining Hamilton and Grassmann

Grassmann ultimately did leave his legacy through Josiah Gibbs, a professor at Yale in the United States of America. Gibbs was inspired by Maxwell's *Treatise on Electricity and Magnetism* (1873). Gibbs was unhappy with quaternions and how they were being applied to physical problems. In Europe Grassmann's ideas were noticed by Hermann Hankel and Victor Schlegel.

On the other side of the Atlantic Ocean, English engineer Oliver Heaviside also saw issues with applying quaternions to physical problems. Unlike Gibbs, Heaviside was not university educated, but a self-taught mathematical scientist in his own right. These two men were independent discoverers of, what is called today, vector analysis.

### i. Gibbs: a well-educated American

Josiah Willard Gibbs (February 11, 1839 – April 28, 1903) is the same Gibbs that chemistry and physics students encounter when they learn about 'Gibbs free energy' in thermodynamics classes.

Gibbs was a professor of mathematical physics at Yale. He originally did his PhD in electrical engineering. He was the first in Yale to receive a PhD. in that subject. Gibbs was aware of Grassmann's work and wanted to make it more useful for his own scientific research. In a letter to Victor Schlegel in 1888, Gibbs makes it clear about how his ideas about vectors were inspired by Maxwell's *Treatise on Electricity and Magnetism* (1873).

> "…where Quaternion notations are considerably used, I became convinced that to master those subjects, it was necessary for me to commence by mastering those methods. At the same time I saw, that although the methods were called quaternionic, the idea of the quaternion was quite foreign to the subject. I saw that there were two important functions (or products) called the vector part & the scalar part of the product, but that the union of the two to form what was called the (whole) product did not advance the theory as an instrument of geom. investigation." (Crowe, 2002, pp.12-13)

Due to misunderstandings about quaternions versus vectors Tait, Hamilton's favorite student of quaternions, became a rival of Gibbs. Tait in his preface to the third edition of *Elementary Treatise on Quaternions* (1890) says that, "Even Professor Gibbs must be ranked as one of the retarders of Quaternion progress, in view of his pamphlet on Vector Analysis, a sort of hermaphrodite monster, compounded of the notations of Hamilton and Grassmann." (Crowe, 1967, p.150; Pritchard, 2010, p.239). Although it is unclear exactly how Tait viewed Gibbs work as a 'hermaphrodite monster'; in the arguments that occurred between Tait and Gibbs, Gibbs focused mainly on the use of sums, scalar and vector products in order to solve physical problems. He found that "As fundamental notions there is a triviality and artificiality about the quaternionic product and quotient, he argued" (Pritchard, 1998, p.239). The arguments "went on with the quatenionists emphasising algebraic simplicity and mathematical elegance and the vector analysts giving weight to naturalness and ease of comprehension". (Pritchard, 1998, p.240).

Gibbs also found quaternions limiting in their scope since they could not be used to analyze more than three dimensions. This bothered Tait since he did not see any use for more than three dimensions, as did most scientists of his day. He responded in the journal *Nature* "What have students of physics, as such, to do with space of more than three dimensions?" (Tait , 1891, p.512). The need for more than three dimensions would become clear in the early 20$^{th}$ century, but may have not have been so obvious in the late 19$^{th}$ century.

Gibbs found that what could be done using vectors could be done, in principle, using quaternions, but with more effort. Thus the controversy became less about vectors versus quaternions, but more about vectors with or without the concept of quaternions as an integral part of its methodology.

### ii. Heaviside: the self-taught English electrician

The other scientist who is credited for bringing vectors into the scientific lexicon was an English electrician Oliver Heaviside (May 18, 1850 –February 3, 1925). Heaviside, like Gibbs, learned about quaternions when he was studying Maxwell's theories. He found that setting up the equations using quaternions required a lot of work. For him they were "…very inconvenient. Quaternionics was in its vectorial aspects antiphysical and unnatural." (Heaviside, n.d.b, p.136) One of the outcomes was that when Heaviside, like Gibbs removed, what they considered the difficult parts, they inadvertently also removed what would may considered the more conceptually interesting parts of quaternions. By doing this Heaviside 'opened the doors', along with Gibbs, for the controversy with Tait and his followers. This controversy became part of *Nature* from around 1890 – 1894 (Wisnesky, 2004, p.14). Heaviside said later about his own use of quaternions "I dropped out the quaternion altogether, and kept to pure scalars and vectors, using a very simple vectorial algebra in my papers from 1883 onward." (Heaviside, n.db, p.136)

Heaviside had a very different background then Gibbs. Heaviside was born in a rough neighborhood in London.  His mother's sister married Charles Wheatstone. This marriage became the 'saving grace' for Oliver Heaviside and his brothers. Wheatstone was a professor of physics at Kings College London. This is the same Wheatstone that the 'Wheatstone bridge' of elementary physics and engineering that students learn today. The 'Wheatstone bridge' is a way to measure an unknown electrical resistance. Wheatstone was also one of the inventors of the telegraph.  He introduced the Heaviside boys (there were three of them Charles, Arthur and Oliver) to the world of electricity. Oliver and his brother Arthur decided to take advantage of their connection to Wheatstone to pursue careers related to the telegraph. This allowed them access to some of the 'state of the arts' electrical laboratories of the era.  The two brothers were involved in laying cable lines for telegraphs. When Oliver reached his mid-twenties he decided that he was more interested in doing research in electrical theory then pursuing a career the telegraph business. He decided to go back home and live with his parents. His parents had an 'extra room' where he lived for the rest of his life. His brother Arthur supported his research interests while remaining in the telegraph business (Hunt, 2012, p.49).

In 1873 Heaviside published *On the Best Arrangement of Wheatstone's Bridge for Measuring a Given Resistance with a Given Galvanometer and*

*Battery* (Heaviside, 1892, pp.3-8). This work attracted the attention of William Thomson (known also as Lord Kelvin). It also was noticed by James Clark Maxwell. Maxwell was impressed enough by Heaviside that he cited him in the second edition of *Treatise on Electricity and Magnetism* (Hunt, 2012, p.50). Since Heaviside was noticed by Maxwell, Heaviside was able to secure a copy of Maxwell's *Treatise* as soon as it came out. Maxwell was encouraged by Tait to incorporate quaternions into the *Treatise.*

Heaviside became aware of Gibbs work from an unpublished pamphlet version of *Vector analysis 1881-4*. Heaviside concluded that "Though different in appearance, it was essentially the same vectorial algebra and analysis to which I had been led." He saw Tait's adherence to a 'quaternions only' attitude as somewhat "…extremest conservatism. Anyone daring to tamper with Hamilton's grand system was only worthy of a contemptuous snub." Heaviside remarked later in his introduction that, in time, Tait did come around and 'soften' a bit (Heaviside, n.db, p.136).

It has been noted by Michael J. Crowe in his book *A History of Vector Analysis-The Evolution of the Idea of a Vectorial* that in our modern understanding of vector analysis

> "It is not possible to argue that the quaternion system is the vectorial system of the present day; the so-called Gibbs-Heaviside system is the only system that merits this distinction. Nor is it legitimate to argue that the quaternion system will be the system of a future day. Both of these alternatives are unacceptable; nonetheless it can be

> argued … that Hamilton's quaternion system led by a historically determinable path to the Gibbs-Heaviside system and hence to the modern system.” (Crowe, 1985, p.19)

It is the opinion of the author of this thesis that Crowe tends to look at the history of vector analysis too much on side of Gibbs and Heaviside, and does not give Grassmann his due. The Gibbs-Heaviside system is not the only vectorial system deserving merit. Grassmann was an essential part of this development and also deserves more merits in this distinction then it appears Crowe is willing to give.

Another observation is the historical importance of quaternions in the development of rotation groups SU(2), SO(4) used in theoretical physics today. Essentially these rotation groups developed with the same conceptual flavor as quaternions, but with different notation.

### E. The Joining of Quaternions with Grassmann algebras : William Kingdon Clifford

William Kingdon Clifford (1845-1879) lived a relatively short life; he died of tuberculosis on March 3, 1879 at the age of 35. Clifford had a prodigious background. He went to Kings College London at only 15 years old. When he was 18 he went on to Trinity College in Cambridge. After graduating from Trinity he became professor of applied mathematics at the University College of London when he was 23. (MacFarlane, 1916, pgs. 49-50)

Clifford, like Gibbs and a few others discovered Grassmann’s work, and was very impressed by it. As a result he published *Applications of Grassmann's Extensive Algebra* in 1878. In this paper Clifford looks for a simpler and more general way to look at algebras in higher dimensions. He noticed that Grassmann algebras and quaternions are not really in conflict with each other as may have been previously believed. Thus with a little tweaking Clifford was able to resolve the apparent issues that existed between quaternions and Grassmann’s algebras. Clifford put his motivation for writing this paper in the following quote:

> “…thereby explaining the laws of those algebras in terms of simpler laws. It contains, next, a generalization of them, applicable to any number of dimensions; and a demonstration that the algebra thus obtained is always a compound of quaternion algebras which do not interfere with one another.” (Clifford, 1878, p.350)

It should be noted that when Hamilton discussed ‘vectors’, he was not using them in the way that they are understood today. Hamilton understood ‘vectors’ to mean the non-scalar part of the quaternion equation.

Clifford’s paper *On The Classification of Geometric Algebras* (Clifford, 1882, p.397) was not discovered until after his death. This paper became the basis for Clifford algebras as they are known today. This is perhaps Clifford’s greatest contribution to mathematics, unfortunately he never finished this paper and it was not found in good condition (Diek, n.d.,p.4).

What essentially Clifford did was to make the connection between Grassmann's algebra and Quaternions by first noting their differences:

> "The system of quaternions differs from this (Grassmann's approach), first in that the squares of the units, instead of being zero, are made equal to - 1 ; and secondly in that the ternary product $\iota_1\iota_2\iota_3$, is made equal to - 1. : The interpretation is at the same time extended to three dimensions, but with this restriction: that whereas the alternate units represent any three points in a plane, and the system deals primarily with projective relations, Hamiltonian units represent three vectors at right angles, and the system is the natural language of metrical geometry and of physics." (Clifford, 1882, p.399)

Clifford makes it clear that his work was a way to connect Grassmann's and Hamilton's ideas. This is what Clifford's work on Geometric Algebras was about. Clifford also had no intension of changing or breaking away from quaternions. What he wanted to do was to 'streamline' them. He wanted to give a more complete and simpler presentation of the ideas that both Hamilton and Grassmann presented in their work. His intention was to make Hamilton's ideas more palatable to physical scientists. This he did by making a method that could more easily be used for calculation while not having to give up the conceptually interesting parts of either quaternions or Grassmann's algebras. Geometric algebras, unlike vectors, are not merely a 'special case' of quaternions, but incorporate the deep conceptual nature of quaternions that was lost in vectors.

Clifford in 1878 essentially reinvented and generalized Hamilton's quaternions. As noted Clifford incorporated ideas from both Grassmann's extensive algebras and Hamilton's quaternions. According to David Hestenes Clifford's system was way to overlap Grassman's and Hamilton's ideas. Clifford did not claim that these ideas were his. This becomes a problem when trying to attribute Geometric Algebra to one specific founder. Hestenes goes onto say "Let me remind you that Clifford himself suggested the term Geometric Algebra, and he described his own contribution as an application of Grassmann's extensive algebra" (Hestenes, 1993, p.2). Clifford did not consider this work 'original', but a synthesis of the best of both the Grassmann's extensive algebras and Hamilton's quaternions.

The depth and subtlety of Clifford's work is not 'easy' to understand, but once understood can be very satisfying.

## F. Sophus Lie

The history of the rotation groups being discussed in this thesis cannot be appreciated fully without having some understanding of Lie Groups and Algebras, since these structures have had a major impact on modern theoretical physics. These structures are named after the mathematician who discovered them, Sophus Lie (1842-1899).

Sophus Lie was born December 17 1842, in the rural village of Nordfjordeide in Norway to a Lutheran pastor and his wife. He was the youngest son of six children. Lie went to the standard elementary and high school schools that existed in Norway at the time (Hawking, 1994, p.6).

Lie went to Christiania University (now University of Oslo), where he completed his PhD with the dissertation in 1871 on *Uber eine Classe geometrischer Transformationen* (About a class of geometric transformations). This dissertation covered issues about what is called today differential geometry.

Up until Lie completed the PhD. he earned his living tutoring. He really didn't find anything in the main stream mathematics curriculum that captured his mathematical or intellectual passions until he read Jean-Victor *Traité des propriétés projectives des figures* (Treaty of projective properties of figures) (1822) and *Théorie des polaires réciproques* (Theory of Reciprocal Polars) published in Crelle's Journal 1829. While reading Poncelet around 1868, Lie was introduced to complex numbers in projective geometry. Lie also read Julius Plücker's *System der Geometrie des Raumes in neuer analytischer Behandlungsweise* (System of the geometry of space in new analytical method of treatment) published in Crelle's Journal 1846. Plücker's used the displacement points in space to represent lines, curves, and surfaces. This inspired Lie to publish his first paper in 1869 *Repräsentation der Imaginären der Plangeometrie* (Representation of imaginary numbers in plane geometry). This paper covered essentially his mathematical interests, rather than incorporate original ideas that would be usually associated with a research paper. None the less, despite the superficial nature of the paper it managed to win him a fellowship to the University of Berlin. This opened doors to meet other mathematicians and students of mathematics. During his stay in Berlin he met fellow student Felix Klein. Even though Klein was seven years younger than Lie they had a deep and productive mathematical relationship. Lie and Klein shared similar mathematical interests they both wanting to take Plücker's ideas and develop them further. In order to do this they decided that they would go to Paris in the spring of 1870. Their intentions was to expose themselves to the latest French mathematical fashions of the day (Helgason, 2002, p.4).

About 3½ years after his rendezvous in Paris with Felix Klein, Sophus Lie discovered his theory of continuous groups. It is this discovery that Lie is most remembered for, especially in physics. This was the beginning of what is called today ‘Lie Group Theory’. Lie decided to take a special type of transformation group, and use these groups to solve differential equations. The first compact Lie group that he discovered is called SU(2). This group is closely connected to quaternions. It was also anticipated by Rodrigues two years before Lie was born.

Lie was disappointed that other mathematicians didn’t take much notice to his work. He wrote to his friend Adolf Mayer in 1884: "If only I knew how to get mathematicians interested in transformation groups and their applications to differential equations. I am certain, absolutely certain, that these theories will sometime in the future be recognized as fundamental. When I wish such a recognition sooner, it is partly because then I could accomplish ten times more." (Helgason, 2002, p.14).

Noticing how isolated Lie was mathematically his friend Klein from his Berlin days sent one of his students to work with him starting in 1884. Friedrich Engel was 22 years old and stayed for nine months. According to Engel this was the happiest and most productive period of his life (Helgason, 2002, p.15).

When Klein vacated his position in the University of Leipzig in 1886, Lie became his successor. This catapulted Lie from his quite isolated life in Christiania University to mainstream mathematical life of Leipzig University. Despite the recognition that he received in Leipzig he still felt unappreciated, although during this time Lie and Engel worked on transformation groups and produced *Theorie der Transformationsgruppen* (Theory of Transformation Groups) (O'Connor & Robertson, 2000).

This was not a small undertaking. The work consisted of three volumes. It was published between 1888 and 1893. Lie was in poor health during this time, so Engel did most of the ‘real’ work. As it turned out Lie would sketch out the problem or proofs and Engel would fill in the details. (O'Connor & Robertson, 2000).

Due to what we would call today 'depression' Lie returned to his alma mater, Kristiania University September1898. Lie was 56 when he died of pernicious anemia where the body is unable to make healthy red blood cells. This disease is caused by a lack of vitamin B12 in one's diet (O'Connor & Robertson, 2000; Gale, 2005-2006).

So far this thesis has given is a historical overview of the mathematical 'zeitgeist' of the times and some of the mathematicians who were involved in these discoveries. Lie Algebras are often employed along with Grassmann's Algebras and Clifford's algebras as a main part of the theoretical physics intellectual diet. The question that remains is how and why the revival of quaternions and their relatives came about. In the next chapter the author will go deeper into some of the mathematical ideas introduced in this chapter and, hopefully, offer an answer or at least some thought to this question in the final chapter.

# Chapter III

## MATHEMATICAL DEVELOMENT

In the previous chapter of this thesis a historical overview was given about where quaternions came from and some of the mathematicians who discovered and worked with quaternion type groups, but that was only part of the story. In this chapter the mathematics that was developed will be discussed in more depth. The following will not be a comprehensive investigation into the mathematics of these rotation groups, but merely an overview of some of the more important and/or interesting ideas that came out of their mathematical investigations. It is important to see how the mathematics developed, and how the techniques evolved and became incorporated into the lexicon of modern theoretical physics.

### A. Algebra of Quaternions

Some of the basic properties of quaternions were introduced at the beginning of this dissertation. A more extensive analysis of quaternions will be given in this section.

Most of the following definitions were retrieved from Goldstein's Mechanics third edition (Goldstein, 2000, p.310), Penrose's book *Spinors and Space-time Volume 1* (Penrose,1984, pp.21-24), *Quaternions and Rotations in 3-Space: How it Works* (Chi, 1998, pp.1-10), various web pages on quaternions including Wolfram MathWorld web pages (Weisstein,1999-2014;Weisstein, 2002), to name a few of the references used. These are standard definitions that can be found in most books and papers that cover this subject.

In order to add quaternions let $P = (p_0+ip_1+jp_2+kp_3)$ *and* $Q = (q_0+iq_1+jq_2+kq_3)$ be two quaternions. Addition is defined component wise; it is both commutative and associative. That is $P+Q = (p_0+ip_1+jp_2+kp_3) + (q_0+iq_1+jq_2+kq_3) = (p_0+q_0)+i(p_1+q_1)+j(p_2+q_2)+k(p_3+q_3)$. For example, if $P$

$= 3 - i + 2j + k$ *and* $Q = 2 + 4i - 2j - 3k$, then $P + Q = 5 + 3i - 2k$ (Chi, 1998, 2)

For multiplication, recall $pq = (p_0 + p_1i + p_2j + p_3k)(q_0 + q_1i + q_2j + q_3k) = p_0q_0 + p_0q_1i + p_0q_2j + p_0q_3k + p_1q_0i + p_1q_1i^2 + p_1q_2ij + p_1q_3ik + p_2q_0j + p_2q_1ji + p_2q_2j^2 + p_2q_3jk + p_3q_0k + p_3q_1ki + p_3q_2kj + p_3q_3k^2$ applying the "quaternion rules" letting $i^2 = j^2 = k^2 = -1$ *and* $ij=-ji=k, jk=-kj=i, ki=-ik=j$: $pq= p_0q_0 + p_0q_1i + p_0q_2j + p_0q_3k + p_1q_0i - p_1q_1 + p_1q_2k - p_1q_3j + p_2q_0j - p_2q_1k - p_2q_2 + p_2q_3i + p_3q_0k + p_3q_1j - p_3q_2i - p_3q_3.$ In standard form the 'scalar part' is written first and the 'vector part', second in the order *i, j, k* as $pq = p_0q_0 - (p_1q_1 + p_2q_2 + p_3q_3) + (p_0q_1 + p_1q_0 + p_2q_3 - p_3q_2)i + (p_0q_2 \;\; p_1q_3 + p_2q_0 + p_3q_1)j + (p_0q_3 + p_1q_2 - p_2q_1 + p_3q_0)k$ (Chi, 1998, p.3).

Quaternions can be written in the form $q_0 + \mathbf{q}$. The real number $q_0$, is referred to as the scalar part, and $\mathbf{q}=iq_1+jq_2+kq_3$ is referred to as the vector part of the quaternion. Quaternion multiplication can be compared to the modern dot and cross products is by following calculation. Let P = $p_0 + \mathbf{p}$ and Q = $q_0 + \mathbf{q}$ where $\mathbf{p} = p_1i + p_2j + p_3k$ and $\mathbf{q} = q_1i + q_2j + q_3k$ continuing in a straightforward manner it can be shown that PQ = $p_0q_0 - \boldsymbol{p}\cdot\boldsymbol{q} + p_0\boldsymbol{q} + q_0\boldsymbol{p} + \boldsymbol{p} \times \boldsymbol{q}$ (Weisstein, 2002, pp.2446-2448).

In order to restrict Quaternions to rotations, their conjugates have to be defined. They are defined in the same way that complex numbers are. Recall a complex number $z = a+bi$ . The complex conjugate is defined as $\overline{z} = a -$ *bi*. Extending this idea to quaternions, let $Q=(q_0+iq_1+jq_2+kq_3)$ where its conjugate is defined as $Q^* = (q_0-iq_1-jq_2-kq_3)$. Here quaternion conjugation is defined by ignoring the vector part of the quaternion $(Q^*)^*=Q$, $(P + Q)^* = P^* + Q^*$, $(PQ)^* = Q^*P^*$ *and* $QQ^*=Q^*Q$. It should be pointed out that multiplication of a quaternion and its conjugate commute. (Ho Ahn, 2009)

Conjugation by Q is also referred to as double multiplication. This means that if Q is a unit quaternion and $q=(s,v)$ where s be the scalar part of the quaternion and v the vector part. The magnitude of v remains unchanged after conjugation by Q. For example if $QqQ^{-1}$ is defined where $Q^{-1}=Q^*$ . A unit quaternion is defined as |Q|=1. Then $QqQ^*=i(i+j+k)(-i)=(-1 + k-j) (-i)=i-j-k$. (Ho Ahn, 2009). What conjugation does is rotate a vector around the axis of Q by the right hand rule, but at twice the angle of Q. (Vilis, 2000). In order to show rotations are preserved the following Theorem is used:

*Theorem:* If $p=[s, v]$, where s is a real number part of the quaternion and v, the vector part, and $p' = qpq^{-1}$, then $p'=(s, v')$ where |v|=|v'|(Ho Ahn, 2009)

*Proof*:

1. Using the fact that the scalar transforms trivially under group transformations implies that scalar multiplication is commutative. Therefore if the scalar part of p is represented by $p=[s, 0]$, then
   $qpq^{-1} = q[s,0]q^{-1}=[s,0]qq^{-1}=[s,0]$

2. If p has a vector part represented by p = [0, v],then the scalar part $qpq^{-1}$ and the scalar part S(q) of the quaternion can be taken out using the formula $2S(q) = q + q^*$, where 2S($qpq^*$) = ($qpq^*$) + ($qpq^*$)* = $qpq^* + qp^*q^*$. Applying this to get $2S(qpq^{-1}) = qpq^{-1} + (qpq^{-1})^*$. Since $q^{-1} = q^*$ if $|q|=1$ (i.e. a unit quaternion). It can now be written as $qpq^* + (qpq^*)^*$ . Apply conjugation to this to get $qpq^* + qp^*q^*$. Use, again, the distributive property to get $q(p + p^*)q^*$, but 2S($p$) = $p + p^*$. This can be rewritten as $q(2S(p))q^*$. The scalar is commutative so this can be written as $(2S(p))qq^*$. From an earlier definition of the unit quaternion

   $qq^*=qq^{-1}=1$ *if* $|q|=1$, so $2S(p)=0$, since $S[0,v]=0$

   Therefore $qpq^{-1}=[0,v']$

3. If $p = [s,0] + [0,v]$., then by a straight-forward calculation $qpq^{-1}$ = $q([s,0] + [0,v])q^{-1}$. Use the distributive property, and steps 1 and 2 this can be rewritten as $q[s,0]q^{-1} + q[0,v]q^{-1} = [s,0]\ q\ q^{-1} + q[0,v]q^{-1} = [s,0] + q[0,v]q^{-1} = [s,0] + [0,v'] = [s, v']$.

   In the case of the scalar parts $p = p'$ so the norm of p' is using the norm property so that $|p'|=|\ q\ pq^{-1}|= |q||p||q^{-1}|$. From the fact that $|q| = |q^{-1}|=1$ it can be concluded that indeed the $|p|=|p'|$ and therefore $|v'|=|v|$ ■ (Ho Ahn, 2009; Shoemake, n. d., 4 )

The norm can be seen as being analogous to a vector in 4-space as follows:

$$|Q| = \sqrt{Q^*Q} = \sqrt{q_0^2+q_1^2+q_2^2+q_3^2} = |Q^*|$$

Where the norm of a quaternion is multiplicative, meaning that the norm of the multiplication of many quaternions is equal to the multiplication of the norms of quaternions:

$|PQ| = \sqrt{PQ(PQ)^*} = \sqrt{PQQ^*P^*} = \sqrt{P|Q|^2P^*} = \sqrt{PP^*|Q|^2} = \sqrt{|P|^2|Q|^2} =$ $|P||Q|$ (Ho Ahn, 2009).

Quaternions have an inverse, $Q^{-1}$. *If* $Q=q_0+iq_1+jq_2+kq_3$ then its inverse is defined to be $Q^{-1} = \frac{(q_0-iq_1-jq_2-kq_3)}{q_0^2+q_1^2+q_2^2+q_3^2}$ . Quaternions have multiplicative inverses defined by $Q^{-1} = \frac{QQ^*}{|Q|^2} = \frac{|Q|^2}{|Q|^2} = 1$ , similarly $Q^{-1}Q = \frac{Q^*Q}{|Q|^2} = \frac{|Q|^2}{|Q|^2} = 1$ (Ho Ahn, 2009).

Recall that unit quaternions have the property that $|Q| = 1$. This is a straight-forward calculation, and left to the reader. This shows that the conjugate of a unit quaternion is the same as its inverse. This fact was used in the earlier proof on rotation preservation, where it was assumed that a unit quaternion is defined as $Q^{-1} = \frac{Q^*}{|Q|^2} = Q^*$ (Ho Ahn, 2009).

Quaternions, like Real and the Complex numbers, form a group under addition. The non-zero quaternions form a group under multiplication. By definition a group consists of a set G satisfies following conditions:

1. Closure, means if *a*, *b* are elements of G then when they are multiplied, *a*b* this is also an element of G.
2. Associativity, this means that if *a,b,c* are elements of the group G then *a(bc) = (ab)c*.
3. Existence of an identity element, often denoted by *e*, where *ae = ea = a* for all elements of G.
4. Existence of inverses, this means that for all *a* elements of G then there exists an element $a^{-1}$ that is also in G where $aa^{-1}=a^{-1}a=e$ (Herstein, 1996, p.40).

The quaternions form a group under addition, and the non-zero quaternions (all the quaternions except for 0) form a group under multiplication.

A multiplication table for the quaternions can be constructed (Table 1). The quaternion group has four basis elements and includes their additive inverses. These form a non-commutative group of order 8.

| | **1** | **−1** | **i** | **−i** | **j** | **−j** | **k** | **−k** |
|---|---|---|---|---|---|---|---|---|
| **1** | 1 | −1 | i | −i | j | −j | k | −k |
| **−1** | −1 | 1 | −i | i | −j | j | −k | k |
| **i** | i | −i | −1 | 1 | k | −k | −j | j |
| **−i** | −i | i | 1 | −1 | −k | k | j | −j |
| **j** | j | −j | −k | k | −1 | 1 | i | −i |
| **−j** | −j | j | k | −k | 1 | −1 | −i | i |
| **k** | k | −k | j | −j | −i | i | −1 | 1 |
| **−k** | −k | k | −j | j | i | −i | 1 | −1 |

**Table 1:** Multiplication Table for the Quaternion Group. Also called a Cayley Group Table. Conversion order: Row entry first followed by column entry.(Adapted from Weisstein, 1999-2014)

An interesting way to remember the multiplication of quaternions is by using the Fano plane analogy (Figure 2). To use this analogy for quaternions, for example, multiplying ij, go clockwise in the diagram to k. This means that ij=k. But when multiplying ji, go counterclockwise, which is the negative direction, i.e., ji = -k (Beaz, 2001, p.151).

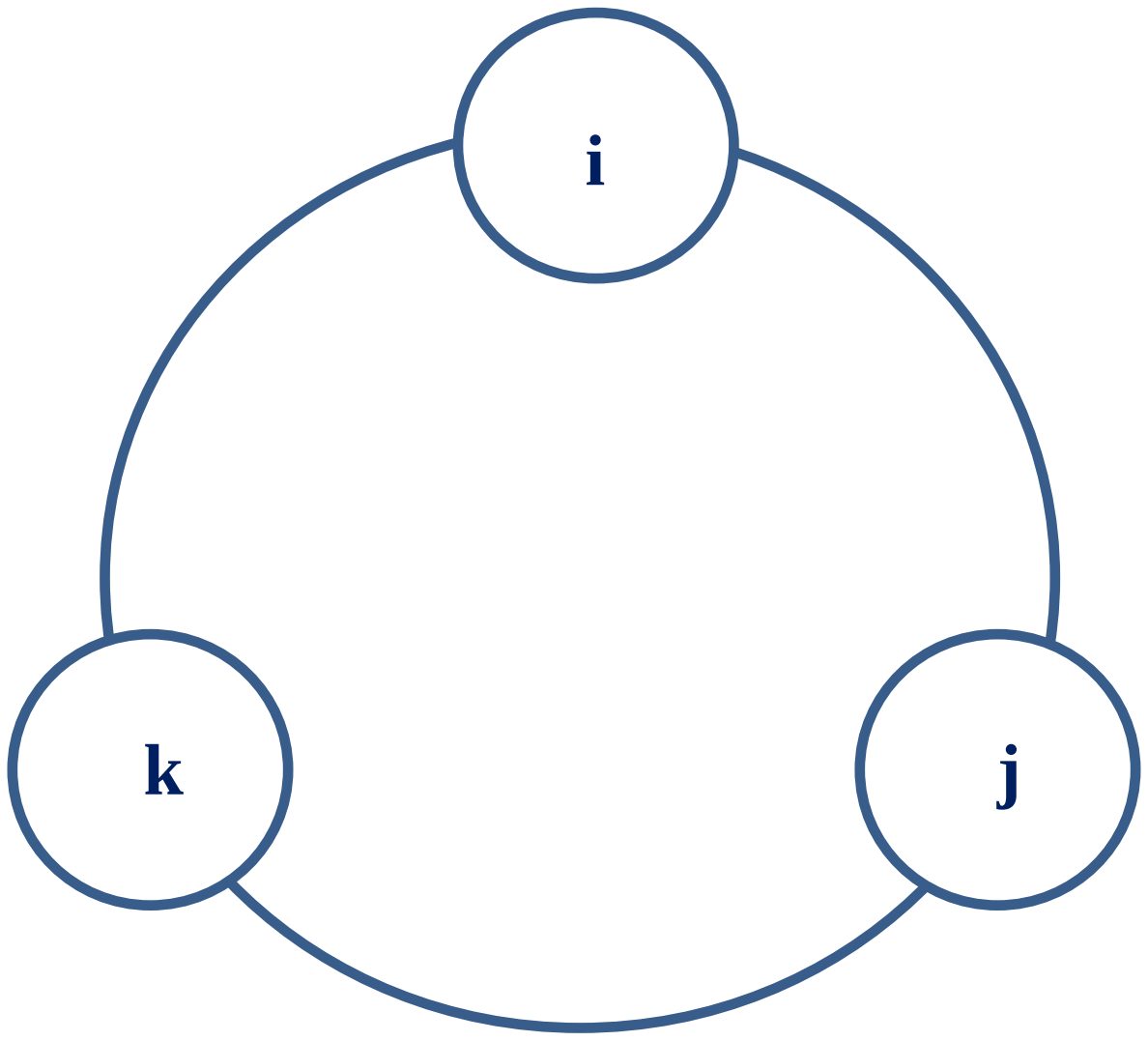

**Figure 2:**
The Fano plane for quaternions (Adopted from Baez, 2001, p.152)

Quaternions form a mathematical object called a division ring or division algebra this means that they closed under the operations of addition and multiplication. According to the CRC of Mathematics often the words division ring and division algebra are used interchangeably. A division ring has the property that all its elements have a multiplicative inverse but this does not imply that multiplication is commutative (Weisstein, 2002, p.803).

Thus by a Theorem: The ring of real quaternions is a division ring. (Byrne, 2013,17). In general multiplication and addition of the real numbers form a field. If multiplication in the field is non-commutative then the field is called a division ring. Using modern terminology this means that the complex numbers can be viewed as a two dimensional vector space over the real numbers; with its multiplication, it forms a field.

The question that Hamilton struggled with for over 10 years was whether or not multiplication could be defined in a 3-dimentional or higher vector space over the real numbers, and would result in a field. It turns out that in 4 dimensions the best that can be done is to define multiplication as non-commutative by doing this the result cannot be a field, but is referred to as a division ring (i.e. the quaternions) (Byrne, 2013, p.15 ).

**B. Vectors analysis and Quaternions**

Quaternions and vectors were both being developed at about the same time. In some sense they had to be developed from each other. As would be expected there are similarities and differences between the two systems. It is these two approaches that will be discussed in more detail in this subsection.

Essentially what Hamilton found is that there was one product formula that involved two quaternions where $ij=k$, $jk=i$, $ki=j$. This was how he described $i$, $j$, $k$ quantities where the product is non-commutative. The two parts of quaternions were discussed earlier as the ‘scalar’ part and ‘vector’ part. Hamilton observed this and introduced his notation for the 'scalar' part S and 'vector' V part of the quaternion. By multiplying two quaternion 'vectors' $\sigma = iD_1 + jD_2 + kD_3$, and $\rho = iX + jY + kZ$ he found that $\sigma\rho = -(D_1X + D_2Y + D_3Z) + i(D_2Z - D_3Y) + j(D_3X - D_1Z) + k(D_1Y - D_2X)$. Hamilton separated these parts by calling the 'scalar' $S.\sigma\rho = -(D_1X + D_2Y + D_3Z)$ and the 'vector' $V.\sigma\rho = i(D_2Z - D_3Y) + j(D_3X - D_1Z) + k(D_1Y - D_2X)$. By doing this Hamilton defined close analogs of the modern dot and cross products. (Tai, 1995,p.7; Bork , 1965, p.202)

Over time many different types of notation were adapted by different authors. For example Gibbs in his original *Vector Analysis* defined vectors to be α,β; scalar products to be α.β, and vector products to be α×β. Gibbs also had the dyadic in 3-dimentional space αβ (Gibbs,1901, pp.17 - 50, pp.50 – 90). The idea of the dyad or the dyadic is equivalent to today's tensor. According to Joseph C. Kolecki of the Glenn Research Center in Cleveland, Ohio a dyad is a tensor of rank 2. This means that it is a system that has magnitude with two directions connected to it (Kolecki, 2002, 4). In general a tensor is an object that obeys specific transformation rules.

What is called a tensor today is essentially a generalization of a scalar and vector, where a scalar is a tensor of rank zero, and a vector is a tensor of rank one. The rank of a tensor defines the number of directions it has, where a dyad has 2 directions thus it is a rank 2 tensor. This means that mathematically that it can be described by 9 entries in a 3×3 matrix.

Where a dyad was used to distinguish the dot product (scalier) from the cross product (vector); today this is symbolically noted as a tensor $\boldsymbol{a} \otimes \boldsymbol{b}$ or (**ab**). A linear combination of dyads, or tensors, would be written today as $\boldsymbol{B} = \boldsymbol{a} \otimes \boldsymbol{b} + \boldsymbol{c} \otimes \boldsymbol{d}$. Although *Vector Analysis* uses somewhat different notation to denote dyads, the definitions would be familiar to a modern reader, where the properties of the dyad and dyadic are the same as those for tensor analysis (Wilson, 1901, 265-281).

Using the notation of Michael Spivak's Calculus on Manifolds.

Let S and T denote two vector spaces and $S \otimes T$ is the tensor product. The order of S and T is important since $S \otimes T$ and $T \otimes S$ are not the same thing. Tensors obey the following rules:

1) $(S_1 + S_2) \otimes T = S_1 \otimes T + S_2 \otimes T$
2) $S \otimes (T_1 + T_2) = S \otimes T_1 + S \otimes T_1$
3) $(aS) \otimes T = S \otimes (aT) = a(S \otimes T)$

For a third vector space, say U, $(S \otimes T) \otimes U$ and $S \otimes (T \otimes U)$ is usually written as $S \otimes T \otimes U$ (Spivak, 1965, p. 31)

Gibbs textbook *Vector Analysis* was published in 1901 by Edwin Bidwell Wilson. It was based on lectures that Gibbs gave in Yale. In this book triple vectors are used. The concept of the dyad is introduced in Chapter V *Linear Vector Functions*.

Gibbs starts by discussing the dot and cross products. He probably noticed that if he wrote the vector product as just **ab** this didn't mean anything, but **ab · c** did mean something, namely **a(b · c).** The parentheses were not needed to express this relationship since there is only one way to understand this expression. Seeing this he may have thought, perhaps, that the expression **ab** could be denoted by its dot-product by a vector such as **c**, meaning that **(ab) · c** means the same as **a(b · c)**, where both expressions could be written without the parentheses. By seeing this it gave meaning to **ab.** This is what Gibbs called a dyad. To summarize formally a dyad can be defined as a pair of vectors say **a, b** where the dyad is, D(**ab)** ≡ **ab**, and its the scalar or dot product is defined to be **a·bc** ≡ **(a·b)c** and **ab·c** ≡ **a(b·c).** (Weisstein, 2003,841)

The next logical step would be to try to add dyads. The most obvious way to do this is by using a distributive law i.e. (**ab** + **cd**) · **e** = **ab** · **e** + **cd** · **e**, and think of this as addition: (**ab** + **cd**). Unfortunately, in general, this is not a dyad. This means that for an arbitrary **a, b, c, d** there isn't an **f, g** where **ab** · **e** + **cd** · **e** = **fg** · **e** for every **e**. So he called (**ab** + **cd**) a dyadic, where dyadics are closed under addition, and dyads are not.

*Vector Analysis* uses somewhat different notation to denote dyads, and dyadic. Although the concepts are the same today, most modern readers are used to using matrices in order to understand these concepts. Going back to the example earlier discussed, a tensor **T** of Rank 2 exists in 3-dimensions this can written as:

$$\begin{pmatrix} T_{11} & T_{12} & T_{13} \\ T_{21} & T_{22} & T_{23} \\ T_{31} & T_{32} & T_{33} \end{pmatrix}$$

A vector **a** can be written as row matrix ($a_1 \quad a_2 \quad a_3$), or as a column matrix, $\begin{pmatrix} b_1 \\ b_2 \\ b_3 \end{pmatrix}$. The dot-product of these can be written as the row-column product:

$$(a_1 \quad a_2 \quad a_3)\begin{pmatrix} b_1 \\ b_2 \\ b_3 \end{pmatrix} = a_1b_1 + a_2b_2 + a_3b_3$$

The dyad, **ab**, can be written as the column-row product:

$$\begin{pmatrix} a_1 \\ a_2 \\ a_3 \end{pmatrix}(b_1 \quad b_2 \quad b_3) = \begin{pmatrix} a_1b_1 & a_1b_2 & a_1b_3 \\ a_2b_1 & a_2b_2 & a_2b_3 \\ a_3b_1 & a_3b_2 & a_3b_3 \end{pmatrix}$$

Here **T** looks like a tensor, and it is a special kind of tensor, since, in general tensors like **T** cannot be expressed as column-row products, but any **T** can be made by adding together at most three such products. When dyads are added together the outcome is something that is not a dyad. Due to this Gibbs called the result a dyadic.

Heaviside used the symbols $\bar{a}, \bar{b}$ for vectors; $\bar{a}.\bar{b}$ for scalar products, $V\bar{a}\bar{b}$ for vector products in his *Electromagnetic theory* (Heaviside, 1893) and Wilson who wrote *Vector Analysis* based on Gibbs's Lectures used the notation **A**, **B** to denote vectors; **A.B** and **A**×**B** for scalar and vector multiplication respectively (Wilson, 1901 pp.265-281). Today dot product (**A·B**) is used for scalar multiplication and cross product (**A×B**) is used for vector multiplication. It was Clifford in 1877 in his book *Elements Dynamic* in the section *Product of Two Vectors* who introduced the notation that is familiar today (Crowe, 2002, p.12). Thus during the formative development of vector analysis the notation was not as transparent as it is today.

### C. Geometry of Quaternions

In this section a geometric interpretation of quaternions will be introduced. The idea of a division ring may not have been important for understanding vectors in the way Gibbs and Heaviside understood them, but they are important in order to understand the geometry of quaternions.

Usually the geometry of quaternions is understood in terms of rotations as illustrated in figures 3 and 4. The geometry of rotations will be explored here and their relationship to the Rodrigues' Rotation Formula will be explored here.

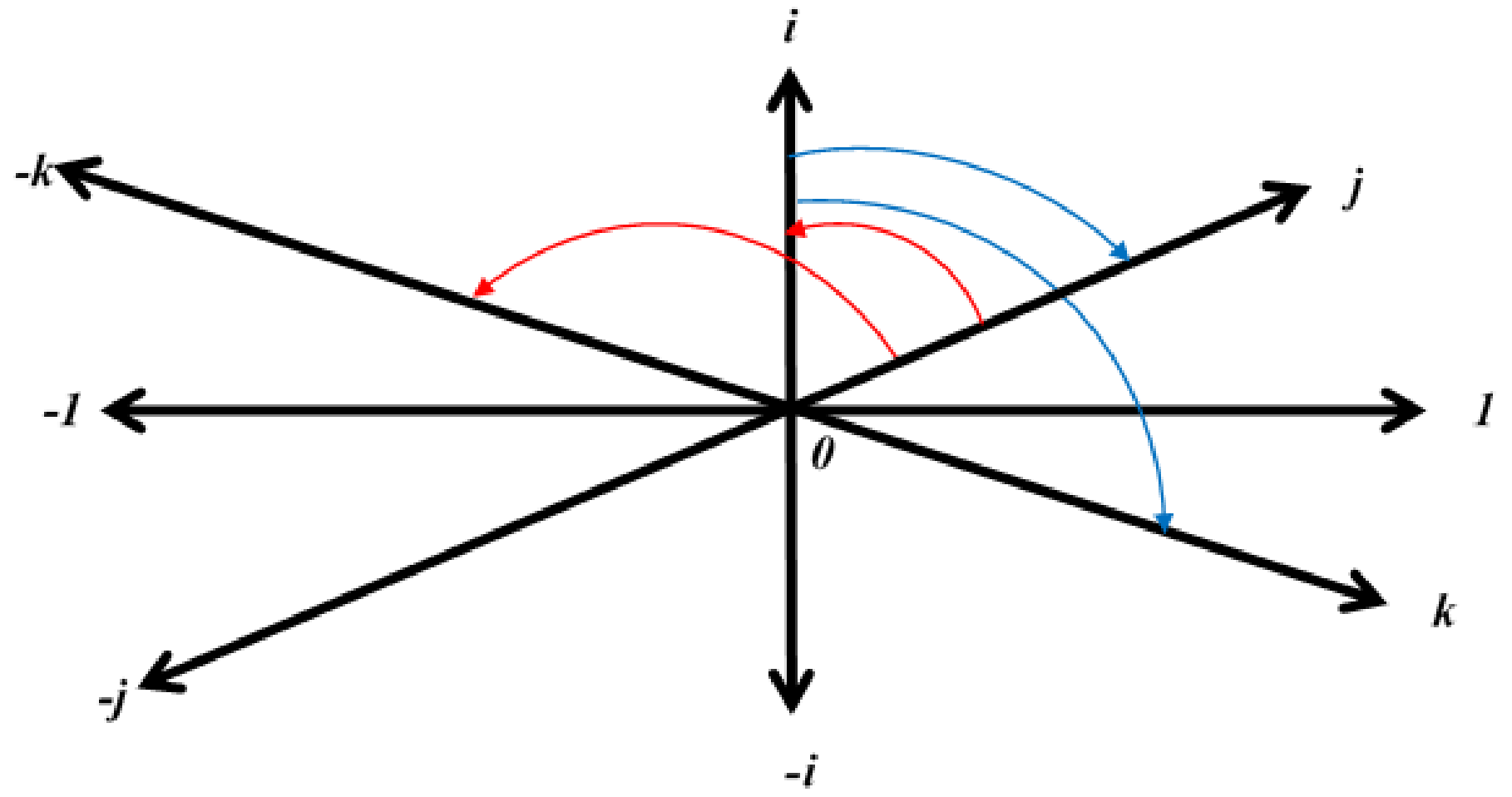

**Figure 3** – Graphical representation of quaternion units product as 90º rotation in 4-dimentional space. This shows the non-commutative nature of quaternions

ij = k
ji = -k
ij = -ji

Adapted from http://en.wikipedia.org/wiki/Quaternion

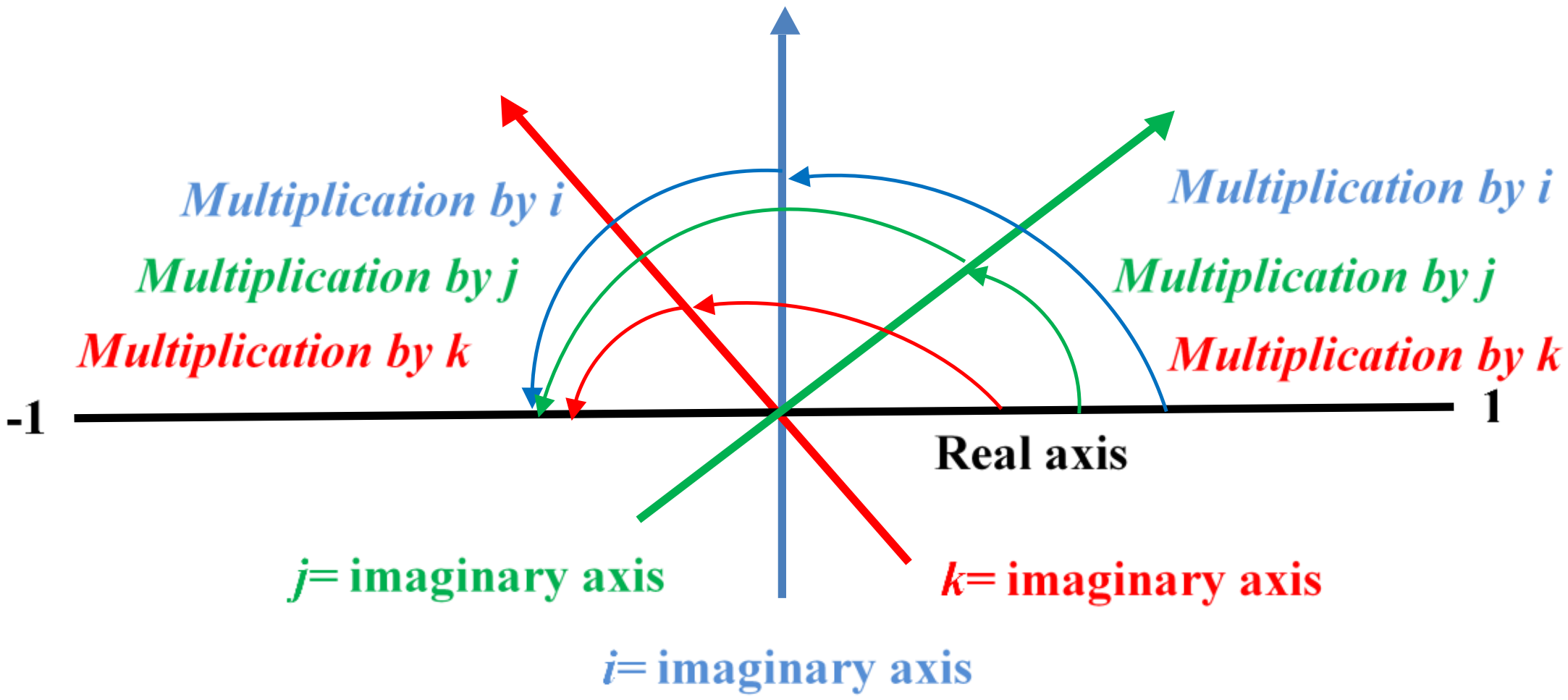

**Figure 4** –Quaternion rotations through the complex plane

http://www.euclideanspace.com/maths/algebra/realNormedAlgebra/quaternions/

### i. Euler's Rotation Theorem

Euler (1707 – 1785) was a very prolific mathematician who studied a wide variety number of mathematical fields. In this thesis the author refers to only one of the many contributions that Euler made in his long and varied mathematical career: his rotation theorem.

The reason that this theorem is important is that it will help to understand a sequence of rotations in 3-dimensions. This means that any rotation can be understood as a combination of a vector and a scalar.

Like most mathematicians of his day he wrote in Latin:

Theorema: Quomodocunque sphaera circa centrum suum convertatur, semper assignari potest diameter, cuius directio in situ translato conueniat cum situ initiali Demonstratio. (Euler,1776, p.189)

Johan Sten translated Euler's Rotation theorem as follows:

Theorem: In whatever way a sphere is turned about its center, it is always possible to assign a diameter, whose direction in the translated state agrees with that of the initial state. (Sten, n.d., p.19)

What this means is that if there is a sphere that is allowed to move freely about this fixed point inside of it, that no matter what orientation it takes, there are two places opposite each other (antipodal points) that have the same initial position. Today the tendency is to describe Euler's rotation theorem in terms of rotation matrices.

Euler's Theorem on the Axis of a Three-Dimensional Rotation. If R is a $3 \times 3$ orthogonal matrix (RTR = RRT = I) and R is proper (detR = +1), then there is a nonzero vector v satisfying Rv = v (Palais, 2009, p.892).

Unfortunately one of the issues that come up when looking at the modern form in terms of rotations is that the idea of a rotation is often confused with a rotation about an axis.

### ii. Quaternion Rotations in 3-Dimensions

When something is rotated one way sometimes it is necessary to go back to the original rotation state. In order to get a rotation back to its original state the rotation has to be what can be called 'undone'. This concept is similar to tying a knot and then untying it. The 'undoing' or 'untying' is what the inverse does, geometrically speaking. The author of this thesis will call the rotation 'doing' and bringing the rotation back to its original state 'undoing'.

One way to understand how complex numbers relate to rotations is to use the Euler formula:

$$e^{i\varphi}=\cos \varphi + i \sin \varphi$$

where a complex number is written $z = x + iy$ in the form $Re^{i\varphi}$ where R is the magnitude of z and φ is the angle that the vector (x,y) makes with the positive x axis, and z can rotate through the angle θ by multiplying $z$ by $e^{i\theta}$ since $ze^{i\theta} = Re^{i\varphi}\, e^{i\theta} = Re^{i(\theta+\varphi)}$ . So $ze^{i\theta}$ has the same magnitude as z but it makes an angle of θ+φ with the x axis, but

$z' = (x + iy)(e^{i\theta}) = (x+iy)(\cos\theta + i \sin\theta) = x \cos\theta - y \sin\theta + i(x \sin\theta + y \cos\theta)$. This can be written as a 2 x 2 matrix (omitting i) as:

$$\begin{pmatrix} x' \\ y' \end{pmatrix} = \begin{pmatrix} \cos\theta & -\sin\theta \\ \sin\theta & \cos\theta \end{pmatrix} \begin{pmatrix} x \\ y \end{pmatrix}$$

(Ho Ahn, 2009b).

For Rotations in 3-dimensional space, unit quaternions are used. Recall that unit quaternions satisfy the property $|Q|=|Q^*|=|Q^{-1}|=1$. Two quaternion factors are needed in order to do an entire rotation θ. Each of the quaternion factors carries an angle θ/2. The Euler rotation formula $e^{i\varphi} = \cos\varphi + i \sin\varphi$, can be replaced by a quaternion form where $\boldsymbol{Q} = e^{\mathbf{v}\theta/2}$ so that $e^{\mathbf{v}\theta/2} = \cos(\theta/2) + v \sin(\theta/2)$ and $\mathbf{v} = \mathbf{i}v_x + \mathbf{j}v_y + \mathbf{k}v_z$. This is called a 'pure quaternion' vector of unit norm. This means that the scalar part of the quaternion is zero.

Instead of $z' = ze^{i\varphi}$ as would be for the Euler rotation formula, in quaternion form this can be written as P' = $QPQ$ where P, P' are 'pure quaternions' or the vector part of the quaternion. These can be written as P, P', where P is the vector that will be rotated into P'.

Let $P = 0+\mathbf{p}$. This forms the vector part of the quaternion, denoted in bold face. Let $P'=QQ^*$, P' be a pure vector rotation. This converts quaternion notation into angular notation, θ being the angle, and $\boldsymbol{v} = (v_x,v_y,v_z)=(v_x i + v_y j + v_z k)$ its axis. Using Euler's formula it can be written:

$Q = e^{\theta/2(vx\boldsymbol{i}+vy\boldsymbol{j}+vz\boldsymbol{k})} = cos(\theta/2) + sin(\theta/2)(v_x i+v_y j+v_z k) = cos(\theta/2), sin(\theta/2)\boldsymbol{v}$

Or simply $Q = [cos(\theta/2), sin(\theta/2)\boldsymbol{v}]$ (Mason, 2012, p.11).

A combination of rotations can be given by, say, $Q_1$ is followed by $Q_2$ means that $Q = Q_2Q_1$, and $Q_2(Q_1PQ_1^*)Q_2^*=(Q_2Q_1)P(Q_2Q_1)^*=QPQ$. When going from the axis angle to the Quaternion angle, respectively, this is the process 'doing' the quaternion:

Start with the axis angle going into the Quaternion:

$Q = [cos(\theta/2) + sin(\theta/2)\boldsymbol{v}]$

'undoing' a quaternion:

Start with the quaternion going into the Axis Angle:

$\theta=2tan^{-1}(|\boldsymbol{q}|/q_0)$ and $v=\boldsymbol{q}/|\boldsymbol{q}|$ where $\theta\neq 0$

(Mason, 2012, p.15; Van Verth, 2013).

In order to develop quaternions further as rotations, the concept of 'pure quaternions' is needed. Recall that the pure quaternions are the vector part of quaternions without the scalar part. These vectors are in 1-1 correspondence with $\mathbf{R}^3$. Thus a vector in $\mathbf{R}^3$ corresponds to the quaternion 'pure vector' $Q = 0 + q_1i+q_2j+q_3k$. In order for quaternions to rotate in 3-dimensionsl space the rotation operator $L_Q$ is established, where $L_Q$: $\mathbf{R}^3 \rightarrow \mathbf{R}^3$. If $\mathbf{v} = q_1i+q_2j+q_3k$ then $L_Q(\mathbf{v}) = QvQ^*$ where $\mathbf{v}$ is an element $\mathbf{R}^3$ and Q is a unit quaternion, then $\mathbf{v}$ is also an element of $\mathbf{H}_0$ where $\mathbf{H}_0$ is the set of all 'pure quaternions'. The following proposition is used to develop the quaternion rotation operator:

*Proposition:* If for all **v** that are elements of $H_0$ and for all Q that are elements of **H**, then $Q\mathbf{v}Q^* = (Q_0^2 - |\mathbf{Q}|^2)\mathbf{v} + 2(\mathbf{Q} \cdot \mathbf{v})Q + 2Q_0(Q \times v)$ => $Q\mathbf{v}Q^*$ is also an element of $\mathbf{H}_0$■ (Gravelle, 2006, p.8)

The rotation operator $L_Q$ is also a linear operator that preserves the length of the vector. It can be shown:

*Theorem 1:* For any unit quaternion $Q = q_0 + \mathbf{q} = \cos\theta + \mathbf{v}\sin\theta$ and for any vector $\mathbf{v} \in \boldsymbol{R}^3$ the action of the operator $L_Q(\mathbf{v}) = Q\mathbf{v}Q^*$ on **v** may be interpreted geometrically as a rotation of the vector **v** through an angle 2θ about **q** as the axis of rotation. ■ (Kuipers, 2000, p.131)

This can be interpreted as the rotation of vector **v** via angle 2θ about Q. Where Q is the axis of rotation. This means that the process of rotating a point through an arbitrary axis using quaternions is a rather straight forward For example if a 'pure quaternion' if there is a vector r that is an element of $\mathbf{R}^3$ where r = 2i + 2j +2k where to rotate the point r about the vector **s** = 4i +4j at an angle θ=π/5. By using *Theorem 1* the calculation can be done as follows:

Let Q =cos (θ/2)+**s**/|**s**|sin(θ/2) then by substitution it can be written:

$$Q = \cos\frac{\pi/5}{s} + \frac{4i+4j}{(4)^2+(4)^2}\sin\frac{\pi/5}{2} = \cos\frac{2\pi}{5} + i\frac{2}{32}\sin\frac{2\pi}{5} + j\frac{2}{32}sin\frac{2\pi}{5}$$

$$= cos\frac{2\pi}{5} + i\frac{1}{16}sin\frac{2\pi}{5} + j\frac{1}{16}sin\frac{2\pi}{5}$$

Where $q_0 = cos\frac{2\pi}{5}, q_1 = i\frac{1}{16}sin\frac{2\pi}{5}, q_2 = j\frac{1}{16}sin\frac{2\pi}{5}$ $and\ q_3 = 0$

(Gravelle, 2006, p.10-12).

From *Thorem 1* The the Rodrigues' rotation formula can also be derived. The Rorigues' rotation formula as mentioned earlier was a precurser to quaternions. Let $\mathbf{H_1}$ = the set of *all* unit quaternions. Let the 'pure quaternions' be r = (0, **v**). Let a unit quaternion $Q = (q_0, \boldsymbol{q})$ and let $P' = QPQ^* = QPQ^{-1}$ so P' can be writen in terms of pure vectors and unit quaternions can be written as

$$\mathrm{P}' = (0, [q_0^2 - \mathbf{q} \cdot \mathbf{q}]\mathbf{v} + 2\mathbf{v} \cdot \mathbf{q}\mathbf{q} + 2q_0\mathbf{q} \times \mathbf{v})$$

For an angle θ and a unit vector **a**, the scalar and vector can be broken into two parts as follows:

$$q_0 = \cos\frac{1}{1}\theta \text{ and } \mathbf{q} \equiv (\mathrm{q}_1, \mathrm{q}_2, \mathrm{q}_3) = \sin\frac{1}{2}\theta\mathbf{v}$$

$$\mathrm{P}' = (0, [1 - \cos\theta]\mathbf{aa} \cdot \mathbf{v} + \cos\theta + \sin\theta\,\mathbf{a} \times \mathbf{v})$$

The vector part is essencially the Rodrigues' rotation formula. This is going to be discussed in the following section, although the notation will be slightly different. (Heard, 2006, pp.23-24)

### iii. Rodrigues' Rotation Formula

As mentioned in the previous chapter Rodrigues had a much stronger understanding of the algebra of rotations than Hamilton did. It is for this reason that the author has set aside an independent discussion of these rotations.

These rotations were discovered by an obscure French-Jewish mathematician named Benjamin Olinde Rodrigues, prior to Hamilton. Rodrigues is known for two discoveries. The first one was part of his doctorate in mathematics for the Faculty of Science of the University of Paris in 1816. In his dissertation he derived a formula for what is called today Legendre polynomials. His discovery, as it turned out, was an earlier version of Legendre polynomials making him an independent discoverer. (O'Connor, 2006).

The other major contribution of Rodrigues, and the one of concern here, is referred to as Rodrigues rotation formula. He published this in 1840. The Rodrigues rotation formula is an algorithm for the composition of successive finite rotations using geometric methods. This really is essentially the same idea as the composition of unit quaternions. (O'Connor, 2006).

The Rodrigues' rotation formula had two distinct versions:

The exponential formula and the vectorial formula.

The exponential Rodrigues rotation formula is given by $\mathbf{R}_\mathbf{v}(\phi)=e^{v\phi}$. This is an efficient way to map, so(3) → SO(3), where big SO(3) is the group of all rotations in three dimentions, and little so(3) is all 3 x 3 anti-symetric real matrices. This is also called a Lie algebra of SO(3) (Weisstein, 2002, p.2581).

What Rodrigues' formula gives is an algorithm to compute the exponential map from the Lie Group to its corrsponding rotation group without having to compute the full exponential matrix. The other Rodrigues rotation formula is the vectorial form; this is the one that is more useful for calculations. In this form let the vector rotation corresponds to an angle, θ. Let this rotation be counterclockwise about a fixed z-axis. Let this vector be **v** where ***v***=*(a,b,c)* where *a,b,c* are elements of **R** , ***v***=*ax* + *by* + *cz* and *a* corresponding vector ***v'*** = *ax'* + *by'* + *cz* is the image vector where *x'* and *y'* These are rotated by angle *θ* where *x* and *y* are vectors in the *x, y* plane. The rotation formula in 2-dimentions is given by $x' = x \cos\theta - y \sin\theta$; $y' = x \sin\theta + y \cos\theta$. Use this to get $v' = \cos\theta(ax + by) + \sin\theta(ay - bx) + cz$.

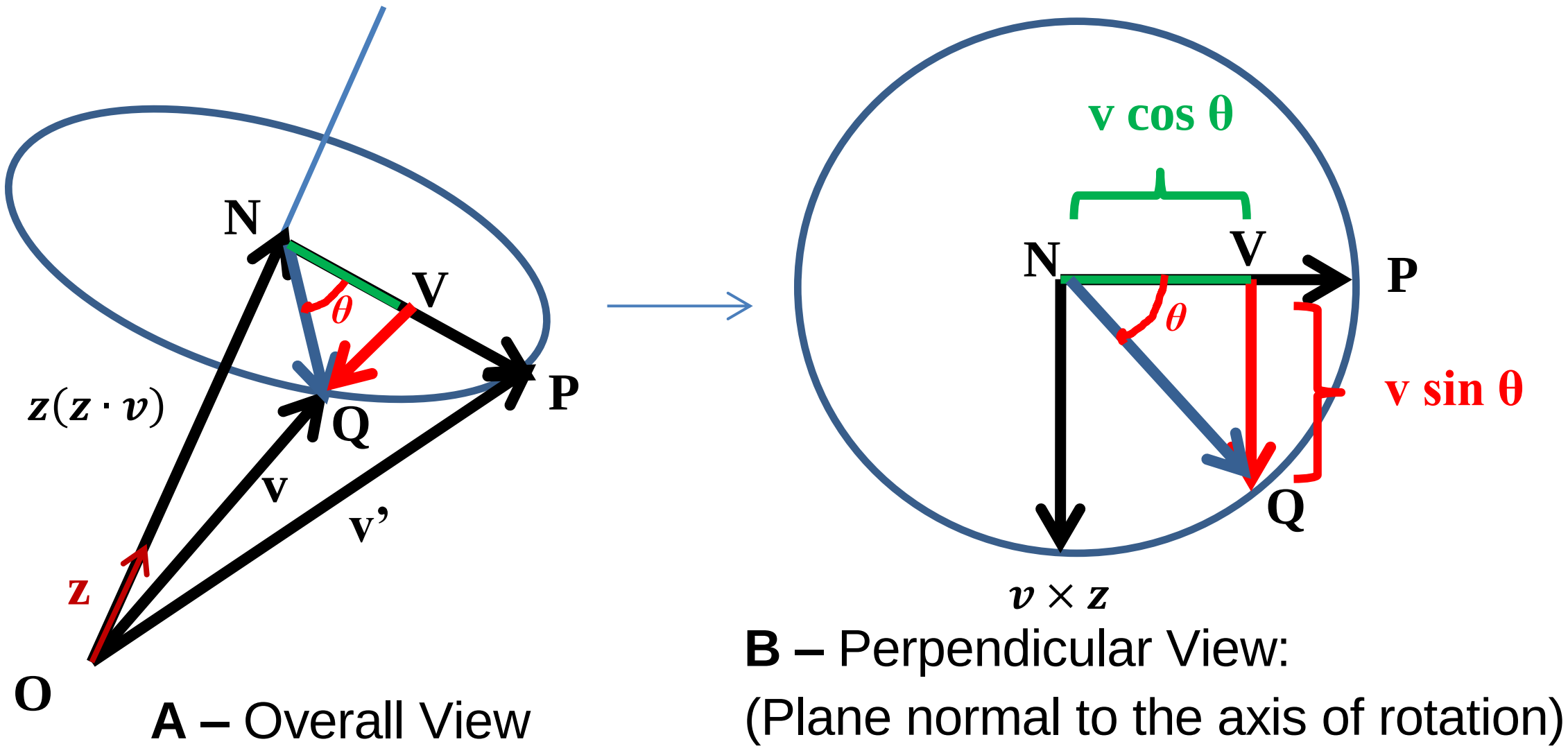

**Figure 5 –** A vector diagram for the derivation for the rotation formula where **v'**=OQ= ON+NV+VQ = **z(z·v)** + (**v**– **z(z·v)**)cosθ + (**v** x **z**)sinθ (Adapted from Goldstein, 2000, figure 4.8; Weisstein, 2002, p.2596)

Let **x** and **y** be orthogonal in the plane. Then vector $a\mathbf{x} + b\mathbf{y}$ is a projection of **v** onto the **x**, **y** plane, and $a\mathbf{y} - b\mathbf{x}$ is a 90° rotation. Where $\mathbf{v} - (\mathbf{v} \cdot \mathbf{z})\mathbf{z} = \mathbf{v} - c\mathbf{z} = a\mathbf{x} + b\mathbf{y}$ and $\mathbf{z} \times \mathbf{v} = a(\mathbf{z} \times \mathbf{v}) + b(\mathbf{z} \times \mathbf{y}) + c(\mathbf{z} \times \mathbf{z}) = a\mathbf{y} - b\mathbf{x}$. Substituting this into the equation for **v'** to get:

$\mathbf{v}' = \cos\theta(a\mathbf{x} + b\mathbf{y}) + \sin\theta(a\mathbf{y} - b\mathbf{x}) + c\mathbf{z} = \cos\theta(\mathbf{v} - (\mathbf{v} \cdot \mathbf{z})\mathbf{z}) + \sin\theta(\mathbf{z} \times \mathbf{v}) + c\mathbf{z}$ can be rewritten *as* $\mathbf{z}(\mathbf{z} \cdot \mathbf{v}) + (\mathbf{v} - \mathbf{z}(\mathbf{z} \cdot \mathbf{v}))\cos\theta + (\mathbf{v} \times \mathbf{z})\sin\theta$ as in figure 5 This is the same Rodrigues rotation formula encountered earlier, although the notation is slightly different (Chang, 2013, p.1).

There are many different ways to derive Rodrigues' rotation formula; this was only one of them. The vectorial formula can also be derived from the exponential formula. This was done in an elegent way in notes by Laura Downs and Alex Berg from Berkeley (Downs, n.d.). This may have been the way Rodrigues first derived this formula, since it involves tools that were available to him at the time, such as Taylor's series.

In this derivation it will be necessary to evaluate the matrix exponential $e^{A\theta}$. This means that an n x n matrix **A** (real or complex) can be found using the exponential Taylor serise as follows:

$$e^x = I + \frac{x}{1!} + \frac{x^2}{2!} + \frac{x^3}{3!} + \frac{x^4}{4!} + \cdots$$

Where if **A** is substuted for x the serise becomes

$$e^{A} = I + \boldsymbol{A} + \frac{\boldsymbol{A}^2}{2!} + \frac{\boldsymbol{A}^3}{3!} + \frac{\boldsymbol{A}^4}{4!} + \cdots$$

Forming an n x n matrix that converges for all complex matrices (Artin, 1991, p. 138-139)

In order to evaluate **A** needs to be created. **A** is the linear transformation. This linear transformation is used to compute the cross product of the unit vector, **a** with another vector, **v**. For example:

$$\boldsymbol{a} \times \boldsymbol{v} = \begin{pmatrix} a_y v_z & -a_z v_y \\ a_z v_x & -a_x v_z \\ a_x v_y & -a_y v_x \end{pmatrix} = \begin{pmatrix} 0 & -a_z & a_y \\ a_z & 0 & -a_x \\ -a_y & a_x & 0 \end{pmatrix} \begin{pmatrix} v_x \\ v_y \\ v_z \end{pmatrix} = \boldsymbol{Av}, where\ \boldsymbol{A}$$

$$= \begin{pmatrix} 0 & -a_z & a_y \\ a_z & 0 & -a_x \\ -a_y & a_x & 0 \end{pmatrix}$$

To write the rotation matrix in terms of **A**: $Q = e^{\boldsymbol{A}\theta} = I + \boldsymbol{A} \sin(\theta) + \boldsymbol{A}^2 [1 - \cos(\theta)]$. Expand $e^{\mathbf{A}\theta}$ as a Taylor series as follows:

$$e^{A\theta} = I + \frac{(A\theta)^2}{2!} + \frac{(A\theta)^3}{3!} + \frac{A\theta^4}{4!} + \cdots$$

Looking back on how **A** was constructed $-\mathbf{A}=\mathbf{A}^3$ , each additional application of **A** turns the plane in the approperate direction, to get:

$$e^{A\theta} = I + \left[A\theta - \frac{A\theta^3}{3!} + \frac{A\theta^5}{5!} + \cdots\right] + \left[\frac{A^2\theta^2}{2!} - \frac{A^2\theta^4}{4!} + \frac{A^2\theta^6}{6!} + \cdots\right]$$

$$= I + A\left[\theta - \frac{\theta^3}{3!} + \frac{\theta^5}{5!} + \cdots\right] + A^2\left[\frac{\theta^2}{2!} - \frac{\theta^4}{4!} + \frac{\theta^6}{6!} + \cdots\right]$$

From this the Taylor expansions of *sin (θ)* and *cos(θ)* , $e^{\boldsymbol{A}\theta}$ becomes obvious, where the Euler formula can be writen as $I$ + $\boldsymbol{A}$ *sin(θ)* + $\boldsymbol{A}^2$ *[1 - cos(θ)]*. Where

$$\boldsymbol{A} = \begin{pmatrix} 0 & -a_z & a_y \\ a_z & 0 & -a_x \\ -a_y & a_x & 0 \end{pmatrix}$$ As before. (Downs, n.d.)

### D. Octonions

Hamilton came up with quaternions. He wrote to his friend John Graves about his discovery and revelation on Broom Bridge. Graves wrote back to Hamilton "If with your alchemy you can make three pounds of gold, why should you stop there?" (Beaz, 2001 , p.146). Three months later on December 26th Graves wrote to Hamilton about his discovery of a kind of 'double-quaternion' that he called 'octives', today are they are known as octonions. Hamilton said that he would announce Graves discovery to the Irish Royal Society; this was the way that mathematical ideas were published at the time. Unfortunately for Graves, Hamilton never got around to it, and in 1845 Arthur Cayley independently discovered the octonions, which became known as 'Cayley numbers'. Thus due to Hamilton's fumble Cayley beat Graves to publication and got the credit for them initially (Penrose, 2004, p.202).

Like quaternions octonions form a division algebra, but unlike quaternions they are not associative. They are related to geometries in 7 and 8 dimensions. This wasn't very useful in the 19th century physics. Nineteenth century physics could barely deal with 4-dimensions, so octonions were more of a mathematical curiosity than anything else. Octonions it appears were just a mathematical ‘fluke’. It wasn’t until the advent of modern particle physics that octonions would find a place in theoretical physics. This took over a century for scientists and mathematicians to find a use for these numbers (Baez &Huerta, 2011, p.63). What is truly amazing is how an apparently whimsical discovery in the 19th century has become so important in 21st century theoretical physics.

Mathematically, octonions are the largest normed division algebras. This means that they satisfy $|ab|^2=|a|^2|b|^2$. When Hamilton discovered quaternions had to give up commutative property in order to preserve the norm; now the associative property has to be given up in order to preserve the norm for the octonions. A multiplication table for octonions can be made like quaternions as follows:

| | $e_1$ | $e_2$ | $e_3$ | $e_4$ | $e_5$ | $e_6$ | $e_7$ |
|---|---|---|---|---|---|---|---|
| $e_1$ | -1 | $e_4$ | $e_7$ | $-e_2$ | $e_6$ | $-e_5$ | $-e_3$ |
| $e_2$ | $-e_4$ | -1 | $e_5$ | $e_1$ | $-e_3$ | $e_7$ | $-e_6$ |
| $e_3$ | $-e_7$ | $-e_5$ | -1 | $e_6$ | $e_7$ | $-e_4$ | $e_1$ |
| $e_4$ | $e_2$ | $-e_1$ | $-e_6$ | -1 | $e_7$ | $e_3$ | $-e_5$ |
| $e_5$ | $-e_6$ | $e_3$ | $-e_2$ | $-e_7$ | -1 | $e_1$ | $e_4$ |
| $e_6$ | $e_5$ | $-e_7$ | $e_4$ | $-e_3$ | $-e_1$ | -1 | $e_2$ |
| $e_7$ | $e_3$ | $e_6$ | $-e_1$ | $e_5$ | $-e_4$ | $-e_2$ | -1 |

**Table 2:** Multiplication table for Octonions. (Baez, 2001c)

During the latter part of the 19th century Killing and Cartan classified what are referred to as the ‘simple’ Lie groups. They also discovered the ‘exceptional’ Lie groups: $F_4$, $G_2$, $E_6$, $E_7$ and $E_8$, all of which can be constructed using octonions. These exceptional Lie groups are the ones that are important to string theory, especially $E_8$. (Baez, 2001b). John Baez of University of California, Riverside, has studied octonions extensively.

Some of the properties of octonions as they relate to quaternions are the multiplication table for octonions, Table 2, where $e_1$,$e_2$, and $e_3$ represent the quaternions which are a subset of the octonions. (Beaz, 2001a, p.150)

The Fano plane analogy for Octonions from table 3 of Cayley and Graves is shown in Figure 17. The Fano plane is smallest projective plane. This plane is order 2. Each of the lines on the triangular diagram contains three points. The diagram has a total of 7 points. Each of the triples has an oriented cycle ordered by the arrows that tell how to multiply them. Although Fino planes do not normally have arrows. The arrows were added to make the analogy easer to follow.

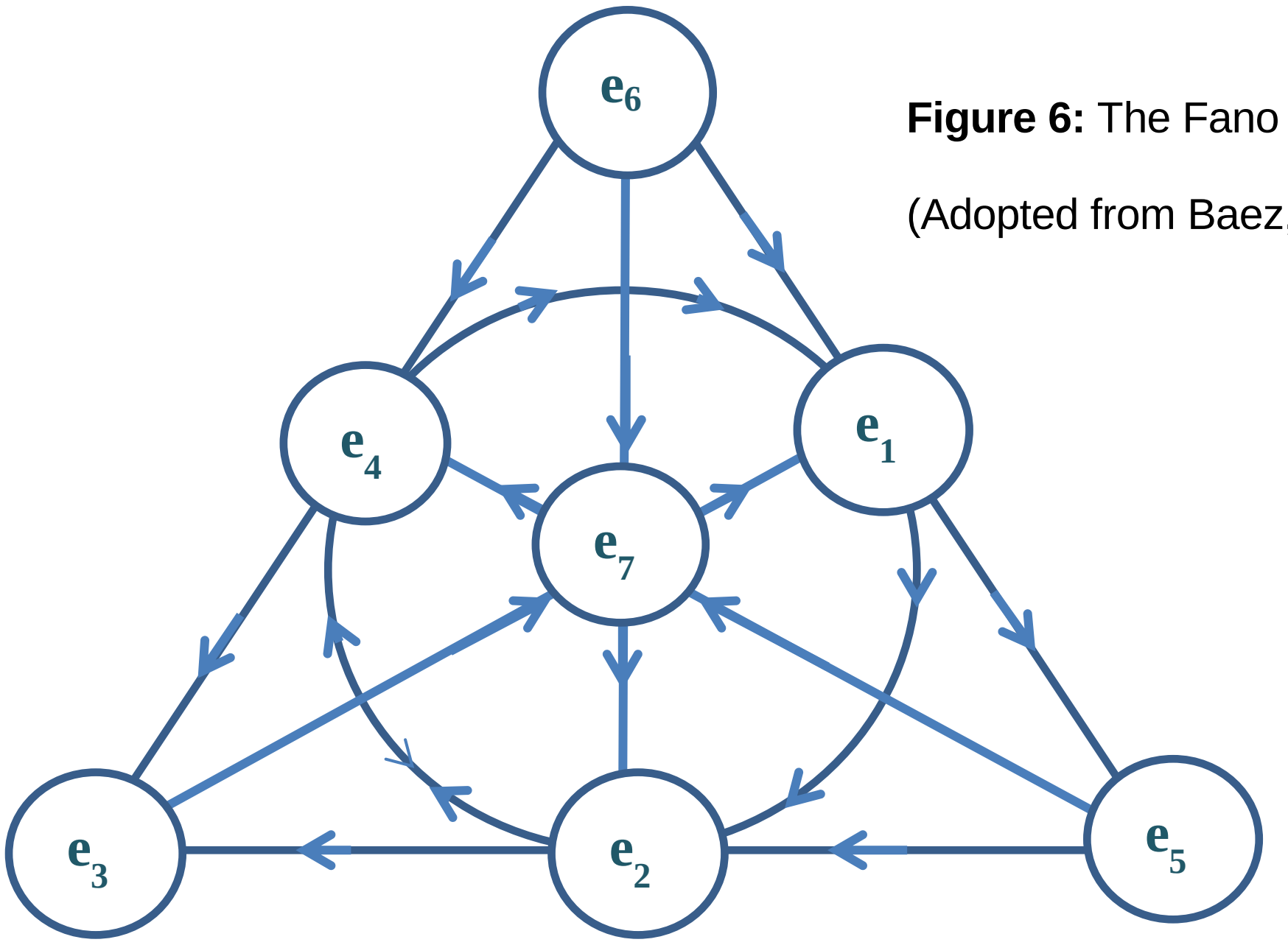

**Figure 6:** The Fano plane for octonions

(Adopted from Baez, 2001, p.152)

For example: $e_i$, $e_j$, and $e_k$ can be represented as follows: $e_ie_j = e_k$; $e_je_i = -e_k$ . Let 1 be the multiplicative identity. Let each circle in the diagram be represented by the relationship $e_i^2 = -1$. (Beaz, 2001a, p.152; Richter, n.d.)

**E. Grassmann Algebras**

Grassmann was a person who was truly ahead of his times. Not only was his work done in n-dimensions, which is truly a modern concept. This is one of the reasons that Grassmann's algebras or exterior algebras are considered to be more general than vector analysis. Grassman's algebras also came before both vector and tensor analysis and overlap both these subjects (Browne, 2001, 2). For this reason a more detailed mathematical description of Grassmann's work will be discussed in this section.

The idea of the exterior product is similar to what is referred to today as linear independence. This is when points or vectors are non-zero the exterior product of this is considered to be linearly independent. To understand more about Grassmann's reasoning let a line as 2 connected points, the plane as 3 connected points and so on. For example the $\mathbf{R}^2$ is a vector space with a basis with unit vectors $e_1=(1,0)$, $e_2=(0,1)$. Let $\boldsymbol{v} = v_1e_1 + v_2e_2$, $\boldsymbol{w} = w_1e_1 + w_2e_2$ are a pair of vector, also called 'bivectors' or '2-blades' in $\mathbf{R}^2$, where **v** and **w**, for example, can be represented as two of the sides a parallelogram. Their area is $A = |det[\mathbf{v} \quad \mathbf{w}]| = |v_1w_2 - v_2w_1|$. This is called the exterior product or wedge product of **v** and **w**. Where the wedge product of **v** and **w** are $\boldsymbol{v} \wedge \boldsymbol{w} = (v_1\boldsymbol{e}_1 + v_2\boldsymbol{e}_2) \wedge (w_1\boldsymbol{e}_1 + w_2\boldsymbol{e}_2)$. The distributive law holds where $v_1w_1\boldsymbol{e}_1 \wedge \boldsymbol{e}_1 + v_1w_2\boldsymbol{e}_1 \wedge \boldsymbol{e}_2 + v_2w_1\boldsymbol{e}_2 \wedge e_1 + v_2w_2\boldsymbol{e}_2 \wedge \boldsymbol{e}_2$. The wedge product alternates, which means that $\boldsymbol{e}_2 \wedge \boldsymbol{e}_1 = -\boldsymbol{e}_1 \wedge \boldsymbol{e}_2$, or more generally $\boldsymbol{e}_i \wedge \boldsymbol{e}_j =$
$-\boldsymbol{e}_j \wedge \boldsymbol{e}_j$,. Using this idea, canceling and collecting like terms to get $(v_1w_2 - v_2w_1)\ \boldsymbol{e}_1 \wedge \boldsymbol{e}_2$. The signs of **v** and **w** determine the orientation of the vertices of the parallelogram being constructed: positive for clockwise, negative for counterclockwise as in Figure 6. (Browne, 2001, p.3).

To Grassmann the shape of the geometric figure being defined wasn't important. The reason that a parallelogram is being used in this example is because of its simplicity. In order to generalize some of Grassmann's ideas the example using a parallelogram can be used to discuss Grassmann's ideas without losing a sense of the geometric intuition. Perhaps rounder shapes like the oval or ellipsoid would be more easily imagined when applying these ideas to tides, rather than a parallelogram or parallelepiped as shown in Figures 7a and 7b respectively.

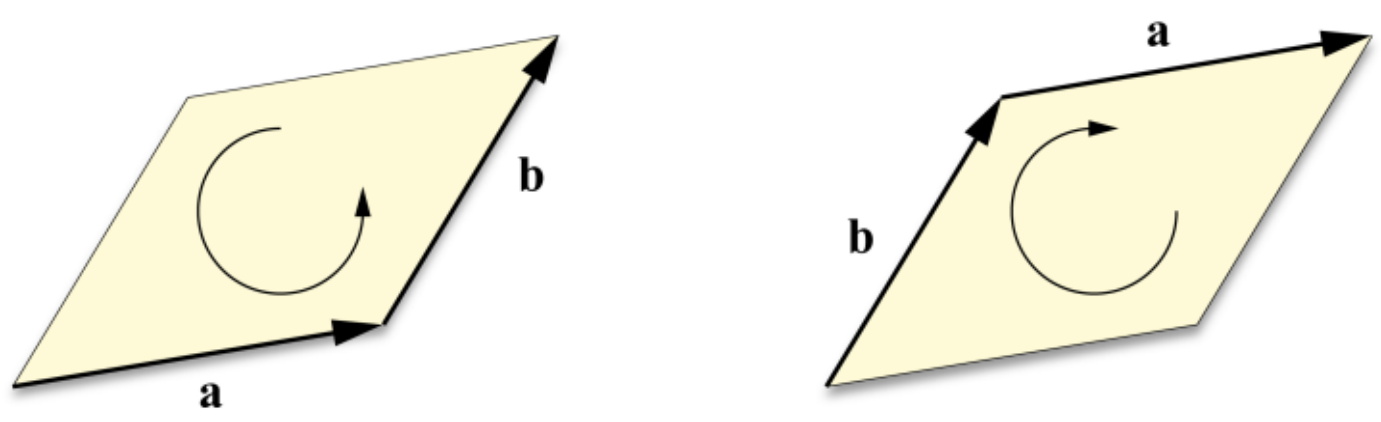

**a∧b** **b∧a**

**Figure 7:** Geometric interpretation of **b∧a = -a∧b**

(From : Lengyel, 2012, Slide 21; Lundholm, 2009, figure 1.3)

Grassmann's algebras make it relatively easy to go from lower dimensions to higher dimensions. To do this take the lower dimension and multiply it by the new element in a higher dimension as shown. For example this is shown in Figure 8 of the parallelepiped as follows (Browne, 2001, p.3):

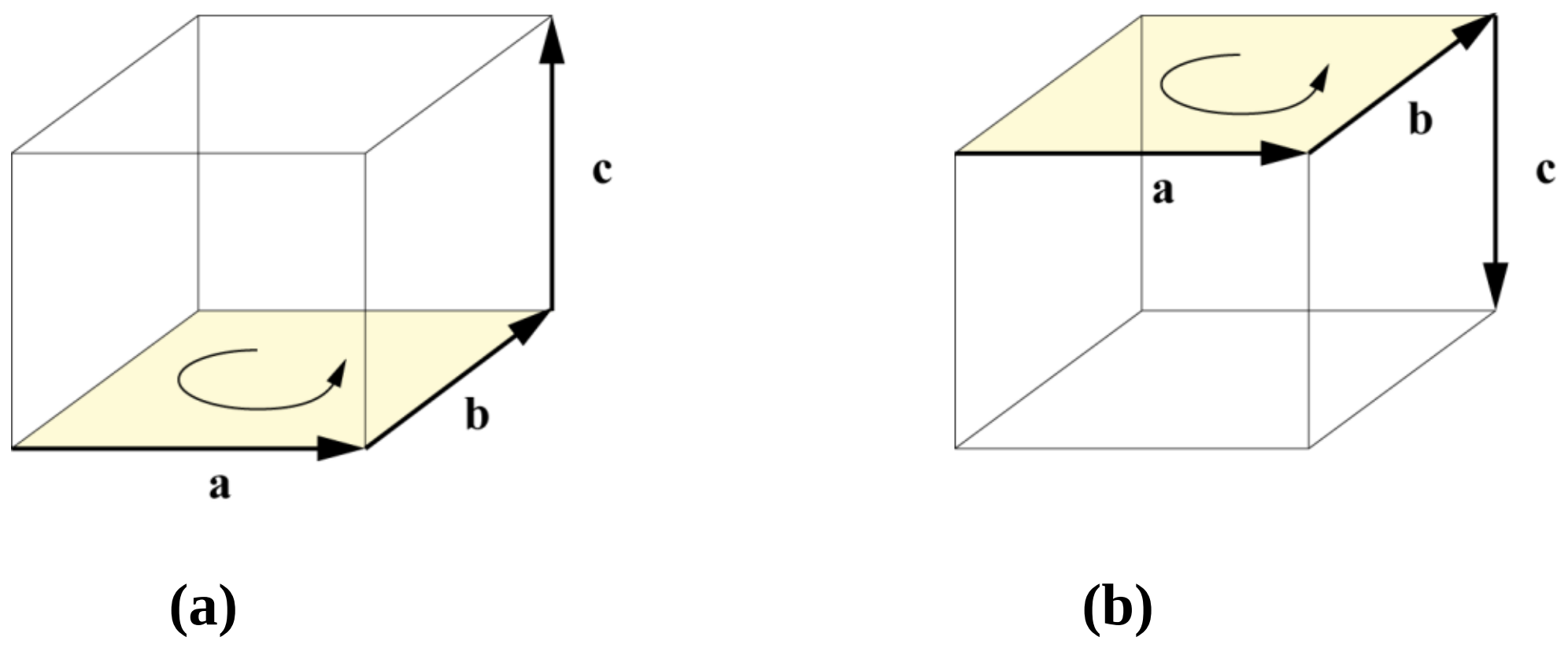

**Figure 8**: **a∧b∧c**

(from Lengyel, 2012, Slide 24; Lundholm, 2009, figure 1.5)

Let P(**v,w**) be area of the parallelogram formed by a pair of vectors **v** and **w.** Then

1. Rescale each side by any ordinary real number, say, s and t then $P(s\mathbf{v},t\mathbf{w}) = st\,P(\mathbf{v},\mathbf{w})$ . This means that the area is rescaled by the same amount, for example $s \wedge t = t \wedge s = st$ where $s \wedge \mathbf{v} = \mathbf{v} \wedge s = s\mathbf{v}$.
2. To reverse the orientation of the parallelogram, one of the sides of the parallelogram has to have its direction switched. This means $P(\mathbf{w},\mathbf{v}) = -P(\mathbf{v},\mathbf{w})$; reversing the sign reverses the direction.
3. Non-commutative of exterior products is due to the fact that $P(\mathbf{v},\mathbf{v}) = 0$. This means that $\mathbf{v}\wedge\mathbf{v} = 0$, for example let $(\mathbf{v}+\mathbf{w}) \wedge (\mathbf{v}+\mathbf{w}) = 0$ solving this gives $\mathbf{v}\wedge\mathbf{v} + \mathbf{v}\wedge\mathbf{w} + \mathbf{w}\wedge\mathbf{v} + \mathbf{w}\wedge\mathbf{w} = \mathbf{v}\wedge\mathbf{w} + \mathbf{w}\wedge\mathbf{v}$ so that $\mathbf{v}\wedge\mathbf{w} = -\mathbf{w}\wedge\mathbf{v}$ as in Figure 11.

4. Adding or multiplying the vectors **v** and **w** does not affect the area of the parallelogram $P(\mathbf{v} + a\mathbf{w},\mathbf{w}) = P(\mathbf{v},\mathbf{w})$,
5. The unit square is defined as one where $P(e_1, e_2) = 1$

(Lengyel, 2012, 11-12).

Another type of product that Grassmann discovered, but is rarely mentioned in most books, the 'anti-wedge' product written $\overline{e_1} \vee \overline{e_2}$, this has the same properties as the exterior product but operates in a dual basis as 'antivectors' in an 'antivector' space. These represent an absence of geometry and behave by removing vectors. This means essentially that the 'wedge' is used to build up 'grades' the 'anti-wedge' is used to tear them down, or to reduce them. For example Let there be an n-dimensional Grassmann algebra. In this algebra if $s$ has a grade S and $t$ has a grade T than $s \wedge t$ has grade S + T for t, but for the anti-wedge $s \vee t$ = S + T – n (Lengyel, 2012, slides 60-62).

The cross product mentioned earlier is defined only in three dimensions, while the exterior product is defined in all dimensions. The overlap between the cross product and the exterior product can be seen by letting $\{e_1, e_2, e_3\}$ be a basis for the vector space. Define the exterior product of a pair of vectors $\boldsymbol{u} = u_1e_1 + u_2e_2 + u_3e_3$ and $\mathbf{v} = v_1e_1 + v_2e_2 + v_3e_3$. The exterior product can be calculated as $u\wedge\mathbf{v} = (u_1v_2 - u_2v_1)(e_1\wedge e_2) + (u_1v_3 - u_3v_1)(e_1\wedge e_3) + (u_2v_3 - u_3v_2)(e_2\wedge e_3)$, $\{e_1 \wedge e_2, e_1 \wedge e_3, e_2 \wedge e_3\}$ form the basis for 2-forms on three-dimensional space. The components $(u_1v_2 - u_2v_1)$, $(u_1v_3 - u_3v_1)$, and $(u_2v_3 - u_3v_2)$ are the same components that are produced by doing the cross product $\mathbf{u}\times\mathbf{v}$. Let $\mathbf{w} = w_1e_1 + w_2e_2 + w_3e_3$, be another vector where the exterior product of the 3 vectors can be calculated as follows:

$$\mathbf{u}\wedge\mathbf{v}\wedge\mathbf{w}= (u_1v_2w_3 + u_2v_3w_1+ u_1v_3w_2 - u_1v_3w_2 - u_2v_1w_3 - u_3v_2w_1)\,(e_1\wedge e_2\wedge e_3)$$

This coincides with the definition of the triple product.

The cross product and triple product have both geometric and algebraic meanings. The cross product **u** × **v** can be represented as a vector that is perpendicular to both **u** and **v.** These vectors determine the magnitude of the parallelogram. Geometrically the triple product of **u, v**, and **w** is a volume. This is the same idea that was seen earlier when using the analogy of the parallelogram to discuss the axioms of the exterior products (Schulz, 2012, p.61, p.65).

Higher products of the cross product are not associative $(\mathbf{u}\times\mathbf{v})\times\mathbf{w} \neq \mathbf{u}\times(\mathbf{v}\times\mathbf{w})$**.** This is the reason why the exterior product is defined in all dimensions and the cross product is only defined in 3-dimentions. In summary $\mathbf{u}\wedge\mathbf{v}\wedge\mathbf{w}$ has the same magnetude as $(\mathbf{u}\times\mathbf{v})\times\mathbf{w}$**,** but the exterior product can be extended to other dimensions where the cross product cannot.

Today mathematics and physics occurs in multidimensional space. This has become the norm making Grassmann's approach more viable today than when it was invented 150 years ago.

## F. Clifford Algebras

Recall that Clifford found a way to take the best of quaternions and Grassmann algebras, incorporating them into one type of mathematics he called Geometric Algebra. Here the mathematics itself will be briefly discussed. This discussion will not do this subject the justice it deserves, but will aid as a brief introduction to the subject.

Grassmann discovered, but did not develop, the fact $uv = u{\cdot}v + u \wedge v$, meaning that the product of 2 vectors is the sum of a scalar and a '2-blade'. This is called the geometric product and is analgous to the real and imaginary parts of a complex number.

To understand Clifford Algebras consider an orthonormal basis $\{e_1,e_2\}$ of the Euclidian space $\mathbf{R}^2$. Let 2 vectors be defined where $\boldsymbol{u} = u_1e_1 + u_2e_2$ and $\boldsymbol{v} = v_1e_1 + v_2e_2$ such that $e_1 \cdot e_1 = e_2 \cdot e_2 = 1$ and $e_1 \cdot e_2 = 0$.

Parellel vectors commute in the following way: $e_1\, e_1 = e_1 \cdot e_1 + e_1 \wedge e_1 = 1$, but orthonormal (perpendicular) ones do not: $e_1\, e_2 = e_1 \cdot e_2 + e_1 \wedge e_2 = -e_2 \wedge e_1 = -e_2\, e_1$. Unit vectors have a negative square: $(e_2 \wedge e_1)^2 = (e_1\, e_2)(e_1\, e_2) = e_1\, e_2(-e_2\, e_1) = -e_1\, e_1 = -1$ (Lasenby, 2003, slides11-13 )

What Clifford did was replace Grassmann's rules $e_j \wedge e_j = 0$ by the rules $e_j\, e_j = 1$, which could be either +1 or -1, according to Clifford this is the identity. As Clifford put this change "the Grassmann algebra is nilpotent, and only homogeneous forms occur; while this is idempotent, and admits of odd forms and even forms, which are not in general homogeneous". Clifford also replaces $e_j \wedge e_k = -e_k \wedge e_j$ for $j \neq k$ by $e_j\, e_k = -e_k\, e_j$ for $j \neq k$. (Clifford, 1876 , pp.397-398; Diek, n.d. , p.2)

This means that for a vector **u** , $\boldsymbol{u}^2 = \boldsymbol{uu} = \varepsilon|\boldsymbol{u}|^2$ , where $|\boldsymbol{u}| \geq 0$. This is called the magnetude of **v**. The signiture ε could be positive if ε = 1, negitive if the signiture is $\varepsilon = -1$, or null if |**u**| = 0 (Hestenes, 2011 , p.250)

For any two vectors, say **u** and **v** the inner (scalar) and outer (vector) product can be written as: $\mathbf{u} \cdot \mathbf{v} = ½ (\mathbf{uv} + \mathbf{vu}) = \mathbf{v} \cdot \mathbf{u}$ and $\mathbf{u} \wedge \mathbf{v} = ½ (\mathbf{uv} \wedge \mathbf{vu}) = -\mathbf{v} \wedge \mathbf{u}$. Geometrically the product of two vectors is defined as being: $\mathbf{uv} = \mathbf{u} \cdot \mathbf{v} + \mathbf{u} \wedge \mathbf{v}$, where the geometric product is both associative and distributive respectively as: **u(vw)=(uv)w=uvw** and **u(v+w)=uv+uw**. Squaring the vectors produces a scalar: $(\mathbf{u}+\mathbf{v})^2=\mathbf{u}^2+\mathbf{v}^2+\mathbf{uv}+\mathbf{vu}$.

The outer or exterior product for matrix multiplication is $\mathbf{uv}^{\mathbf{T}}$, where T represents the transpose of matrix **v**. The outer product can written as $\mathbf{u} \otimes \mathbf{v}$; for example let **u** be an m x 1 column vector. Let **v** be an n x 1 column vector, further more let m = 4 and n = 3. These can be represented as follows:

$$\boldsymbol{u} \otimes \boldsymbol{v} = \boldsymbol{u}\boldsymbol{v}^{\boldsymbol{T}} = \begin{bmatrix} u_1 \\ u_2 \\ u_3 \\ u_4 \end{bmatrix} [v_1 \quad v_2 \quad v_3] = \begin{bmatrix} u_1 v_1 & u_1 v_2 & u_1 v_3 \\ u_2 v_1 & u_2 v_2 & u_2 v_3 \\ u_3 v_1 & u_3 v_2 & u_3 v_3 \\ u_4 v_1 & u_4 v_2 & u_4 v_3 \end{bmatrix}$$

The next object that needs to be defined are called 'grades'. A scalar is considered to be 0-grade, a vector is a 1-grade, a bivector a 2-grade etc, where even grade objects form complex numbers. In general the outer product is an anti-symmetric geometric product of vectors that gives k-blades, where k is any integer. Blades here are defined the same way that Grassmann defined them. The outer product of k implies $u_1 \wedge u_2 \wedge \ldots \wedge u_k \equiv U$ iff $[u_1 u_2 \ldots u_k] \equiv [U]$, where k is called the grade of U (Hestenes, 2011 , p.249).

According to Clifford's original paper *The Classification of Geometric Algebras* he let **u** and **v** be defined as scalars. Clifford considered n-unit vectors $e_1, e_2, \ldots, e_n$, where all $(e_j)^2 = 1$ and $e_j e_k = -e_k e_j$ *for* $(j \neq k)$, and every product consists of basic terms that are either of odd order ( $e_j$, $e_j e_k e_l$, ... such that $j \neq k \neq l$ ), or basic terms of even order ($1$, $e_j e_k$ , ... where $j \neq k$ ) . There can be no term higher than the n[th] order since "…we have altogether one term of order 0, n of order 1, ½ n (n -1) of order 2, ... one of order n; meaning $1 + n + \frac{1}{2} n (n - 1) \ldots + n + 1 = 2^n$ terms" (Clifford, 1876, p.400).

So far it is clear how geometric algebras incorporate Grassmann's algebras, but what about quaternions; recall that in 2-dimentions, vectors can be rotated using the following relationships: $v = e^{i\theta} u$, $u = e_1 \mathbf{x}$, $v = e_1 \mathbf{y}$

Let $e_1 \wedge e_2 = I$ where $y = e_1 \mathbf{v} = e_1 e^{I\theta} e_1 \mathbf{x}$, but $e_1 I e_1 = e_1 e_1 (e_1 e_2) e_1 = e_2 e_1 = -I$. So a rotation can be done for $y = e^{(-I\theta)} x = x e^{(I\theta)}$ as in Figure 9

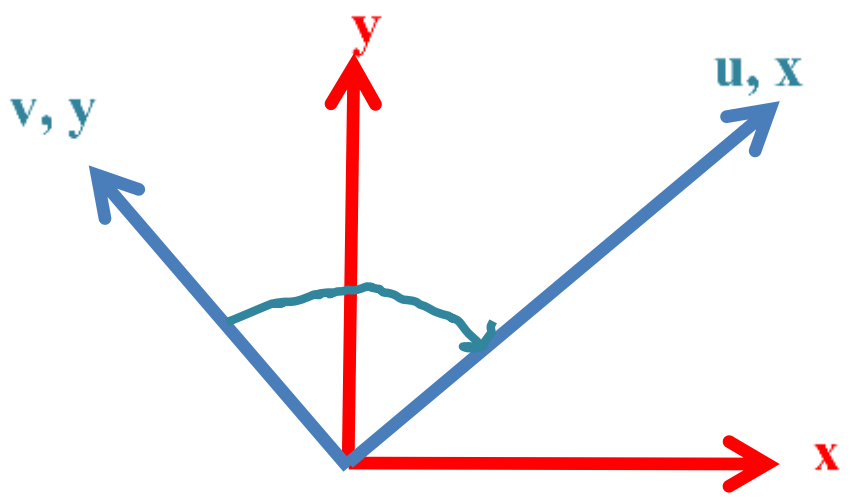

**Figure 9 –** Vectors rotated by complex phase rotations

(Adapted from Dorian,,slide 14)

Let $e_3$ be a third unit vector where all 3 vectors are non-commutative,

$e_1\, e_2 = -\, e_2\, e_1$ etc.. These vectors form 3-blades as follows in figure 10:

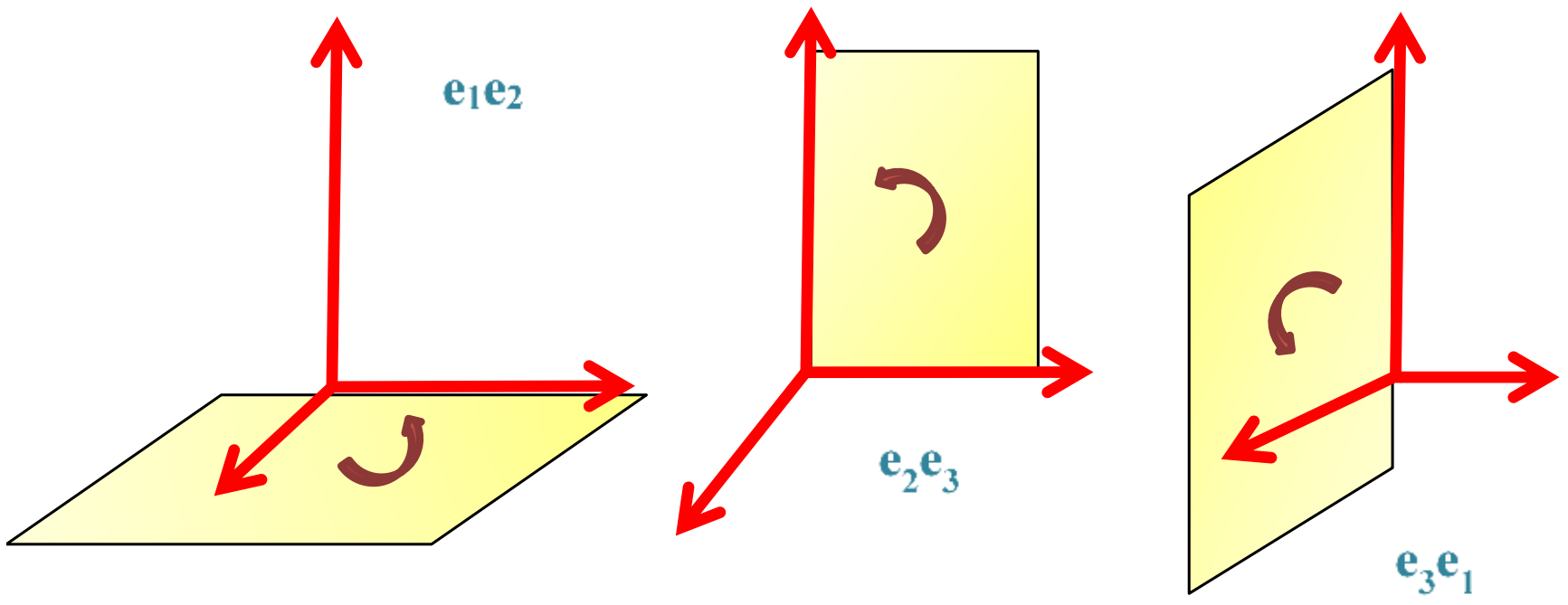

**Figure 10 –** Three 2-blades (Adapted from Dorian, 2003, slide 14)

These three 2-blade can form the following geometric objects:

1. The product of a vector and a bivector (2-blade): $e_1(e_1e_2)=e_2$ or $e_1(e_2e_3)= e_1e_2e_3=I$

2. The product of 2 perpendicular bivectors (2-blades): $(e_2e_3)\ (e_3e_1)= e_2e_3e_3e_1= e_2e_1 = -e_1e_2$

3. In order to get back the quaternion set the the following are applies: $i = e_2e_3$ , $j = -e_3e_1$ , $k = e_1e_2$, where $i^2 = j^2 = k^2 = ijk = -1$.

(Lasenby, 2003, slide14-16)

In this section the author of this thesis has given a very basic mathematical discussion on how Clifford proposed to unify the worlds of Grassmann and Hamilton.

## G. Lie Groups, Lie Algebras

Lie groups and Lie algebras are formally two different subjects. Lie himself thought of his work in terms of the infinitesimal group structures. Lie groups and algebras have grown into a deep and complex enterprise far too vast for the interests and scope of this dissertation. Rather than going into too many mathematical details the author will just briefly discuss some facts about the Lie groups and algebras of a few relevant non-commutative rotation groups that are important to theoretical physics in this chapter.

Lie groups can be commutative or non-commutative. There are really only a few Lie groups that most physicists will encounter. The groups discussed here will be restricted to the ones that are related to quaternions, namely the ones that are non-commutative and complex, with the exception of SU(3) which is an octonion.

By definition the Lie group G is a continuous group of symmetries (Evans, 2013, p.7). Symmetries are very important in theoretical physics. The fact that they are continuous is also important. Continuous means that the manifold is 'smooth', and differentiable everywhere. This means that calculus can be done on a Lie group. The maps $\rho: (x, y) \mapsto xy, G \times G \to G$ are 'smooth' and its inverse $\sigma: x \mapsto x^{-1}, G \to G$ is also 'smooth'. Every Lie group can be associated with a Lie algebra through an exponential mapping (Evens, 2013, p.54).

A simple example of a Lie group is the circle SO(2). This is not a quaternion group but it will serve to illustrate what a rotation group is and what can be done with one. SO(2) exists in Euclidian space on the 2-dimentional plane. It rotates about the origin at angle θ Figure 11.

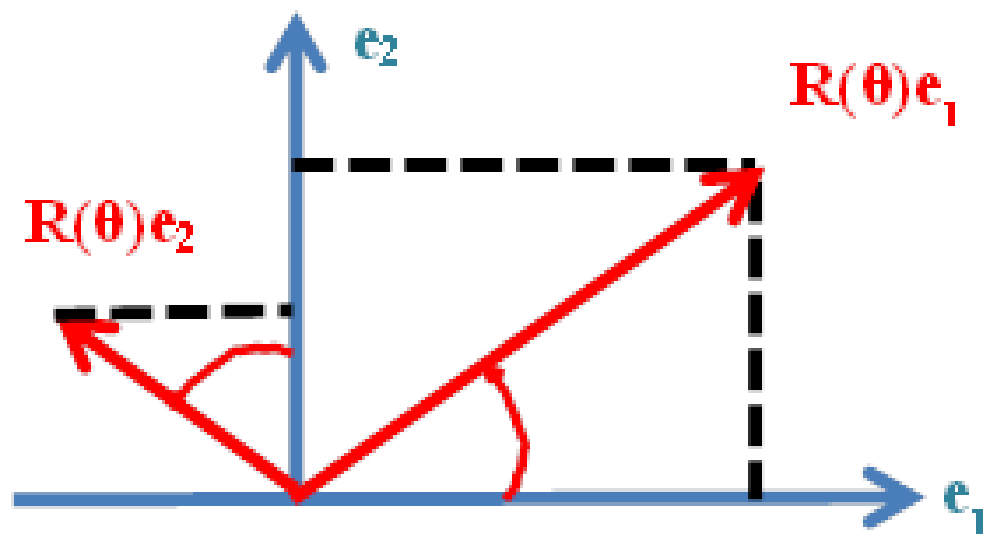

**Figure 11** **Rotation on a plane** (Adapted from, Tung, n.d.,fig 6.1)

Where $R(\theta)e_1 = e_1 \cos \theta + e_2 \sin \theta$ and $R(\theta)e_2 = -e_1 \sin \theta + e_2 \cos \theta$. Its rotation matrix can be written as:

$$R(\theta)e_1 R(\theta)e_2 = (e_1 e_2)\begin{pmatrix}\cos\theta & -\sin\theta\\ \sin\theta & \cos\theta\end{pmatrix}$$

In general $R(\theta)e_i = e_j\, R(\theta)_i^{\,j}$ , the rotation matrix is written

$$R(\theta) = \begin{pmatrix}\cos\theta & -\sin\theta\\ \sin\theta & \cos\theta\end{pmatrix}$$

The rotation preserves the length of the vectors. This means that the length of the vector remains the same after the rotation (Tung, n.d., p.2).

Topologically SO(2) is a circle. It is also abelian (commutative).

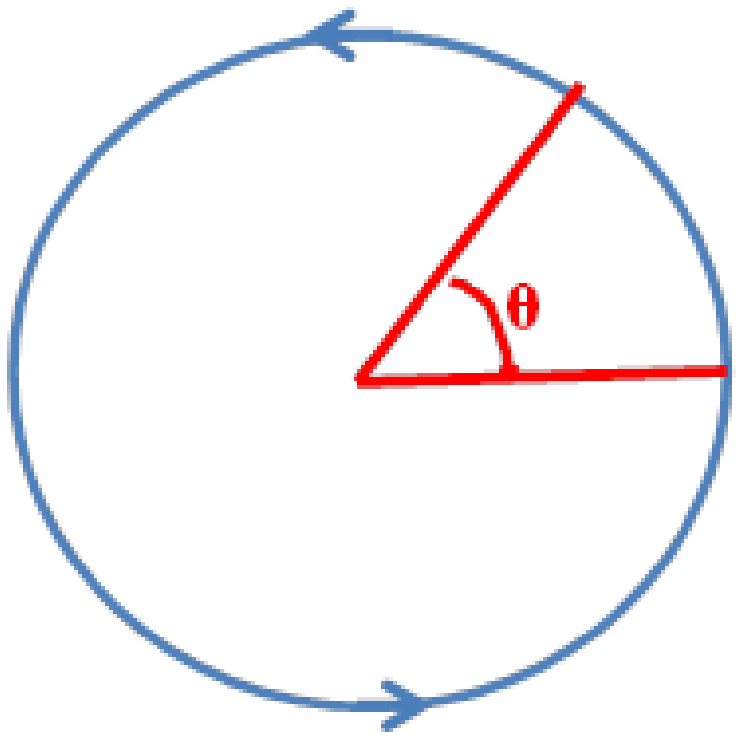

**Figure12:**

SO(2)

(Adapted from, Tung, n.d. fig. 6.2)

As mentioned earlier, SO(2) forms the simplest Lie group. SO(2) has one generator. This means that it is a Lie group of 1. It is a Lie group because all the properties of a Lie group can be defined on it, meaning it satisfies the definition of what a Lie group is: it has a group structure, it and its inverse are 'smooth' i.e. differentiable everywhere. SO(2) is also called a 1-sphere where when mapped to itself forms a double covering where $\mu$:SO(2)→ SO(2) is defined by $r_\theta \rightarrow r_{2\theta}$ . This is illustrated in Figure 13 (Artin, 1991, p.277)

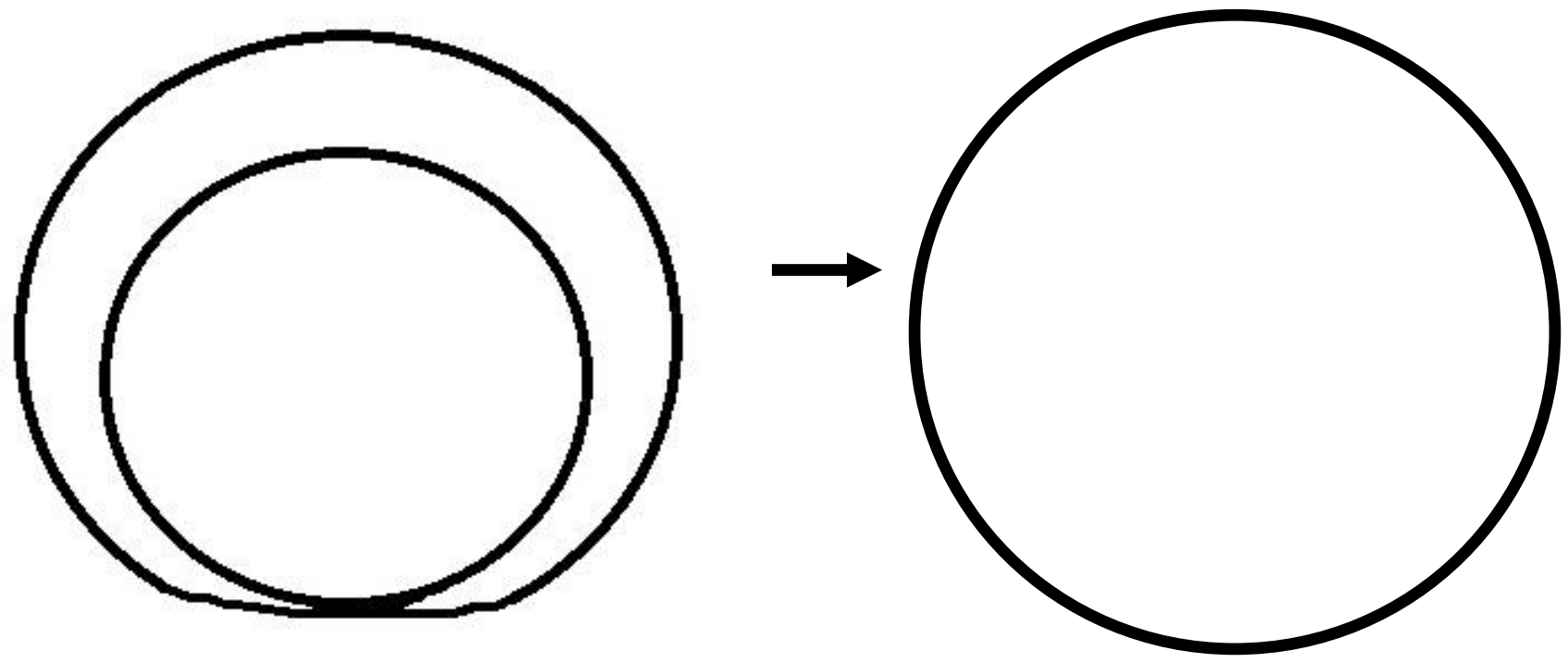

**Figure 13:**
Double cover of SO(2) (Adapted from: Artin, 1991, p.277))

Countless number of books have been written on this subject of Lie Groups. What will be given here is a very superficial way to define Lie groups, Lie algebras and how they are related to each other, without going into too many technical details. This is all that needs to be understood for now.

## H. Quaternions and Rotation Groups and their associated Lie groups.

There are a number of different rotation groups that show up in theoretical physics, amongst them are the unitary groups, special unitary groups and the special orthagional groups. The two special unitary groups that are associated with quaternions are SU(2) and SU(3). SU(2) in physics these are referered to as the Pauli matrices and SU(3) as the Gell-Mann matrices. The special orthogonal group of 3 x 3 matrices is SO(3), and the special orthogonal group of 4 x 4 matrices is SO(4). These rotation groups are central to modern theoretical physics and mathematics. The adjective 'special' indicates that the determinant of the matrices in these groups is equal to +1. These groups form the backbone for many applications in physics from classical physics to string theory. Here the author will discuss how the rotation groups SU(2), SO(3), SU(3) and SO(4) are related to each other, to quaternions, and Lie groups.

Recall the quaternion rules: $i^2 = j^2 = k^2 = -1$, $ij = -ji = k$ , $jk = -kj = i$ , $ki = -ik = j$ . This means the quaternion group, $Q = \{-1, -i. –j, -k, 1, i, j, k\}$, is generated by 2 elements (Lang, 2002, p.9, p.723).

An interesting result that comes from the fact that quaternions are non-commutative is that polynomial equations over the quaternions can produce infinitely many quaternion solutions. Geometrically these infinitely many solutions form a two-dimensional sphere that is embedded in the ‘pure vector’ or ‘pure quaternion’ subspace. The sphere is centered at zero and intersects the complex planes at the antipodal points, i, j, k and – i, -j,-k respectively (Ballif, 2008, p5) . For example let **H** be the division ring of real quaternions, i, j, and k are each roots of the polynomial $z^2 + 1 = 0$. From this, infinitely many conjugates that can be formed using, $z = bi + cj + dk$ and letting $b^2 + c^2 + d^2 = 1$. Each of these infinitely many conjugates of $i$ is a root of $x^2 + 1$. (Ballif, 2008, p.4)

Recall that complex numbers are points on the complex plane. This represents a 2-dimentional space. One of the motivations behind quaternions was to develop a system that would generalize complex numbers in higher dimensions (Francis, 2011)

### i. The rotation group SU(2) and its associated Lie group and algebra

SU(2) is the group of all 2 x 2 unitary matrices with determinant 1. In general let $a$ and $b$ be two complex parameters where a matrix can be formed by $a, b$ : $M = \begin{pmatrix} a & b \\ -b^* & a^* \end{pmatrix}$ ,where $|a|^2 + |b|^2 = 1$ form the the group of unitary matricies with determinate 1. This is the group SU(2) ('t Hooft, 2007, 33). This is what is referred to also as a ‘simple’ Lie group. A simple group in general is defined to be a group that is not the trivial group, meaning that G $\neq\{1\}$ and doesn't contain any subgroups other than {1} and itself. (Artin, 1991, 201). A simple Lie group is different from the usual simple abstract group that is encountered in a college modern algebra course in that a simple Lie group can contain discrete normal subgroups (Agol, 2009, 2).

A normal subgroup is defined as being a subgroup, H, where for every element x in the group, G, H is said to be a normal subgroup of G, if $xHx^{-1}=H$. The subgroup is trivial when x = identity (Weisstein, 2002 , p.2037).

*Proposition* The longitudes of SU(2) are the conjugate subgroups QTQ* of subgroup T. (Artin, 1991, 275)

This can be understood pictorially by viewing SU(2) as a 3-sphere which shows a few of the latitudes and longitudes in SU(2) .

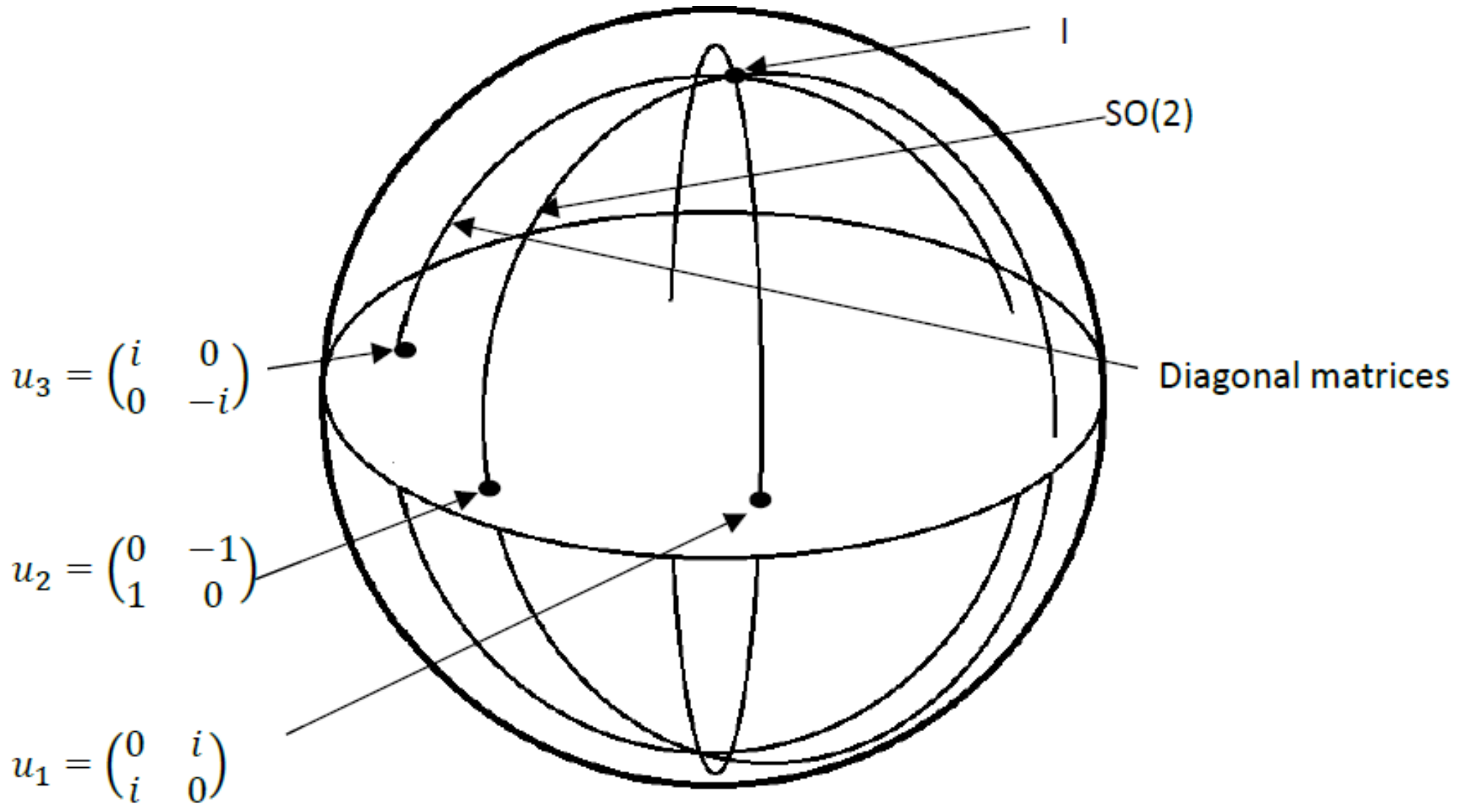

**Figure 14:** the 3-sphere SU(2) projected from $\mathbf{R}^4$ onto the unit disc in the plane. (Adopted from: Artin, 1991, p.275 figure 2.15)

$$u_1 = \begin{pmatrix} 0 & i \\ i & 0 \end{pmatrix}, u_2 = \begin{pmatrix} 0 & -1 \\ 1 & 0 \end{pmatrix}, u_3 = \begin{pmatrix} i & 0 \\ 0 & -i \end{pmatrix},$$

These matrices are called 'anti-Hermitian'. If a matrix is Hermitian this means that the matrix is self-adjoint or that it is equal the transpose of its conjugate i.e. A=A* where $A^* = (\bar{A})^T = \overline{A^T}$(Weisstein, , pgs.1357, 2650). If it is anti-Hermitian this means that A= -A*. In physics A* is often written as $A^\dagger$, A dagger. These are related to the Pauli Matrices that are used in quantum mechanics. The Pauli spin matrices are:

$$\sigma_1 = \begin{pmatrix} 1 & 0 \\ 0 & -1 \end{pmatrix}, \sigma_2 = \begin{pmatrix} 0 & 1 \\ 1 & 0 \end{pmatrix}, \sigma_3 = \begin{pmatrix} 0 & -i \\ i & 0 \end{pmatrix},$$

The Pauli matrices are Hermitian, $\sigma^{\dagger} = \sigma$, and are related to $u_1$, $u_2$, and $u_3$ where $u_1 = i\,\sigma_1$, $u_2 = -i\,\sigma_2$ and $u_3 = i\,\sigma_3$. Furthermore they are also related to quaternions in the following way: Recall that quaternions are defined to be $\mathbf{H} = \{q_0 + q_1\boldsymbol{i} + q_2\boldsymbol{j} + q_3\boldsymbol{k} \mid q_0, q_1, q_2, q3 \in \mathbf{R}\}$ and $\boldsymbol{i}\cdot\boldsymbol{j} = -\boldsymbol{j}\cdot\boldsymbol{i} = \boldsymbol{k}$, $\boldsymbol{j}\cdot\boldsymbol{k} = -\boldsymbol{k}\cdot\boldsymbol{j} = \boldsymbol{i}$, $\boldsymbol{k}\cdot\boldsymbol{i} = -\boldsymbol{i}\cdot\boldsymbol{k} = \boldsymbol{j}$, $\boldsymbol{i}^2 = \boldsymbol{j}^2 = \boldsymbol{k}^2 = -1$. Where $\boldsymbol{i} = i\,\sigma_3$, $\boldsymbol{j} = i\,\sigma_2$, $\boldsymbol{k} = i\,\sigma_1$ when represented as 2 x 2 matrices (Nakahara , 2003, xxii)

Quaternions are also isomorphic to Pauli matrices. One way to show this is by recalling that quaternions can be viewed as having a scalar ($q_0$) and a vector ($\mathbf{q}$) part. A quaternion can be represented as a 2 x 2 matrix in the following way:

Let $Q = [\,q_0, \boldsymbol{q}\,] = q_0\sigma_0 - i\boldsymbol{q}\cdot\boldsymbol{\sigma}, where\ \boldsymbol{q}\cdot\boldsymbol{\sigma} := \sum_{a=1}^{3} q_a\sigma_a,$ where $\sigma_0$ is the identity matrix, and $\sigma_a$ for , $a = 1,2,3,$ are the Pauli matrices $\sigma_1$, $\sigma_2$, and $\sigma_3$

The product rule for the Pauli matrices can be written:

$$\sigma_a\sigma_b = \delta_{ab}\sigma_0 + i\epsilon_{abc}\sigma_c \ \ where\ a,b,c = 1,2,3$$

where $\delta_{ab}$ is the Kronocker delta and $\varepsilon_{abc}$ is the Levi-Civita symbol. The Levi-Civita symbol is anti-symmetric and $\varepsilon_{abc}=1$. The product rule can be written explicitly as $\sigma_1^2 = \sigma_2^2 = \sigma_3^2 = \sigma_0$ and $\sigma_1\sigma_2\sigma_3 = i\sigma_0$, and are invariant under cyclic permutations of {1,2,3}.

Let the quaternion basis be {1,i,j,k} and let the real linear span of the Pauli matrices be $\{\sigma_0,\sigma_1\sigma_2,\sigma_2\sigma_3,\sigma_3\sigma_1\}$, where $\sigma_0$, $\sigma_1\sigma_2$, $\sigma_3\sigma_1$ form a closed multiplicative group so there is no need for complex coefficients and all the

determinates =1. By mapping $1 \mapsto \sigma_0, i \mapsto \sigma_1\sigma_2, j \mapsto \sigma_3\sigma_1, k \mapsto \sigma_2\sigma_3$, this gives a way to describe Pauli matrices in terms of SU(2). Unit quaternions are isomrphic to SU(2) making Pauli matrices and quatertnions isomorphic (Holms, 2013, pgs.19 -20; Goldstein, 1980, p.156).

SU(2) evolved from the unit quaternions. These structures are an essential part of the mathematical foundations of modern physics. In physics SU(2) = Spin (1), this means that the group is made up of all 1 x 1 quaternion matrices. Their length, **H** = **R**$^{\mathbf{4}}$, is preserved under multiplication. In general unit quaternions that are related to SU(2) and Spin(1) as follows:

$$Spin(1) = \left\{ \begin{pmatrix} a+id & -b-ic \\ b-ic & a-id \end{pmatrix} \middle| \, a^2+b^2+c^2+d^2 = 1 \right\} = SU(2)$$

(Savage, n.d. , p.24).

**ii. The rotation group SO(3) and its associated Lie group and algebra**

Another important group that comes up is SO(3). This is the group of all 3 x 3 orthogonal matrices with real elements. Each element is specified by 3 continuous parameters, and its determinant ≠1. Like SU(2), SO(3), is topologically compact, meaning they have the property of being closed with bounded subsets of the real line (Royster, 1999).

To see how SO(3) works as a rotation group, let it be rotated by angle $\phi$ about the direction **n** = $(\theta,\psi)$, where **n** is a unit vector, then $R_{\mathbf{n}}(\phi) = R(\phi,\theta,\psi)$ where $0 \leq \theta \leq \pi$ *and* $0 \leq \psi < 2\pi$ as in Figure 15 (Tung, 1985, p.96)

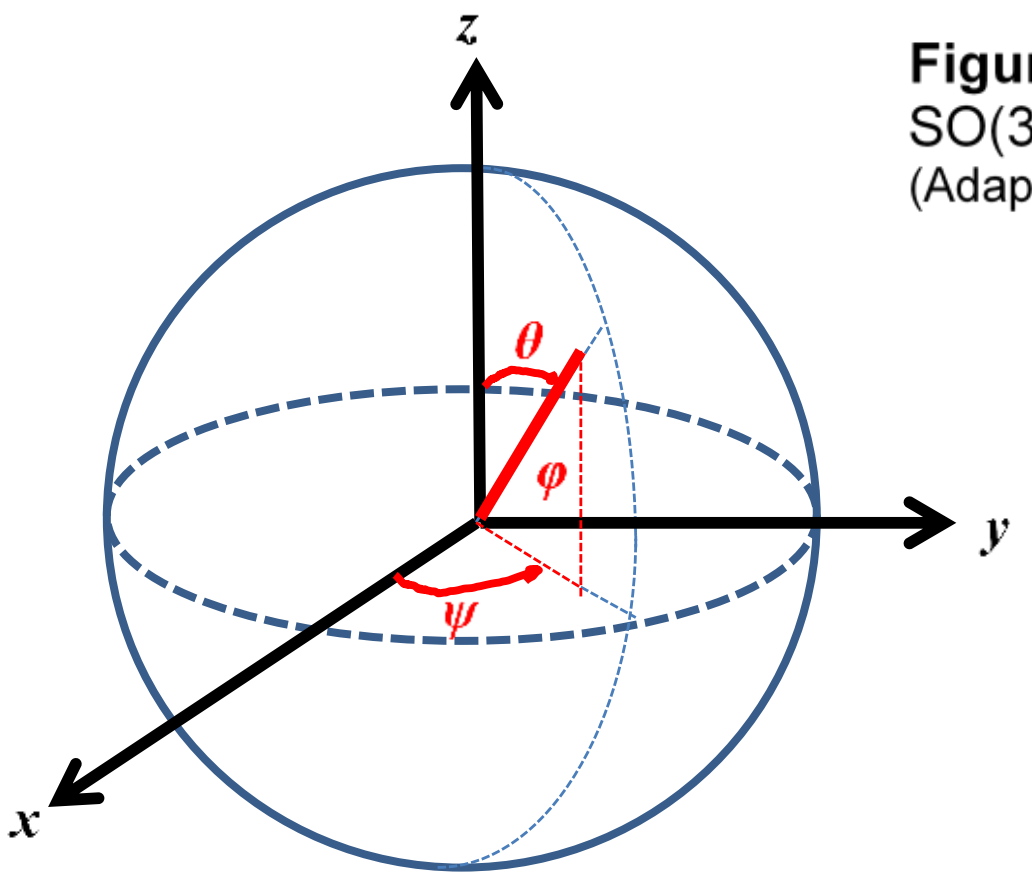

**Figure 15:**
SO(3)
(Adapted from, Tung,1985, p.96)

Recall that SO(2) is a circle, thus SO(3) it is a sphere. SO(3) in pure mathematical terms is called a doubly connected non-commutative rotation group. (Tung, 1985, p.96); Gallier, 2011, p.282). Groups that are simply-connected are characterized by the property that any closed loop is contractible. This means that the group SO(3) is not simply connected. This is due to the fact that it is doubly connected meaning that there are two topological classes of loops on this structure which depend on whether the net angle of rotation is an even or odd multiple of 2π. (Evans, 2013, p.1).

For example a torus is not simply connected. This means that there can be loops made on the torus that can be shrunk to a point without braking it as in loop a and b in Figure 16. These loops, a and b, can be shrunk to a point without breaking them, but loops c and d cannot be shrunk to a point without braking the loop. This is because loops c and D include the torus's central hole.

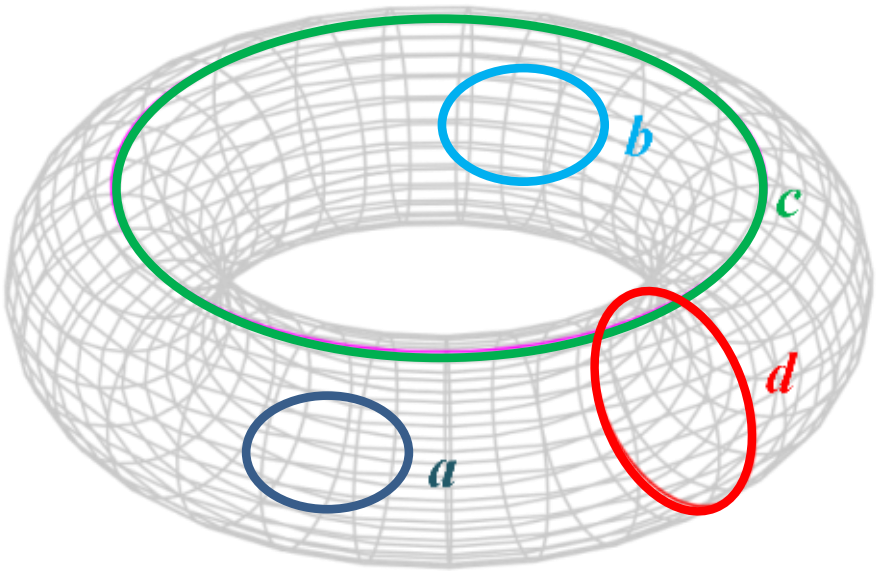

**Figure 16:**
Torus illustrating 2 types of closed loops:
a and b are contractable to a point.
c and d are not.
(Adapted from,(2015) Encyclopædia Britannica, Inc.)

It turns out that there are also 2 types of closed curves on SO(3) (figure 17 see curves (a) and (b) ). A contractible curve is a curve that can be shrunk a point as shown in Figure 17(a). This curve can be continuously deformed an even number of times. Curve (b) is closed and non-contractible. This means that it cannot be shrunk to a point, the two end points (P) on the line correspond to the same points on the manifold, meaning that the end points are fixed. This type of curve winds around the sphere an odd number of times. (Tung, 1985, p.97)

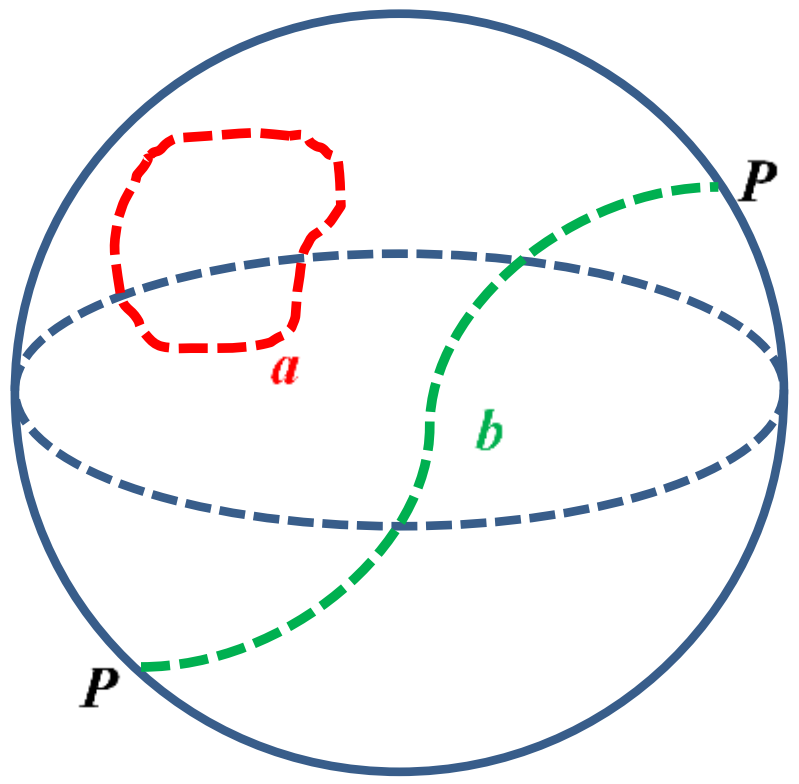

**Figure 17:**

SO(3) Double Cover

(Adapted from, Tung, n.d.,p.4)

One of the reasons why the Lie group of SO(3) is so important to physics is that the basis of its Lie algebra is related to angular momentum:

$$J_x = \begin{pmatrix} 0 & 0 & 0 \\ 0 & 0 & -1 \\ 0 & 1 & 0 \end{pmatrix}, J_y = \begin{pmatrix} 0 & 0 & 1 \\ 0 & 0 & 0 \\ -1 & 0 & 0 \end{pmatrix}, J_z = \begin{pmatrix} 0 & -1 & 0 \\ 1 & 0 & 0 \\ 0 & 0 & 0 \end{pmatrix}$$

These are also the generators of SO(3) (Costa & Fogli , 2012, 32).

In order to form the Lie algebra recall that what connects Lie groups to Lie algebras is the exponential matrix that was discussed in section D on Rodrigues' Rotation Formula. In general the exponential matrix is of the form

$$e^A = 1 + \sum_{k=1}^{\infty} \frac{1}{k!} A^k$$

The exponential matrix of $J_x$, $J_y$, $J_z$ can be written as rotation matrices as follows:

$$R_1 = e^{-i\theta J_x}, R_2 = e^{-i\theta J_y}, R_3 = e^{-i\theta J_z}$$

In general, a rotation through an angle θ in the direction n is written as

$$R = e^{-i\theta J \cdot n}$$

(Kirillov, n.d. , p.35; Costa & Fogli , 2012, 32).

The two groups SO(3) , SU(2) are isomorphic to each other in the following way: $SO(3) \cong SU(2)/\{\mp 1\}$. This means that SU(2) is the ‘double cover’ of SO(3) as shown in Figure 17 (Savage, n.d. , p.16). This does not mean that they are the same. This means that there exists a two-to-one, surjective (onto) homomorphism: $\varphi$ : SU(2) → SO(3) , ker $\varphi$ = $Z_2$ = {±1} Where $Z_2$ is the fundamental group of SO(3). (Evans, 2013, p.4).

A fundamental group is an algebraic topology term where an algebraic space is formed from the loops in the space. The paths in the space start and end at the same point as shown on the Torus in Figure 16 (Hatcher, 2001, 22).

### iii. The rotation group SU(3) and its associated Lie group and algebra

The rotation group SU(3) is a is related to SU(2) in the following way: Where SU(2) is related to the quaternion rotation group, the elements of SU(3) are associated with octonions which are not only non-commutative, but also non-associative.

Since these groups are outside the main body of this thesis, these rotation groups are mentioned in passing here. They play a central role in particle physics. They are also called the Gell-Mann matrices after their founder Murray Gell-Mann who won the Nobel prize for particle classification using SU(3).

SU(3) is the group of all 3 x 3 unitary matrices with det = 0. The SU(3) matrices have 8-dimentions They also have 8 linearly independent generators represented by the following matrices: (Stover, n.d.)

$$\lambda_1 = \begin{pmatrix} 0 & 1 & 0 \\ 1 & 0 & 0 \\ 0 & 0 & 0 \end{pmatrix}, \lambda_2 = \begin{pmatrix} 0 & -i & 0 \\ i & 0 & 0 \\ 0 & 0 & 0 \end{pmatrix}, \lambda_3 = \begin{pmatrix} 1 & 0 & 0 \\ 0 & -1 & 0 \\ 0 & 0 & 0 \end{pmatrix}, \lambda_4 = \begin{pmatrix} 0 & 0 & 1 \\ 0 & 0 & 0 \\ 1 & 0 & 0 \end{pmatrix},$$

$$\lambda_5 = \begin{pmatrix} 0 & 0 & -i \\ 0 & 0 & 0 \\ 1 & 0 & 0 \end{pmatrix}, \lambda_6 = \begin{pmatrix} 0 & 0 & 0 \\ 0 & 0 & 1 \\ 0 & 1 & 0 \end{pmatrix}, \lambda_7 = \begin{pmatrix} 0 & 0 & 0 \\ 0 & 0 & -i \\ 0 & i & 0 \end{pmatrix}, \lambda_8 = \frac{1}{\sqrt{3}} \begin{pmatrix} 1 & 0 & 0 \\ 0 & 1 & 0 \\ 0 & 0 & -2 \end{pmatrix}$$

These matrices were coined by Gell-Mann as the ‘8-fold way’. In particle physics they represent the Lie algebra.

These are essentially the Pauli matrices with an extra column of zeroes. They are often referred to as a generalization of the Pauli matrices and have similar properties.

## iv. The rotation group SO(4) and its associated Lie group and algebra

Another group that is also important in physics and evolved from quaternions is SO(4). This is called the special orthogonal group in 4-dimensional space. This group is isomorphic to *SU(2) x SU(2)*. This is the double cover of *SO(4)* (Baez, 2001 , p.147). This means that the group isomorphism is: $(SU(2) \times SU(2))/\{\mp 1\} \cong SO(4)$ (Savage, n.d., p.21). This isomorphic relationship is a well-known theorem in representation theory. To show this remember that $\mathbf{H} = \mathbf{R}^4$, where pairs of unit quaternions can be defined as (**u,v**) where $(\mathbf{u}, \mathbf{v}) \in \mathbf{H}$ and $\mathrm{Q} \in \mathbf{H}$ since $\mathbf{v}^{-1}$ is a unit quaternion then $\mathbf{v}^{-1} \in SO(4)$ so the action $Q \rightarrow \mathbf{u}Q\mathbf{v}^{-1}$ preserves lengths of vectors and is linear in Q (Savage, n.d. , p.20).

*Theorem 2:* The map

$$\varphi: SU(2) \times SU(2) \rightarrow SO(4)$$

$$\text{where } (v, w) \rightarrow (Q \rightarrow vQw^{-1})$$

is a surjective (onto) group homomorphism with kernel {(1, 1) (-1, -1)}.■

(Savage, n.d., p.21). A group homomorphism is defined by mapping

$\varphi: G \rightarrow H$, if all u and v are elements of G than $\varphi(u\ v) = \varphi(u)\ \varphi(v)$. The group operation on the left is for G. The group operation on the right is for H. This means that the unit quaternion multiplication rotations of a 4-dimentional space act on both the left and right side for all quaternions. Therefore the homomorphism from SU(2) x SU(2) is onto the 4-dimentional rotation group SO(4), where it's The kernel is {(1, 1) (-1, -1)}. The fact that this homomorphism can be formed implies that SU(2) the double cover on SO(4) (Baez, 2005 , p.234).

In this chapter the author has very briefly discussed some very rich and deep mathematical topics. In the following chapter both the historical and mathematical development will be discussed as it relates to physics, both classical and modern. The focus in the next chapter will be how quaternions and its relatives became so central to physics today.

It would be up to the reader, if interested, to go beyond the scope of this chapter and investigate the mathematics in more depth. There are many books and websites devoted to these topics as the author of this thesis has discovered.

# Chapter IV
# APPLICATIONS OF QUATERNIONS IN PHYSICS

So far the history and some of the mathematical ideas and structures that incorporate or are related to quaternions have been discussed. In this chapter the focus will be on how quaternions became assimilated into physics and how these ideas developed and became a part of every physics student's experience without his or her realizing where these ideas came from. Although Hamilton had some 'vague' ideas about their possible applications to physics he never truly developed them.

The first person to truly apply quaternions to physics was Hamilton's student Tait. Tait put quaternions through the test of applied mathematics in his 1867 book *Elementary Treatise of Quaternions.* Crowe said about this book "A noteworthy feature of Tait's Treatise was the extensive attention that he gave (as Hamilton had not) to physical applications. " (Crowe, 2002, p.10)

Quaternions after Tait and Maxwell took a 'back seat' in physics until Pauli and Dirac rediscovered them for quantum mechanics in the theory of spin. They have become even more deeply incorporated into theoretical physics through the application of Clifford algebras and complex non-commutative rotation groups.

### A. Tait and the *Elementary Treatise of Quaternions*

The fashion during the nineteenth century and the industrial revolution was to apply existing mathematics to give greater insight into the physics of the day. Here the author presents a collection of ideas and subjects that Tait discussed in his *An Elementary Treatise of Quaternions.* This is done in order to give the reader an insight into the use of quaternions in physics and some of the highlights of Tait's seminal book.

In his preface to the second edition of *Treatise,* Tait discusses his relationship with Hamilton and Hamilton's insistence that Tait apply quaternions to physics after the publication of the *Elements*. Tait writes that there is a lot in his own work that is unconnected to physics and attributes this to Hamilton himself. He goes on to say that he had made conciderable advances in applying quaternions to physics, but calls this part of the work essentially a collection of problems in which he "… managed (at least partially) to effect the application of Quaternions to line, surface, and volume integrals, such as occur in Hydrokinetics, Electricity, and Potential generally" (Tait,1873, p.xi). He forecasts that his quaternion research "… is certain in time to be of incalculable value to physical science" (ibid, p.xi)

Quaternions were not initially motivated by physical problems. When Hamilton was investigating quaternions he was looking for a way to extend the complex plane, applications became an afterthought. Tait was the real trailblazer in applying quaternions to the physics problems of the day. He defines time and motion in the usual way. Since quaternions can express rotations, Tait starts by introducing quaternion applications to kinematics in Ch. X § 336 of the second edition of *An Elementary Treatise on Quaternions* (ibid, p.195). Tait devotes chapters X and XI to Physics, where X is devoted to Kinematics and XI is devoted to Other Physical Applications. The earlier chapters of the *Treatise* were devoted to purer mathematical concerns, techniques, propositions and proofs.

Browsing through the book it can be seen that the mathematical concerns are mainly geometric and these precede the more analytic interests. Tait says, he is "Keeping always in view, as the great end of every mathematical method, the physical applications, I have endeavored to treat the subject as much as possible from a geometrical instead of an analytical point of view." (ibid, p.vii)

By the third edition of the Treatise he adds Chapter IX. Surfaces of The Second Degree, and rearranges various topics. He enlarges the *Treatise* to include more physics applications then the previous editions (Tait ,1890, p.v). He starts the new chapter on Kinematics, now Chapter XI, connecting quaternions to physics. Claiming that the topics he "…selected for treatment will be those of most direct interest in their physical applications." (Tait,1890, p.279).

In 1896, Arthur S. Hathaway, Professor of Mathematics In The Rose Polytechnic Institute, Terre Haute, Indiana, in his *A Primer of Quaternions* (Hathaway, 1896, p.iii) says in his introduction to *The Elements of Quaternions* by Tait "is the accepted text-book for advanced students." (Hathaway, 1896, p.iii)

**i. Kinematics and Rigid body motion**

One of the most obvious ways to incorporate quaternions into classical physics is through mechanics. Many of Tait's results here would be familiar to today's physics student and more easily obtained using modern vector analysis. According to Tait "All that is contemplated is to treat a few branches of the subject in such a way as to shew the student how to apply the processes of Quaternions."(Tait, 1890, 279).

One of the most obvious issues for today's student when trying to work through Tait's derivations is the language. He starts out looking at the *Kinematics of a Point* in Section 354. Here he introduces the definitions of velocity and acceleration, which are defined in the usual way, $v = \frac{d\rho}{dt}$, $a = \frac{d^2\rho}{dt^2}$ , where ρ is the point vector. From these definitions he uses quaternions to discuss uniform circular motion.

Tait then goes on to discuss the motion of a point in a plane curve using the usual 'polar coordinates', r and θ (Tait, 1890, 280). His approach differs in some respects from the modern method of Gibbs and Heaviside, which most physics students encounter in text books. What these two methods have in common is the idea that a vector can be manipulated according to algebraic rules, without writing out all its components in a fixed coordinate system. Tait's work preceded Gibbs and Heaviside. Thus Tait was the first to apply this principle to problems of kinematics, making his work truly groundbreaking.

In Tait's method every symbol represents a quaternion. All these can be freely added and multiplied according to the usual quaternion rules. Those quaternions that do not have a vector part are called 'scalars', which if they are positive are referred to as 'pure magnitudes'. The quaternions parts do not have a scalar part he called 'vectors'. A unit vector (one with magnitude 1) has square equal to -1. Any vector is uniquely the product of a pure magnitude and a unit vector.

Tait writes the vector that represents the position of the particle at time t as

$$\rho=\rho(t)=r(t)\zeta(t) \ldots\ldots\ldots\ldots\ldots (1)$$

where r and ζ are the magnitude and unit vector factors of ρ. (here making slight changes to Tait's notation.) He defines two constant unit vectors α and β where

$$\beta = \rho(0) \ldots\ldots\ldots\ldots\ldots\ldots\ldots\ldots\ldots (2)$$

is the initial direction of motion in the plane, and α is perpendicular to the plane. Thus the quaternion αβ is a unit vector perpendicular to β. He chooses the sign of α so that β, αβ, α form a right-handed coordinate system

and therefore

$$\varsigma(t) = \beta cos\theta(t) + \alpha\beta sin\theta(t) = (cos\theta(t) + \alpha sin\theta(t))\beta \ldots. (3)$$

Observe that α, being a unit vector, satisfies $\alpha^2 = -1$. This is a property of quaternions not shared by Gibbs and Heaviside. Because of it α satisfies Euler's equation:

$$cos\theta(t) + \alpha sin\theta(t) = e^{\alpha\theta(t)} \ldots\ldots\ldots\ldots\ldots\ldots\ldots\ldots\ldots\ldots\ldots (4)$$

Combining (3) with (4), to get

$$\dot{\varsigma} = \dot{\theta}\alpha\varsigma \ldots\ldots\ldots\ldots\ldots\ldots\ldots\ldots\ldots\ldots\ldots\ldots\ldots\ldots\ldots\ldots\ldots\ldots\ldots..(5)$$

Now one differentiates (1) by the product rule using ζ to get:

$$\dot{\rho} = \dot{r}\varsigma + r\dot{\varsigma} = \dot{r}\varsigma + r\dot{\theta}\alpha\varsigma \ldots\ldots\ldots\ldots\ldots\ldots\ldots\ldots\ldots\ldots\ldots\ldots\ldots.(6)$$

In this exercise Tait is really looking for the formula for acceleration

$$\ddot{\rho} = (\ddot{r}\varsigma + \dot{r}\dot{\varsigma}) + (\dot{r}\dot{\theta}\alpha\varsigma + r\ddot{\theta}\alpha\varsigma + r\dot{\theta}\alpha\dot{\varsigma}) = \ddot{r}\varsigma + 2\dot{r}\dot{\theta}\alpha\varsigma + r\ddot{\theta}\alpha\varsigma + r\dot{\theta}^2\alpha^2\varsigma$$

(7)

Again using $\alpha^2 = -1$, and collecting terms in ζ and αζ, one gets

$$\ddot{\rho} = (\ddot{r} - r\dot{\theta}^2)\varsigma + (2\dot{r}\dot{\theta} + r\ddot{\theta})\alpha\varsigma \quad \ldots\ldots\ldots\ldots\ldots\ldots\ldots\ldots\ldots\ldots..(8)$$

where the term $-r\dot{\theta}^2$ is the centripetal acceleration of a point traveling uniformly in a circle, and $2\dot{r}\dot{\theta}\alpha\varsigma$ is the Coriolis term. (1) can also be used to identify ζ with ρ/r or what in modern notation is called $\hat{\rho}$.

This is something that can only be done using quaternions. It can't be done in quite this way using modern vectors, nor can Euler's formula (4) be applied to perpendicular planes. (4) was precisely what Hamilton was looking for, an analogy to Euler's formula in 3-dimensions. Tait took this idea that is unique to quaternions and applied it to physics problems without the use of coordinates or modern vectors.

Tait continues on in the *Treatise* in a similar way through rigid body motion. Here Tait refines an earlier paper on rigid body motion (Tait, 1869, pp.261 – 303). He continues to develop quaternion applications of rigid body motion in the second and third edition of the *Treatise.* He calls part B of the Kinematics chapter –*Kinematics of a Rigid System*. Tait develops this more clearly in the third edition than the second edition, selecting problems that appeal to him and solving those using quaternions (Tait, 1873, pp.202-218; Tait, 1890, pp.287-308). Quaternions was a new way to do some old problems without the tedium of fixed coordinates. He did this before modern vector analysis was invented.

Although Tait does continue to give physics examples he stresses in the introduction of Chapter XII in the third edition (XI in the second edition) “This Chapter is not intended to teach Physics, but merely to shew by a few examples how expressly and naturally quaternions seem to be fitted for attracting the problems it presents.” (Tait , 1873, p.222;Tait, 1890, p.309 ) Here again Tait uses rigid body motion to illustrate the power of quaternions when applied to actual physical situations. A nice little problem in section 427 is where Tait uses quaternions to derive the equation of motion of a simple pendulum, taking into account the Earth’s rotation (Tait, 1890, p.336). In practice, quaternions are useful in representing rotation orientations. This is how they are still used today, although this particular problem would solved using modern vector analysis. Here there is no real need to employ quaternions explicitly, so vectors as they are used today appear to be adequate.

A. McCaulay discusses Tait’s quaternion applications to elasticity in his 1893 book *Utility Of Quaternions In Physics*, Section III.1. He says,

> “As far as I am aware the only author who has applied Quaternions to Elasticity is Prof. Tait. In the chapter on Kinematics of his treatise on Quaternions, §§ 360–371, he has considered the mathematics of strain with some elaboration and again in the chapter on Physical Applications, §§ 487–491, he has done the same with stress and also its expression in terms of the displacement at every point of an elastic body.” (McCaulay, 1893, p.11)

Thus it appears Tait was the first person to truly apply quaternions to physical problems. Problems that today would be worked out with the vectors of Gibbs and Heaviside were originally worked out by Tait. The brilliance of what Tait did was to work out a system where problems can be set up without using coordinates. Before Tait these problems would rely on tedious calculations using coordinates to solve them. For most problems the modifications of Gibbs and Heaviside is fine, and if everything can be done using vectors there isn't a problem, but sometimes quaternions will get you there faster, as in the example of a particle in a moving system mentioned earlier. It is not that the problem cannot be done using other methods, it is that in some cases with quaternions the result can be deduced more quickly, but at the price of being more difficult to set up. Tait was a 'purist' when it came to quaternions and perhaps didn't like Gibbs and Heaviside modifying them to fit their needs at the time. This misunderstanding was unfortunate since both systems could have learned and grown from each other.

ii. **Electricity and Magnetism**

Tait first explores the applications of quaternions to electricity and magnetism in example 428 of the second edition the *Treatise*. Here Tait sets up the following situation "As another example we take the case of the action of electric currents on one another or on magnets; and the mutual action of permanent magnets." (Tait, 1873, p.249). Tait uses some well-known examples from Ampère and Murphy to test the usefulness of applying quaternions to various known problems. In the 1890 edition the *Treatise* he repeats many of the same problems he did in 1873, but in the newer edition he extends *Physical Applications* part *E. Electrodynamics* giving more examples the he did in his previous edition. Most of these examples can be more easily derived using vectors as is done today.

### iii. The operator "del", and the Laplacian

Hamilton used quaternions to define the dell operator $\nabla = i\frac{\partial}{\partial x} + j\frac{\partial}{\partial y} + k\frac{\partial}{\partial z}$ . This is still used today in mathematics and physics. It is a symbol that every physics student would recognize from their electricity and magnetism courses, and every math major from their advanced calculus courses. With the dell operator only the vector parts, i, j, k, of the quaternions are used.

Young Sam Kim in his 2003 thesis on *Maxwell's Equations and a Historical Study of Incorporating Them into Undergraduate Mathematics and Engineering Education* (Kim, 2003, p.88) discusses the influence that quaternions had on Maxwell, especially in the application of the operator 'dell'($\nabla$) and the operator $\nabla^2$. Tait brings up 'del' for the first time on page 76 of the second edition of the *Treatise* (Tait, 1873, p.76). Tait credits the 'del' to Hamilton calling this "the very singular operator devised by Hamilton" (Tait, 1873, p.173). He first applied this operator to a physical situation when he discussed an example in Kinematics "…it may be interesting here, especially for the consideration of any continuous displacements of the particles of a mass, to introduce another of the extraordinary instruments of analysis which Hamilton had invented" (Tait, 1873, p.215) he goes on to let -ν represent the direction and size of the force at any point he then writes the operator

$$\nabla = i\frac{d}{dx} + j\frac{d}{dy} + k\frac{d}{dz}$$

Using the observation that he discussed in exercise 317 where if ν=∇F then

$$\nabla\nu = \nabla^2 F = -\left(\frac{d^2}{dx^2} + \frac{d^2}{dy^2} + \frac{d^2}{dz^2}\right).$$

In the next exercise he shows "…the effect of the vector operation ∇, upon any scalar function of the vector of a point, is to produce the vector which represents in magnitude and direction the most rapid change in the value of the function"(Tait, 1873, p.216)

By the time the third edition is published the application of quaternions to electromagnetic theory is neither small nor trivial as Kim also observed in his thesis

> "…Tait's major emphasis in Quaternions was in the area of physical applications, and that is also the area where Quaternions attracted the attention of scientists and applied mathematicians who were interested in electromagnetic theory. That in turn led to the creation of a new system of vector analysis derived from the vast structure of Quaternions suitable for their needs in their perspective of the universe in the late nineteenth century." (Kim, 2003, p.89)

Between Tait's second edition (1873) of the *Treatise* he devoted twenty sections to Electrodynamics - §§428-448 (Tait, 1873, pp.249-260) , by the time the third edition was published Tait devoted not only part *E-Electrodynamics*. §§ 453-472, but also added parts *F- General Expressions for the Action between Linear Elements,* §473, *G- Application of* $\nabla$ *to certain Physical Analogies* , §§474 – 478, *H. Elementary Properties of* $\nabla$.§§ 479 – 481, and *K- Application of the* $\nabla$ *Integrals to Magnetic &c. Problems* §§502 – 506 (Tait, 1890, pp.350-374, pp.387-390). This indicates that his interest in applying quaternions to electromagnetic theory became more intense in time. This may, in part, be because he recruited his former classmate, James Clark Maxwell, into the quaternion enterprise, and thereby enticing Maxwell to apply quaternions to his own electromagnetic theory.

**B. Maxwell and Quaternions**

In 1864 Maxwell had worked out the equations for electromagnetism, in his paper *A Dynamical Theory of the Electromagnetic Field* dated 27 October 1864. Before Maxwell used quaternions in his electromagnetic theory he had used complex numbers and coordinates as his mathematical tools. (Maxwell, 1865, pp.467–512) He originally wrote eight equations for electromagnetic fields similar to those in Table 3.

**Table 3** – Maxwell's Equations: the equations on the left reflect Maxwell's equations from his initial 1864 paper *A Dynamical Theory of the Electromagnetic Field*. The equations on the right reflect today's use of Maxwell's equations in modern vector notation.

$$\left.\begin{aligned} p' &= p + \frac{df}{dt} \\ q' &= q + \frac{dg}{dt} \\ r' &= r + \frac{dh}{dt} \end{aligned}\right\} \longrightarrow \left.\begin{aligned} J_1 &= j_1 + \frac{\partial D_1}{\partial t} \\ J_2 &= j_2 + \frac{\partial D_2}{\partial t} \\ J_3 &= j_3 + \frac{\partial D_3}{\partial t} \end{aligned}\right\} \Longrightarrow \boldsymbol{J} = \boldsymbol{j} + \frac{\partial \boldsymbol{D}}{\partial t} \qquad (1)$$

$$\left.\begin{aligned} \mu\alpha &= \frac{dH}{dy} - \frac{dG}{dz} \\ \mu\beta &= \frac{dF}{dz} - \frac{dH}{dy} \\ \mu\alpha &= \frac{dG}{dx} - \frac{dF}{dy} \end{aligned}\right\} \longrightarrow \left.\begin{aligned} \mu H_1 &= \frac{\partial A_3}{\partial y} + \frac{\partial A_2}{\partial z} \\ \mu H_2 &= \frac{\partial A_1}{\partial z} + \frac{\partial A_3}{\partial x} \\ \mu H_3 &= \frac{\partial A_2}{\partial x} + \frac{\partial A_1}{\partial y} \end{aligned}\right\} \Longrightarrow \mu\boldsymbol{H} = \nabla \times \boldsymbol{A} \quad (2)$$

$$\left.\begin{aligned} \frac{d\gamma}{dy} - \frac{d\beta}{dz} &= 4\pi p' \\ \frac{d\alpha}{dz} - \frac{d\gamma}{dy} &= 4\pi q' \\ \frac{d\beta}{dx} - \frac{d\alpha}{dy} &= 4\pi r' \end{aligned}\right\} \longrightarrow \left.\begin{aligned} \frac{\partial H_3}{\partial y} + \frac{\partial H_2}{\partial z} &= 4\pi J_1 \\ \frac{\partial H_1}{\partial z} + \frac{\partial H_3}{\partial x} &= 4\pi J_2 \\ \frac{\partial H_2}{\partial x} + \frac{\partial H_1}{\partial y} &= 4\pi J_3 \end{aligned}\right\} \Longrightarrow \nabla \times \boldsymbol{H} = \boldsymbol{J} \quad (3)$$

$$\left.\begin{aligned} P &= \mu\left(\gamma\frac{dy}{dt} - \beta\frac{dz}{dt}\right) - \frac{dF}{dt} - \frac{d\Psi}{dx} \\ Q &= \mu\left(\alpha\frac{dz}{dt} - \gamma\frac{dx}{dt}\right) - \frac{dG}{dt} - \frac{d\Psi}{dy} \\ R &= \mu\left(\beta\frac{dx}{dt} - \gamma\frac{dy}{dt}\right) - \frac{dH}{dt} - \frac{d\Psi}{dz} \end{aligned}\right\} \longrightarrow \left.\begin{aligned} E_1 &= \mu(H_3 v_2 - H_2 v_3) - \frac{dA_1}{dt} - \frac{d\varphi}{dx} \\ E_2 &= \mu(H_3 v_2 - H_2 v_3) - \frac{dA_2}{dt} - \frac{d\varphi}{dy} \\ E_3 &= \mu(H_3 v_2 - H_2 v_3) - \frac{dA_3}{dt} - \frac{d\varphi}{dz} \end{aligned}\right\} \Longrightarrow$$

$$\boldsymbol{E} = \mu(v \times \boldsymbol{H}) - \frac{\partial \boldsymbol{A}}{\partial t} - \nabla\varphi \qquad (4)$$

$$\left.\begin{aligned} P &= kf \\ Q &= kg \\ R &= kh \end{aligned}\right\} \longrightarrow \left.\begin{aligned} \varepsilon E_1 &= D_1 \\ \varepsilon E_2 &= D_2 \\ \varepsilon E_3 &= D_3 \end{aligned}\right\} \Longrightarrow \varepsilon \boldsymbol{E} = \boldsymbol{D} \qquad (5)$$

$$\left.\begin{aligned} P &= -\zeta p \\ Q &= -\zeta q \\ R &= -\zeta r \end{aligned}\right\} \longrightarrow \left.\begin{aligned} \sigma E_1 &= j_1 \\ \sigma E_2 &= j_2 \\ \sigma E_3 &= j_3 \end{aligned}\right\} \Longrightarrow \sigma \boldsymbol{E} = \boldsymbol{j} \qquad (6)$$

$$e + \frac{df}{dx} + \frac{dg}{dy} + \frac{dh}{dz} = 0 \longrightarrow \rho + \frac{\partial D_1}{\partial x} + \frac{\partial D_2}{\partial y} + \frac{\partial D_3}{\partial z} = 0 \Longrightarrow -\rho = \nabla \cdot \boldsymbol{D} \quad (7)$$

$$\frac{de}{dt} + \frac{dp}{dx} + \frac{dq}{dy} + \frac{dr}{dz} = 0 \longrightarrow \frac{\partial \rho}{\partial x} + \frac{\partial j_1}{\partial x} + \frac{\partial j_2}{\partial x} + \frac{\partial j_3}{\partial x} = 0 \Longrightarrow -\frac{\partial \rho}{\partial t} = \nabla \cdot \boldsymbol{j} \quad (8)$$

(Equations adapted from Waser, 2000, p.2)

Maxwell viewed current flow as analogous to fluid flow. He discusses in this paper how electricity flows in an ether where "…from the phenomena of light and heat, that there is an aethereal medium filling space and permeating bodies, capable of being, set in motion and of transmitting that motion from one part to another, and of communicating that motion to gross matter so as to heat it and affect it in various ways." (Maxwell, 1864, p.460) as was the tradition of the times.

The way that Maxwell initially developed his equations is different than the way that most textbooks write them (Table 3). Today Maxwell equations are usually written as (6)-Ohm's law, (4)-The Faraday-force and (8)-The continuity equation for a changing region. (Waser, 2000, p.3)

Tait and Maxwell knew each other from school and College. They remained friendly until Maxwell's death in 1879. It was probably due to Tait's persuasion that Maxwell decided to incorporate quaternions into his work.

At this time Maxwell became interested in quaternions, and perhaps liked the idea of using them. According to Waser Maxwell did not actually calculate with quaternions (Waser, 2000, p.3). Maxwell wrote a Manuscript on the *Application of Quaternions to Electromagnetism* in November 1870. At about the same time he wrote about the general use of quaternions in a letter to Tait:

> "... The invention of the Calculus of Quaternions by Hamilton is a step towards the knowledge of quantities related to space which can only be compared for its importance with the invention of triple coordinates by Descartes. The limited use which has up to the present time been made of Quaternions must be attributed partly to the repugnance of most mature minds to new methods involving the expenditure of thought ..."

(Maxwell, 1995, p.570). It is evident that during this time Maxwell was enthusiastic about quaternions, but setting up a physics problem using quaternions was often clumsy and tedious.

Cayley, another prominent mathematician at the time remarked that quaternions were analogous to "'pocket maps', concise and compact, but not as easy to read as full scale coordinate maps" (Macfarlane, 1916, p.41). It appears that what Cayley meant by this is that quaternions contained a great amount of information in relatively few symbols, but in order to access all the information that may lay hidden in some problems, full scaled Cartesian coordinate should be used in certain types of calculations (Pritchard, 1998, p.236). Tait did not agree with Cayley's use of coordinates over quaternions and wrote to Cayley in August 28. 1888 that "…no problem or subject is a fit one for the introduction of Quaternions if it necessitates the introduction of Cartesian Machinery" (Knott, 1898, p.159) To Tait it appears that there were certain parts of mathematics that did not lend themselves to quaternions if the use of Cartesian coordinates had to be employed.

By the 1873 edition of the *Treatise*, Maxwell included quaternions. He essentially modified the original equations that he used in his 1865 edition of the *Treatise* into the new edition. Whether Maxwell actually calculated with quaternions is debatable. Maxwell probably used what he found convenient for his needs, without 'ruffling too many feathers'. This would make sense, for example, according to Waser, Maxwell defined the field vectors; $\mathbf{B} = B_1 i + B_2 j + B_3 k$. Here Maxwell used quaternions notation without scalar part. According to Waser there are other instances where Maxwell uses the scalar part of quaternions without vector part (Waser, 2000, p.3).

As for gradient, divergence, curl, and the Laplacian, these terms, which that most students of physics are familiar with, were not Maxwell's original terms for these objects. Maxwell was aware of the Laplacian as it is understand today. He derived what is known today as gradient, divergence, curl, from this. Maxwell called the scalar part S del σ and the vector part V del σ, where σ is a vector function of position. He called the scalar part the convergence, the negative what is called today divergence, and the vector part curl (Maxwell, 1890, p.264). Maxwell illustrated his definitions in his paper *On the*

*Mathematical classification of Physical Quantities* as follows:

**Figure 18** – Maxwell's diagrams (adopted from: Maxwell, 1890, p.265).

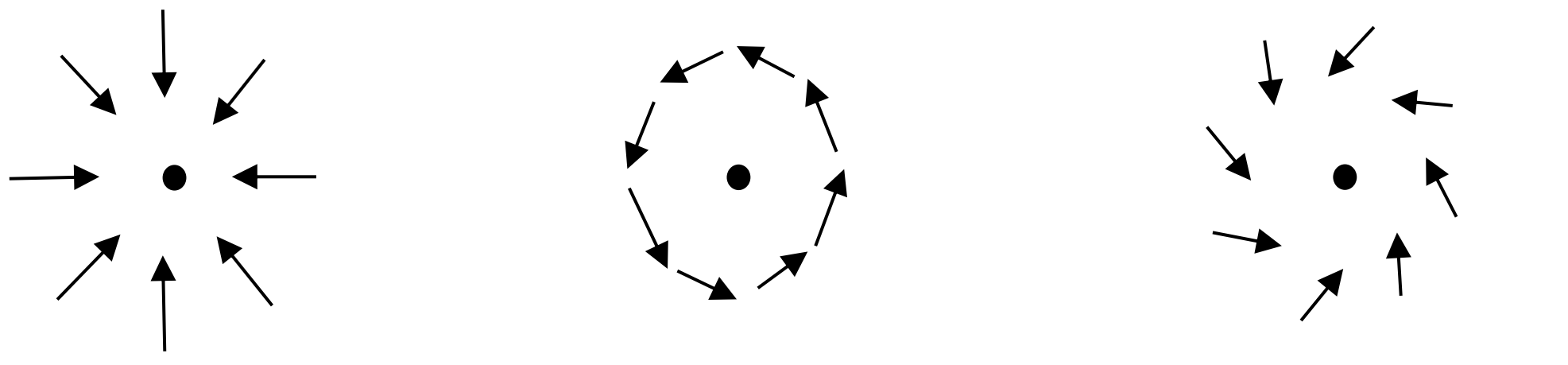

Convergence Curl Convergence and Curl

Maxwell went through some struggles when naming these quantities. He called what today is called divergence, convergence; gradient, slope; and initially he called curl by other names such as twirl, but finally settled on curl. As Maxwell described his struggle with these words "I have sought for a word which shall neither, like Rotation, Whirl, or Twirl, connote motion, nor, like Twist, indicate a helical or screw structure which is not of the nature of a vector at all." (Maxwell, 1890, p.265).

Quaternions was not the best way to express Maxwell's equations due to its awkward notation. Maxwell did not deviate far from the notation used in his previous edition. Here is where Gibbs and Heaviside improved upon what Maxwell presented. They made Maxwell's equations more accessible than how they were written in the 1873 edition *A Treatise on Electricity & Magnetism* that both Gibbs and Heaviside read (Wilson, 1901).

In order to derive Maxwell's equations to correspond to the ones that students encounter in their physics books it is more convenient to start from his original 1865 set of equations, before he incorporated quaternions into his work, as was done in Table 3.

After some reworking these can be derived into a form that most undergraduate and graduate students would recognize as Maxwell's equations, for example in David J. Griffiths *Introduction to Electrodynamics* or John David Jackson's *Classical Electrodynamics*.

In this chapter so far the application of quaternions ware developed and applied to classical physics. Hamilton's student Tait really set quaternions on their way to become an integral part of physics. Even Maxwell, at least for a while, developed his electromagnetic equations using quaternions.

### C. Quaternions and Special Relativity Theory

Hamilton wrote a letter to Graves in 1837, which has almost a prophetic ring. He asked "…what special connexion has the number Four with mathematics generally…?" and answered

> "One general form of answer to this question is the following:
> that in the mathematical quaternion is involved a peculiar synthesis, or combination, of the conceptions of space and time …Time is said to have only one dimension, and space to have three dimentions former is an … The mathematical quaternion partakes of both these elements; in technical language it may be said to be "time plus space", or "space plus time": and in this sense it has, or at least involves a reference to, four dimensions. And how the One of Time, of Space the Three, Might in the Chain of Symbols girdled be." (Graves, 1889, p.635)

One of the quaternion advocates during this time was Macfarlane who in 1891 formulated a space-time in terms of quaternions using a modified version of quaternion multiplication he called ‘hyperbolic quaternions’ (Macfarlane , 1894). The Hyperbolic quaternion multiplication table can be written as follows (Table 4):

| × | **1** | **i** | **j** | **k** |
|---|---|---|---|---|
| **1** | 1 | *i* | *j* | *k* |
| ***i*** | *i* | +1 | *k* | −*j* |
| ***j*** | *j* | −*k* | +1 | *i* |
| ***k*** | *k* | *j* | −*i* | +1 |

**Table 4-**Hyperbolic quaternion multiplication table

(Adapted from Hyperbolic Quaternion , n.d. )

Hyperbolic quaternions, unlike Hamilton’s quaternions that were not only non-commutative, hypolerbolic quaternions are also non-associative, that is *(ij)j* = *kj* = *-i* and *i(jj)* = *i*. This was algebraically problematic. Macfarlane in his 1900 paper *Hyperbolic Quaternions* able to restore the associative property of multiplication by resorting to using biquaternions (Macfarlane, 1900b),.

In his paper *Hyperbolic Quaternions* on page 179 he writes about figure 7 shown on page 181 that “Just as a spherical vector is expressed by r√-1 ξ, so a hyperbolic vector is expressed by rξ, where r denotes the modulus and ξ the axis” (Macfarlane, 1900b, 179). This essentially the same understanding that Poincaré had with respect to the space being a hyperbolic rotation and time as the imaginary coordinate. (Poincaré, 1906). Macfarlane’s diagram 7 (Figure

19) is essentially the same as what is called the Minokwski space today. Macfarlane shows in his figure 7 (Figure 19) the properties of space-time used in special relativity today. Macfarlane came before Minokwsky and special relativity, but is often overlooked in its history, as one of the first to lay the foundations to special relativity theory and space-time using hyperbolic quaternions. Macfarlane thus opened to discussions about linear algebra, vector analysis, differential geometry and relativity theory (Macfarlane, 2010).

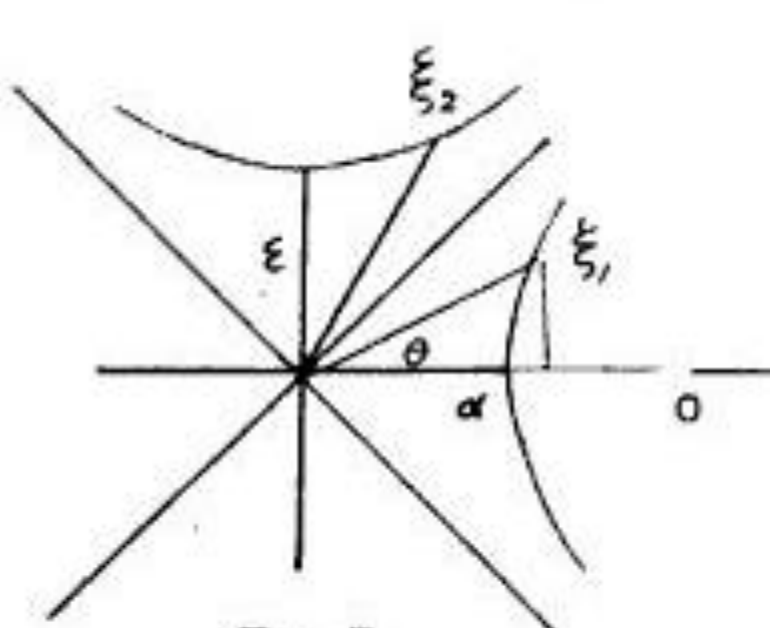

Fig. 7.

**Figure 19**

Macfarlane's Figure 7: $\xi_1$ and $\xi_2$ represent 'sheets' similar to Minokwski space.

(From: Macfarlane, 1900b, p.181)

Arthur W. Conway in his 1911 article *On the Application of Quaternions to some Recent Developments of Electrical Theory*, published in Proceedings of The Royal Irish Academy was one of the first people to see the connection between quaternions and special relativity. A year later Ludwik Silberstein independently recognized this connection, which he published in Philosophical Magazine, *Quaternionic Form of Relativity*. Following Silberstein's article Conway comments, focusing on his own priority in this discovery:

> "The appearance of Prof. Silberstein's paper entitled "The Quaternionic Form of Relativity" in the May issue of the Phil. Mag. is a welcome sign

> that continental mathematicians, who have already largely availed themselves of various systems of vector notation, are perhaps awakening to the suitability of quaternions in such a connexion…". Making reference to his own paper Conway continues "An application of quaternions to the Relativity Principle will be found in a paper, vol. xxix. Section A, No. 1, Proc. Irish Academy (read Feb. 1911)." (Silberstein, 1912, p.208).

Conway goes on to discuss the asymmetric advantaged that quaternions has to offer the theory of special relativity, since nature is not symmetric with respect to space and time. Unfortunately the author of this thesis was unable to find in either Silberstein or Conway's work any reference to Macfarlane and his insights into the ideas discussed using hyperbolic quaternions.

### i. Quaternions and Lorentz transformations

Special relativity was introduced into the scientific community by Albert Einstein in 1905. According to Dr. Martin Erik Horn, who is part of the Institute of Physics Education in Goethe University Frankfurt am Main in Germany, Einstein became familiar with quaternions and applied them to Lorentz transformations many years after he originally introduced special relativity to the world (Einstein, 1932). As to Einstein's 1932 article being used for an introductory physics course, Horn says that "… their general representation is so abstract that an introduction at this level seems hopeless. The fundamental associations however can be reduced to a point didactically so that they can be instituted at the introductory level." (Horn, 2002, p.3).

Although it is not clear to the author of this thesis how Einstein specifically incorporated quaternions into his theory of special relativity one can observe the theory of special relativity takes place in 4-dimensions: 3-space dimensions and 1-time dimension. It is due to this 3-space 1-time relationship that Lorenz transformations became so important in Einstein's theory of special relativity. By definition the Lorentz group is the set of all the Lorentz transformations that form in a Euclidean four-space. The Lorentz group conserves the quantity $c^2t^2 – x_1^2 – x_2^2 – x_3^2$ where c = constant speed of light.

An interesting historical fact is that George Francis FitzGerald was an independent discover of what are usually called Lorentz contractions. In *Science* 1889 he wrote in a short letter about the Michelson and Morley's experiment:

> I have read with much interest Messrs Michelson and Morley's wonderfully delicate experiment attempting to decide the important question as to how far the ether is carried along by the Earth. Their result seems opposed to other experiments showing that the ether in the air can be carried along only to an inappreciable extent. I would suggest that almost the only hypothesis that could reconcile this opposition is that the lengths of material bodies changes, according as they are moving through the ether or across it, by an amount depending on the square of the ratio of their velocities to that of light.(FitzGerald, 1889, p.390)

Although FitzGerald didn't use the word 'contraction' or another synonym, the idea of 'contraction' is implied (Browne, n.d.). Thus it appears that FitzGerald was the first to conceive of the idea of contraction. But it was Macfarlane who showed this mathematically using hyperbolic quaternions. In his understanding about the operator products of hyperbolic quaternions he found that $e^{ar}me^{ar} = (u + vr)e^{2ar} + w(\cosh 2a)s$ the implication of this formula is that a stretch of *cosh2a* in any direction perpendicular to r implies that the untransformed space experiences a contraction relative to the stretch of *cosh2a*. Macfarlane's work in the 3-dimensional hyperboloid model reflects many of the notions that come up in space-time. Where for example a point on the hyperboloid "…represents a frame of reference having a particular velocity with respect to the waiting frame which corresponds to the point 1 + 0i + 0j + 0k" (Macfarlane, 2010). It was Lorentz who worked out the mathematical details without the issues of the non-associativity of multiplication, and perhaps some other algebraic complications, coming up with a cleaner way for doing contractions than Macfarlane.

Quaternions look as though they should be useful for special relativity; unfortunately no one has ever made them work in a clear and practical way that the author of this thesis is aware of.

## ii. Clifford algebras and Lorentz Transformation

Clifford algebras have been more successfully applied to Relativity theory then some other subjects. Hestenes in his famous book *Space-Time Algebra* written in 1966 made the connection between Clifford's Algebra and the Theory of Special Relativity and Quantum Theory. Hestenes was one of the first to apply Clifford algebras rigorously to special relativity theory.

(Hestenes, 1986, p.1). He called this new subject STA (Space-Time algebra). In his lecture notes Baylis writes that "Much of the power of Clifford's geometric algebra in relativistic applications arises from the form of Lorentz transformations." (Baylis, 2004, p.91). Here the Minkowski metric is put directly in Clifford algebra form, $C\ell_{1,3} = C\ell_{3,1}$ representing the 3-space and 1-time dimensions of space-time. In order to generate the basis of $C\ell_{1,3}$ the results of the basis vectors $\gamma_\mu$ are used, where $\mu = 0, 1, 2, 3$. These satisfy the relation:

$$\frac{1}{2}\left(\gamma_\mu\gamma_\nu + \gamma_\nu\gamma_\mu\right) = \eta_{\mu\nu} = \begin{cases} 1, \mu = \nu = 0 \\ -1, \mu = \nu = 1,2,3 \\ 0, \mu \neq \nu \end{cases}$$

(Baylis, 2004, p.94)

STA has its own specialized language as noted by ' $C\ell_{1,3}$'. To develop the mathematical-physics, and all the associated definations further would go beyond the scope of this thesis.

### D. Quaternions and Quantum Mechanics

Pauli matrices, named after Wolfgang Pauli, was discussed earlier as fundamentally being quaternions. In an interesting quote from Feynman:

> "Take Hamilton's quaternions: the physicists threw away most of this very powerful mathematical system, and kept the part – the mathematically trivial part - that became vector analysis. But when the whole power of quaternions *was* needed, for quantum mechanics, Pauli re-invented the system I n a new form. Now, you can look back and say that Pauli's spin matrices and operators were nothing but Hamilton's quaternions" (Feynman, 1999, pp.200-201)

It was discussed earlier that Pauli matrices are isomorphic to quaternions but they are not exactly the same thing. To make them identical they must be multiplied by i. Also Pauli matrices do not form a division algebra, since an inverse cannot be formed, thus they are not 'true quaternions' (Baez,1997).

### i. Spinors

By the time Einstein wrote Special Relativity vector analysis was well established within the physics community. Quaternions applied to relativity theory, as such, were largely ignored by the scientific establishment except by a few, as mentioned earlier.

It was the Austrian-Swiss physicist, Wolfgang Pauli, who introduced spin matrices to quantum mechanics. It was through his work that quaternions really made a comeback. Pauli matrices or spinners are very similar to quaternions (Edmonds, 1972, pp. 205; Lambek, 1995).

Dr. David Hestenes, the same one who made the connection between special relativity, quantum mechanics and Clifford algebras, points out

> "… quaternions have been popping up in quantum mechanics for a long time, but often are disguised as matrices, spinors etc. so they were not recognized as quaternions, but conceptually these objects, despite notational changes, and perhaps improvements, are still related to the original ideas of quaternions as Hamilton originally envisioned." (Hestenes, 2007)

### ii. The Heisenberg Uncertainty Principle

One of the most important relationships in quantum mechanics is the Heisenberg Uncertainty Principle. The Heisenberg Uncertainty Principle says that when two quantum mechanical operators do not commute that the more practice an observable is measured the less precise another observable can be measured. Thus two observables cannot be measured simultaneously to arbitrary precision. By observables what is meant is something that can be measured. Any attempt to do so will yield results more or less uncertain for each observable in the system. The best that can be done is the product of the two uncertainties that is approximately related to the degree of noncommutation (the commutator) of the two operators(Stack, 2013, pgs. 3-5).

In physics the usual examples that is given of two observables and how they are related by the Uncertainty Principal is:

(i) any component of position and the corresponding component of momentum; and

(ii) any two components of the classically defined angular momentum **r** x **p**.

Many of the familiar elementary particles (electron, proton, neutron, neutrino, quark, and many others) have spin ½. This means that the particle has an intrinsic angular momentum that must be added to **r** x **p.** This must be done in order to maintain the conservation of angular momentum, and the operators' $s_x$, $s_y$, $s_z$ corresponding to the x, y, z components of the spin angular momentum. These are the Pauli matrices multiplied by ½. The Pauli matrices are used to represent an observable. These matrices have commutation relations resembling those of the three components of **r** x **p**. This is so the Uncertainty Principle applies smoothly to the whole angular momentum, as well as to the internal and external parts separately (Dorney, 2011).

This principle says if one measures physical information (position or momentum) about the x-axis for $s_x$, all the physical information about the other two, $s_y$ and $s_z$, will be lost. This is partly true of the spin along some intermediate axis, say at an angle between **x** and **z**. Thus non-commutative here means physically that spin cannot simultaneously be measured in more than one direction. (Mitteldorf, 2001).

At the same time, the Pauli matrices are 2 x 2, so that each spin operator, say $s_z$, has just two eigenstates. These eigenstates are commonly referred to as spin up and spin down, and have eigenvalues +½ and -½. For example if it is determined that $s_z = +\frac{1}{2}$ , then the value of $s_x$ becomes completely undecided between +½ and -½. The interplay of the two dimensions occurs in the spinor Hilbert space. The spinor Hilbert space as Dirac put it is "…just a Euclidian with an infinite number of dimensions …" (Dirac, 1974, p.4) Where the three orthogonal spin operators is an outgrowth of the near-isomorphism of SU(2) to SO(3) discussed at the end of Chapter III section H.ii

## E. Quaternions in Particle Physics

### i. Isospin

By the mid-twentieth century particle physics was all the rage. New particles were being discovered constantly; a way to classify them was needed. The structures that ultimately developed grew out of the idea of isospin was introduced as early as 1932. Isospin was introduced by Werner Heisenberg, the same Heisenberg that is connected with the Uncertainty Principle. Isospin, like spin angular momentum, is based on the group SU(2) and can therefore be considered as an application of quaternions to particle physics.

In the early 1920's the proton and electron were known, and the electric charge of the atomic nucleus could be accounted for by supposing it to contain a certain number of protons. But there was a problem, this number of protons fell a good deal short of accounting for the measured atomic mass. So ideas floated around to account for the discrepancy in the atomic mass; for example if there were additional protons in the nucleus, there would have to be electrons to cancel their charge, but it seemed unlikely that electrons could be bound so tightly in the atom in order to do this.

It was discovered in 1930 by Herbert Becker and Walter Bothe that non-ionizing radiation was produced when alpha particles were sent against beryllium. Finally James Chadwick showed in 1932 that in the experiment done by Becker and Bothe really did involve the emission of uncharged particles with mass very close to that of the proton. These particles became known as neutrons. Thus by adding a suitable number of these neutrons to the protons in the nucleus, the atomic masses could be explained (Chadwick, 1935).

Very quickly, Werner Heisenberg noticed the close similarity between the proton and neutron. He supposed that they were two states of the same particle, the nucleon. A natural formalism was copied from that describing spin - ½: the proton was described as the 'up' state and the neutron as the 'down' state, where n denotes the neutron and p denotes the proton it can be written as follows:

$$|n\rangle = \begin{pmatrix} 0 \\ 1 \end{pmatrix} = \left| \tfrac{1}{2} - \tfrac{1}{2} \right\rangle \; and \; |p\rangle = \begin{pmatrix} 1 \\ 0 \end{pmatrix} = \left| \tfrac{1}{2} \tfrac{1}{2} \right\rangle.$$

(Boaz & Huerta, 2010, p.6; Marrone, 2007 , p.9)

By itself, this formalism seems to have little content, apart from saying that the masses are the same. In subsequent years the picture was gradually filled in. The conclusion was that two nuclei tend to be extremely close in mass if the number of protons is exchanged with the number of neutrons, suggesting that the short-range p-p interaction and the n-n interaction are the same. Moreover, p-n scattering strength is very close to n-n scattering strength.

In 1936 Cassen and Condon proposed that the short-range nuclear force is actually invariant under the whole continuous group SU(2), so that all the richness of true angular momentum conservation can be carried over to 'isospin'. This was verified experimentally after the Second World War when the pion or π-meson was discovered. The pion virtually carries the nuclear force as the photon virtually carries the electromagnetic force. By identifying the three differently charged pions $\pi^+$, $\pi^0$, $\pi^-$ as the three states of an isospin-1 particle, the conservation of 'isospin-angular momentum' could be verified nontrivially (Boaz & Huerta, 2010, p.8)

### ii. Higher symmetries

The explosion of new particles generated by accelerators in the 1950's led to the introduction of SU(3) discovered independently by Murray Gell-Mann and Yuval Ne'eman around 1962 (Harari, 2006, 72). This was the prelude to even larger groups such as supersymmetry and the Grand Unification Theory groups, SU(5) and SO(10) (Boaz & Huerta, 2010, p32, p42). Since these groups no longer connected with quaternions and are beyond the scope of this dissertation.

### iii. Electroweak isospin

Quaternions enter once more in the unified theory of electromagnetic and weak interactions often called the electroweak interactions merging into the electroweak force. This model was presented in 1967 by Sheldon L. Glashow, Steven Weinberg and Abdus Salam. The model became known as the Glashow-Weinberg-Salam Model winning them as independent discovers of this model the Nobel Prize in 1979 (Xin, 2007, p.1).

Here the weak interaction is described as making transitions between the up and down members of left-handed isodoublets. This is referred to as a U(1) × SU(2) gauge theory. The fermions live in U(1), fermions are particles with ½ spin, and the weak isospin is SU(2). This means that these doublets are subject to the SU(2) 'electroweak gauge group' which is exact but 'spontaneously broken'. This means that these are symmetries by the standards of the laws of physics, but are not symmetries in a vacuum and need extremely high energies to see these symmetries (Boaz & Huerta, 2010, p. 32).

For example, the neutrino is an 'up' particle whose 'down' partner is the left-handed electron. The W is an isospin-1 object that plays a role analogous to that of the pion in strong interactions. These interactions predicted massive particles W and Z bosons and ultimately led to the prediction famous Higgs boson, sometimes referred to as the 'God Particle'. The W and Z bosons were experimentally verified in 1983 and more recently the Higgs boson in 2012. To go any further into this subject would be beyond the scope of this thesis.

### F. Octonions and String Theory

In Chapter III Section E, the discovery of octonions in the 1840's by John Graves and, independently, Arthur Cayley was discussed. Octonians round out the list of division algebras that can be constructed from finite-dimensional real matrices. The whole list is **R** (reals; 1 x 1), **C**(complex; 2 x 2), **Q**(quaternions; 4 x 4), **O**(octonions; 8 x 8).

Over a century later, these four division algebras were found to play a part in the basic construction of the intriguing and controversial branch of Quantum Field Theory known as String Theory. Only an extremely thin and superficial discussion can be given here.

Quantum Gravity was a way to synthesize quantum mechanics with gravity (General relativity). Unfortunately the localization of a particle to a point in space led to uncomfortable infinities. These could be softened by stretching the point out to a tiny string. Just as a point traces out a 1-dimensional trajectory or 'curve' in space-time, so a string traces out a 2-dimensional manifold or 'sheet' (Rovelli, 2008,4)

Often a torus is used to model this, Figure 20, where the length of the string around the torus corresponds to the particle mass (Zimmerman Jones, 2010, p.17,). String theory is deeper and more complex than this analogy, but this gives a basic idea about how string theory works.

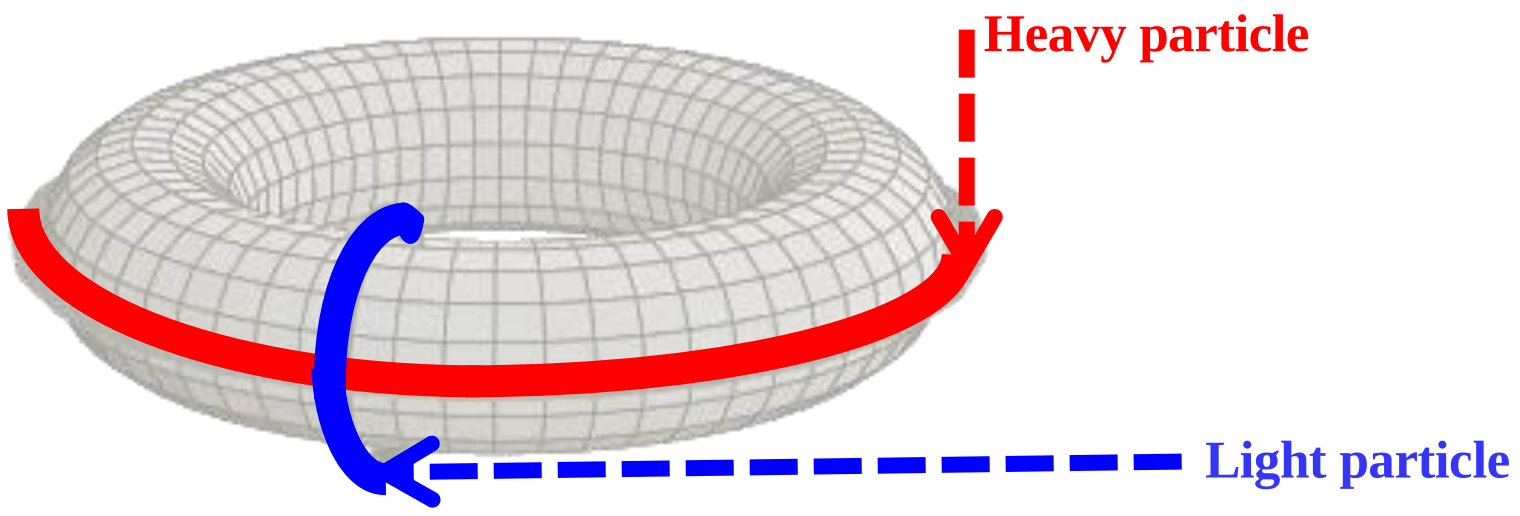

**Figure20:** Two ways a string can wrap around a torus (Adopted from Zimmerman Jones, 2010, p.17)

The rest mass of each kind of particle is to be identified with the frequency of oscillation, sometimes called a vibration, of some mode of deformation of the tiny string. For example if space-time has D-dimensions, then space itself has D - 1, but one of these cannot be used as a deformation because the string is not deformed by sliding it along its own length. Therefore there are D - 2 directions of deformation. This is analogous to the two directions of polarization of light, in our usual space-time where D = 4. For reasons outside this discussion, the amplitude of vibration must be described by one of the four generalizations of real numbers **R, C, Q, O** mentioned earlier. Hence there are only one of the four possibilities: D - 2 = 1; 2; 4; 8 or D = 3; 4; 6; 10. It should also be noted that the possibilities D > 4 are not ruled out because in addition to the 4 dimensions of space-time that present themselves to the physical reality perceived in everyday life there can be additional dimensions that are curled up into tiny finite strings that are not perceived in everyday reality, but large enough to be accommodated by tiny strings.

So far this describes only classical string theory. The process of quantization produces unwanted extra terms in the particle masses. These are known as known as ‘anomalies’. Fortunately there is a classical term of the opposite sign. Since the quantum anomaly receives a contribution from each direction of ‘polarization’, its total value is proportional to D - 2; this raises the possibility of choosing D so that the classical term is exactly cancelled. For strings admitting both bosonic and fermionic vibration modes, the cancellation is exact if D  2 = 8, that is if D = 10. This is the famous 10-dimensional string theory that emerges, and the octonions are the unique division algebra that supports it (Baez, n.d., p.9; Baez & Huerta, 2011, pp. 64-65).

One of the interesting aspects of string theory is how a rather obscure idea such as octonions became so important. So much so that John C. Baez & and John Huerta wrote an article for Scientific American about this calling it *The strangest Numbers in String Theory* (Baez & Huerta, 2011).

Although what has been presented in this chapter is largely superficial with regard to the highly complex and deep physics discussed it is hoped that the contents will show how quaternions and their relatives have found their way into physics. Ideas that were once considered, perhaps, a bit strange during their time of conception have become a major part of modern, and post-modern ideas about how the physical universe works.

# Chapter V

## JURY EVALUATION OF THE SOURCE BOOK

The jury of readers who agreed to judge this source book consisted of six readers who are professors mainly teaching in public urban universities and received their PhD's in applied or pure mathematics from public urban universities. One of these readers did his dissertation on quaternions, but from a purely mathematical prospective. Another is an emeritus physics professor from an urban Ivy League University, who received his PhD from an Ivy

League University physics department. It was with this professor that the author had the deepest and most valuable conversations about this sourcebook and the thesis in general.

The following questionnaire was used as a guide for discussions with the jury members. The readers were initially asked to fill out the questionnaire in writing. It was then followed up by ether an in person or phone interview. The interviews were recorded by the author.

1. Mathematical and scientific accuracy of the monograph.

Here all the readers agreed that the information contained in the sourcebook is accurate, but a few mentioned that more definitions and/or examples were needed in key places and in some places clearer explications were needed in the chapter on the mathematics of quaternions, Chapter III. One reader added that the need for many definitions, "…in order to make the mathematics readable by a large audience".

2. Are there other areas of historical, mathematical or scientific study, which you are aware of that should be included in this monograph?

Here most agreed that given the intended audience the source book is adequate. There was some discussion about some aspects of the physics part that needed to be reworked since initially it was too wordy, and could benefit from a more mathematical exposition. Although the person making this observation believed that the choice of subjects discussed was very good, his concerns were more about details, agreeing that the physics itself was accurate, but could be presented better.

Others made it clear that the choice of topics is one of the strongest aspects of the source book. Rather than a collection of topics, the choice made by the author shows how these topics are connected in one cohesive source. One added in his interview that the source book is, "…a very good source for students to see how the concepts connect."

3. Does the manuscript include appropriate modern applications?

Here again, all agreed that the sourcebook has adequate and appropriate modern applications. As one reader, whose research interests are in group theory, wrote, "Yes. It is certainly a good introduction [with respect to modern applications]". Another added that there are also things that could have been mentioned, for example codes. The author realized that there are a lot of modern applications using quaternions, but, after some reflection thought that details on codes and computer graphics was outside the scope of the source book.

4. Is the connection between the mathematics and the physics explained appropriately for students of mathematics?

For the most part they all agreed that the connection between mathematics and physics was explained appropriately. One reader, a mathematician, particularly liked the section about *Kinematics and Riged Body motion.* He said in his interview, “My more advanced math students, who are also interested in physics, would find the sourcebook’s [Tait’s] use of quaternions in formulating a 3 dimensional version of polar coordinates readable and full of insights.”

5. Is the connection between the mathematics and the physics explained appropriately for students of physics?

Most of the professors questioned are professors of mathematics and received their PhDs from mathematics departments. None seemed to indicate that the mathematics was beyond advanced undergraduate or graduate students of physics, as far as they knew. One mentioned in his interview that “Students of physics should be exposed to the level of math presented.” A professor of physics at an Ivy League University said the physics is, in general, ‘common knowledge’ to most physics students, except for the particulars about Tait’s quaternion applications.

A professor who worked with undergraduates in Mount Holyoke College and Medgar Evers, a four year college in City University of New York (CUNY), felt that the level of the physics was appropriate for the advanced undergraduate physics students that he encountered while teaching mathematics courses in these schools.

6. How appropriate is the manuscript as an ancillary monograph for undergraduate and graduate instructors?

All agreed that it is appropriate, one of the readers wrote, “The manuscript will make a wonderful addition as an ancillary resource for graduate and undergraduate instructors.” Another mentioned in an interview that “The historical part is very useful. Physicists and mathematicians will not encounter how [these subjects] historically were done.” Another mentioned in an interview that this was the strongest part of the source book adding that “There is very little in terms of work like this for instructors” For example, he added that “if someone is teaching an undergraduate course in mathematics …[students] who need something robust”. He then discussed the relationship between mathematics and its history adding that “… the historical detail and some of the other things could be of use [in a course]. This is a really nice resource for instructors in this regard.” He mentioned later in the interview that an instructor “…can parse in ways that may be appropriate for students.”

7. How appropriate is the manuscript as an ancillary monograph for undergraduate and graduate students?

The consensus was that the sourcebook is fine for graduate students, but, depending on the background of the undergraduate, could be challenging. As one of the readers put it, “…in a few sections undergrads might need some guidance to follow all of the ideas.” Another mentioned in an interview that the source book is, “…a good source for students, historically speaking.” Another reader said that this is very good for a graduate student who “are interested and motivated by this kind of reading", but would be more challenging for an undergraduate student. He also added that it could be "…quite useful as a summer research project for an advanced undergraduate student …and extremely useful for graduate students…on issues related to quaternions."

8. How (if any) use would this manuscript be to your department or in other departments or institutions that you know of?

Most of the professors interviewed teach in urban public colleges that are a part of a larger urban public university. The colleges that they are affiliated with do not teach many graduate or advanced undergraduate courses. Most mentioned that it would be good supplementary material for a modern abstract algebra or linear algebra course.

One of the readers mentioned the source book could be used as part of a mathematics education course in The University of Denver where he is presently teaching. He mentioned that it would be useful for this course for its historical content. He elaborated on this in a later interview "…for example a course I am teaching this winter [2015, quarter system] on the use of history, philosophy in mathematics education. It is clear that to elect this [course], it [the source book] really integrates mathematics in its historic context. This is of vital importance in this kind of course" He then started to discuss about the relationship between mathematics and its history saying that "…if one really wants to understand mathematics it is good to know how these questions emerged, in historical context… it [the source book] would be useful in this regard.” He continued, thinking outside his immediate department, he has, over the years been involved in a number of undergraduate summer research projects in various schools and said that the source book would “… serve as a really good resource as an undergraduate student research project on quaternions or more probably quaternion algebras, this would serve as a wonderful resource in that direction…it [the source book] allows for many places of exploration.” He summarized saying that he can see this source book being used for a “…course for a mathematics, period, [pure mathematics] and I can think of a number of variations in between [pure mathematics and physics].” He saw the source book as useful for both undergraduate mathematics and graduate mathematics, but stressed that it can also useful in mathematics education.

9. Based on your experience at the undergraduate and/or graduate level, how appropriate is the material in the monograph for students who may be interested in the subject matter at yours or any institution you are familiar with?

Here most said that the material is accessible, but weaker students may need some guidance in order to properly understand the material. One of the readers suggested that it would make an excellent source for summer reading for a student after finishing their undergraduate degree, before entering graduate school, in mathematics or physics. This person mentioned in an interview that this source book "…can be used to communicate interesting ideas in mathematics" for certain students in mathematics education. He also remarked that there is too little of this type of source book done in mathematics education and the author of this source book is doing something of value not only for mathematics and physics, but also for mathematics education.

The author reworked the source book according to the suggestions made by the readers. After some reworking of the physics part of the sourcebook by the author it was agreed that it would be of interest to physics students who are interested in the history of quaternions in their respected subject. It was also agreed that there was a need for this type of source book to be available for advanced physics and mathematics students and instructors in order to get a sense of the 'bigger picture' of the historical connection with mathematics of quaternions.

# Chapter VI
# SUMMARY, CONCLUSIONS, AND RECOMMENDATIONS

This thesis has focused mainly on historical framework by which quaternions have evolved over some 150 years. Here the author has only given very brief discussions on how quaternions and their relatives have reemerged in recent years, perhaps in new clothing, but conceptually remaining intact.

Quaternions have been making a substantial come back after a long absence. The reemergence of quaternions has not only been in physics, but have found themselves all over the place in such diverse areas from computer animation to aviation. In this thesis the author has tried to explore some of the reasons why and how this has happened, focusing on theoretical physics.

The main focus of this thesis is to provide a sourcebook for a topic that is deeply embedded in the modern theoretical physics consciousness, yet rarely explored beyond its numerous applications. It is not the goal of this thesis to come up with new ideas about how quaternions used in theoretical physics or how to apply them beyond known techniques, but merely give an exposition of how they evolved and became embedded in today's physics.

## A. Summary

This thesis opened with the conception of quaternions by Hamilton and their romantic discovery during a walk on Bloom Bridge on October 16, 1843. The central focus of this thesis has been historical, specifically related how quaternions and their associated rotation groups have been applied to theoretical physics.

Tait was the first mathematician to develop quaternions seriously in physics. His influence opened the doors to almost a cult following of this subject. Tait also influenced Maxwell and the formulations of his famous equations that students still encounter today. Whether Maxwell would have used quaternions in his electromagnetic theory without Tait's influence is hard to say; but this particular application of quaternions did influence the development of vector analysis by Gibbs and Heaviside.

Quaternion concepts were part of the zeitgeist of Grassmann, who was outside the academic establishment, developed exterior algebras. Grassmann's work was an inspiration to a number of mathematicians and scientists including Gibbs. Ultimately it was Clifford who developed geometric algebras by incorporating the best of quaternions and Grassmann's exterior algebras. Although Grassmann was not recognized during his lifetime his work ultimately made a substantial mark on the development of vector analysis.

After their 'cult phase' quaternions gave way to vector analysis around the mid-1880s. Vectors by this time was essentially a 'watered down' version of quaternions, as Feynman's quote (Feynman, 1999, pp. 200-201) implies. The 'vector' or 'pure quaternion' part was useful in physics where the primary focus was on classical problems. Matrices and vector analysis proved to be the right tools for these applications. This pushed quaternions into obscurity into 20th century, (Stephenson, 1966). This was fine until the advent of relativity and quantum mechanics. Here quaternions were rediscovered by Wolfgang Pauli in order to solve problems about spin in quantum mechanics. GUT's and associated theories opened up an explosion of quaternion type rotation groups giving theoretical physicists powerful tools into the 21st century.

This thesis demonstrates how mathematics that was developed in the 19th century has become relevant to today's thinking in theoretical physics. It is for this reason that Chapter's II and III are devoted to the mathematical historical development during the 19th century. These two chapters were meant to provide the mathematical backdrop for Chapter IV. Chapter IV was devoted to the applications of quaternions and their associated rotation groups and algebras to theoretical physics starting with Tait's *Elementary Treatise of Quaternions*. The chapter continues to follow quaternions and their relatives into their recent explosion as an integral part of modern theoretical physics research.

### B. Conclusions

As discussed the goal of this thesis has been to draft a source book that would show how mathematical and physical ideas can be synthesized into powerful theories about how our universe works. Unfortunately the author of this thesis was unable to give more complete discussions of all the topics discussed and some topics were discussed more deeply than others as a result. Thus this thesis became a compilation of many works by other authors of varying disciplines and backgrounds within the history of mathematics, mathematics and physics. This would include those who have done, and still do, research or study these subjects for various reasons. In this sense this thesis was a success: the subjects discussed were varied and some conceptually and technically complicated yet a thread could be drawn to connect them all within a coherent historical and mathematical framework that could give a beginner a ‘bigger picture’ about how all these subjects fit together.

Educational programs have been developed to incorporate quaternions into the mathematics and physics curriculums since their beginning. One of the issues that have come up is where to place quaternions within the mathematics or physics curriculum (Cajorie, 1890, p.293). For example, should they be placed in Geometry, Modern Algebra, or Linear Algebra courses, to name a few possibilities?

A few traditional introductory textbooks do introduce quaternions. For example, Serge Lang’s in *Liner Algebra* introduces them in the section on *Geometric interpretation in dimension 3* (Lang, 1987 , p.287). He also introduces quaternions in his *Algebra* as an example of a group that is generated by 2 elements (Lang, 2002, p.9) and later on develops them in the form of exercises (Lang, 2002 ,p.545,p.723,p.758), similarly in Michael Artin , Algebra (Artin, 1991, p.306), Birkhoff & Mac Lane, *A Survey Of Modern Algebra-fourth edition* (Birkhoff & Mac Lane, 1977 , p.258) to name a few.

In Physics, Herbert Goldstein in *Classical Mechanics 2nd edition* discussed them in the text (Goldstein, 1980, 156). By the 3rd edition a section called

*Quaternion Group* was added to the appendix (Goldstein, Poole & Safko, 2000, 610). Other, more specialized books in mathematical methods in physics also include quaternions. For example George B. Arfken and Hans J. Weber *Mathematical Methods For Physicists Sixth Edition,* the section on *Symmetry Properties* (Arfken & Webber, 2005, pp.203-205), and in the form of spinners and Clifford algebras in the aforementioned book (Arfken & Weber, 2005, p.212).

Due to the recent interest in quaternions they have become part of certain curriculums on a need to know basis. For example if a class in computer animation needs to have students program using quaternions, it is placed in that courses curriculum. As for physics, quaternions are usually introduced in the form of Pauli matrices in an introduction to quantum mechanics course, but often the textbook and even the professor will not recognize these matrices as being a representation of the quaternions).

## C. Recommendations

The purpose of this thesis is to allow students, educators and instructors one cohesive source book to start them on their historical, mathematical or physical journey. As discussed, in the purpose of this study in chapter 1, the source book provides a historical and mathematical backdrop to the development of modern vector analysis in classical physics and the electromagnetic theory of Maxwell. It was discussed that Maxwell's incorporation of quaternions in his electromagnetic theory was the inspiration for the development of Vector analysis by Gibbs and Heaviside. Before quaternions there weren't vectors, or anything like it. Any student who studies classical physics will encounter vector analysis, but never have the notion that the mathematics that they are using in physics problems was inspired by the use of quaternions; even if it is a conceptually 'watered down' version of quaternions to make them more useful for calculations.

Quaternions would not be left in the positions of being just a 'shadow of itself' through vector analysis, but remerged in full force in the 20$^{th}$ century and the advent of quantum mechanics. Modern physics, as it turned out, needed the full power of quaternions to describe its nuances. Thus quaternions and their cousins and other relatives have reemerged in full force all over the landscape of today's theoretical physics, as this source book discussed. Again most students who study quantum mechanics are unaware that the matrices that they use in calculations are connected to the mid-19$^{th}$ century discovery of quaternions. It is for this reason that this source book was written, in order to fill in these gaps, so that a student and instructor can get a sense of the 'bigger picture' with regard to the subject matter they are studying or teaching. In this way this thesis as a source book was a success, and the readers as the 'jury' of this thesis as a source book, in general, also agreed.

This thesis also provides an extensive bibliography. Many of these sources can be retrieved on line, whereby the student or instructor, can use these sources as a starting point to learn more about the topics discussed. This is considered an important part of this thesis since, as my readers demonstrated, that students and instructors from varied interests and backgrounds, outside of theoretical physics, are interested in the subject of quaternions and its history. This would include group theorists, differential geometers, and applied mathematicians to name a few. Thus by having a vast and varied bibliography a reader can pick and choose what works are appropriate for them and their background. This thesis was not meant to be a textbook on quaternions and their relatives, but a resource that could help students and instructors choose texts and papers appropriate to the subject matter and level.

Chapter V was a summary of discussions and interviews that the author had with the readers of the source book. All the readers are professionals in their fields and had some interest or knowledge about quaternions and/or their associated subjects. The author of this thesis had them read, answer a questionnaire and/or comment on the source book part of this thesis. Based on what the readers wrote and said the author rewrote, added or deleted parts based on their comments, and the authors reflections on the discussions with the readers. For the most part the structure of the source book has remained the same. The changes were made primarily in details, presentation and/or the addition of definitions and examples.